\newif\ifpdflatex    
\pdflatextrue           

\documentclass[fleqn,usenatbib,useAMS,twocolumn]{aastex}
\makeatletter
\@ifundefined{NR@label@copy}{}{\let\label\NR@label@copy\let\ltx@label\label}
\makeatother
\usepackage{silence}
\usepackage{amsmath,amssymb,amsbsy,esint}
\usepackage{graphicx}
\usepackage{xurl}
\usepackage{bm}
\usepackage{xspace}
\usepackage{color}
\usepackage{booktabs}
\usepackage[flushleft]{threeparttable}
\usepackage[T1]{fontenc}
\usepackage{ae,aecompl}
\usepackage{upgreek}
\hypersetup{colorlinks=true, linkcolor=blue, citecolor=blue, urlcolor=blue}

\usepackage{afterpage}  

\def\lesssim{\mathrel{\hbox{\rlap{\hbox{\lower5pt\hbox{$\sim$}}}\hbox{$<$}}}}
\def\gtrsim{\mathrel{\hbox{\rlap{\hbox{\lower5pt\hbox{$\sim$}}}\hbox{$>$}}}}

\def\til{\raise.17ex\hbox{$\scriptstyle\mathtt{\sim}$}}

\definecolor{coral}{rgb}{1.0, 0.498, 0.314}

\newcommand{\micronmath}{\mu\mathrm{m}}      
\newcommand{\jwstpipeline}{\texttt{jwst}\xspace} 
\newcommand{\nrsg}{101{,}219 }
\def\apjl{ApJ}%
\def\apjs{ApJS}%
\def\aap{A\&A}%
\shorttitle{EMBERS: RSG Catalog and Physical Properties}
\shortauthors{}

\begin{document}
\title{Extragalactic Multi-Band Exploration of Red Supergiants (EMBERS): Pipeline, Source Classification, and Bolometric Properties of \nrsg Red Supergiants from {\it James Webb Space Telescope} Imaging}
\def\ciera{Center for Interdisciplinary Exploration and Research in Astrophysics (CIERA), Northwestern University, Evanston, IL 60201, USA}
\def\northwestern{Department of Physics and Astronomy, Northwestern University, Evanston, IL 60201, USA}

\author[0009-0005-8230-030X]{Aswin~Suresh}
\affil{\northwestern}
\affil{\ciera}

\author[0000-0002-5740-7747]{Charles~D.~Kilpatrick}
\affil{\ciera}

\author[0000-0002-7374-935X]{Wen-fai~Fong}
\affil{\northwestern}
\affil{\ciera}

\author[0000-0001-7081-0082]{Maria~R.~Drout}
\affil{David A. Dunlap Department of Astronomy \& Astrophysics, University of Toronto, 50 St. George St., Toronto, ON M5S 3H4, Canada}

\author[0000-0002-1125-9187]{Daichi~Hiramatsu}
\affil{Department of Astronomy, University of Florida, Bryant Space Science Center, Gainesville, FL 32611-2055, USA}

\author[0000-0001-9695-8472]{Luca~Izzo}
\affil{INAF, Osservatorio Astronomico di Capodimonte, Salita Moiariello 16,
I-80121 Naples, Italy}

\author[0000-0002-3934-2644]{Wynn~V.~Jacobson-Gal\'{a}n}
\altaffiliation{NASA Hubble Fellow}
\affil{Cahill Center for Astrophysics, California Institute of Technology, MC 249-17, 1216 E California Boulevard, Pasadena, CA, 91125, USA}

\author[0000-0002-4410-5387]{Armin~Rest}
\affil{Space Telescope Science Institute, 3700 San Martin Drive, Baltimore, MD 21218, USA}
\affil{Physics and Astronomy Department, Johns Hopkins University, Baltimore, MD 21218, USA}

 \author[0000-0002-1481-4676]{Samaporn~Tinyanont}
\affil{National Astronomical Research Institute of Thailand, 260 Moo 4, Donkaew, Maerim, Chiang Mai, 50180, Thailand}

\correspondingauthor{Aswin~Suresh}
\email{AswinSuresh2029@u.northwestern.edu}
\shortauthors{Suresh et al.}

\begin{abstract}
The evolution of red supergiants (RSGs), especially in their final stages before exploding as hydrogen-rich Type II supernovae (SNe), remains poorly understood, owing to small samples of well-characterized stars in the Local Group. We present the largest population study of RSGs to date using $\sim10^5$ stars with bolometric parameters inferred from archival {\it James Webb Space Telescope}/Near-Infrared Camera (NIRCam) imaging of 15 galaxies within 20\,Mpc. We also present {\tt jwst123}, a newly-developed general-purpose NIRCam reduction and photometry pipeline, and its application to our set of 15 galaxies. We build a simulation-based parameter inference framework to rapidly fit multi-band photometry of $\sim1.5\,$million luminous stars in milliseconds per source. Using clustering in the effective temperature ($T_{\rm eff}$)--luminosity ($\log L/L_{\odot}$)--dust optical depth ($\tau_V$) phase space, we introduce a novel method to select RSGs with high completeness and purity. Leveraging the scale of our catalog, we find that the dustiest RSGs prefer metal-rich hosts, and RSG luminosities correlate with host specific star formation rate, suggesting that environment shapes RSG evolution. While most selected RSGs agree with theoretical stellar evolution tracks, we identify a new population of $1{,}669$ RSGs that are heavily dust-enshrouded ($\tau_V>4$), overluminous ($\log(L/L_{\odot})>5.55$), or anomalously cool ($T_{\rm eff}<3000\,$K), including analogs of known SN progenitors. Our catalog provides a population-scale baseline for understanding massive stellar evolution and connecting field RSGs to dusty progenitors of Type~II~SNe.
\end{abstract}

\date{Last updated \today; in original form \today}

\section{Introduction}
\label{sec:intro}

Red supergiants (RSGs) represent core-helium burning and later advanced-burning phases in massive stars (zero age main sequence mass $M_{\rm ZAMS} \gtrsim 7\,M_{\odot})$. RSGs dominate the near-infrared (NIR) light of young stellar populations and are the most likely immediate progenitors of ordinary hydrogen-rich Type II-P supernovae \citep[SNe;][]{Gazak2013,Smartt09,Kilpatrick2018,Rui2019,VanDyk2019,VanDyk2025}. RSGs are characterized by low effective temperatures (\til3500--4500\,K), high bolometric luminosities $(\gtrsim 10^{3.5}\,L_{\odot})$, and low surface gravity envelopes. The RSG phase lasts only $\sim$0.1–1 Myr, short relative to the lifetime of a massive star \citep{Meynet2015}.  With over two dozen direct detections of RSG progenitors to Type II-P SNe from archival pre-explosion imaging, RSGs represent a unique class of well-studied SN progenitors and serve as an important testbed for terminal-stage stellar evolution \citep{Smartt2015, VanDyk2025}. 

A major open problem from observations of SNe is elucidating the evolution of massive stars in their final years-to-centuries, characterized by erratic variability and episodic mass loss, while quantifying the impact of binary companions and/or the host environment on these phenomena \citep{Neugent2020,Jacobson-Galan22, Kilpatrick23, deWit2024, deWit25}. How a star evolves and loses mass as an RSG has dramatic implications for the subtype of SN produced, the nature of the resulting compact object, and cosmic dust production \citep{Zapartas2025, Merritt2025}. A significant number of Type II-P SNe show interaction signatures with confined, dusty circumstellar material (CSM) formed through mass loss from the RSG progenitor \citep{Khazov2016, Bruch2021, JacobsonGalan2024b, JacobsonGalan2025}. Evidence from pre-explosion imaging of the progenitor star in recent CSM-interacting Type II-P SNe such as SN\,2023ixf, SN\,2024ggi, and SN\,2025pht points to elevated mass-loss episodes in the final decade before core collapse, creating confined CSM \citep{JacobsonGalan2023, Kilpatrick23, JacobsonGalan2024, Xiang24, Kilpatrick25}. Elevated mass loss $(\dot{M}\gtrsim10^{-3}\,M_{\odot}\,{\rm yr}^{-1})$ is inferred to occur in $M_{\rm ZAMS}\approx10$--$12M_{\odot}$ progenitors \citep{JacobsonGalan2025}, potentially driven by different mechanisms from those that operate in the broader RSG population, which experiences relatively weak mass loss $(\dot{M}\approx10^{-6}\,M_{\odot}\,{\rm yr}^{-1})$. Late-time X-ray, UV, and radio studies of old Type II SNe probe carbon core burning in the progenitor and find mass-loss rates that span two orders of magnitude before explosion \citep{Dwarkadas2014, Nayana2025, JacobsonGalan2025, Ferdinand2026}.

Serendipitous pre-explosion observations of SN progenitors also reveal another surprising trend: RSGs with luminosities $\gtrsim10^{5.2}\,L_{\odot}$, or equivalently $M_{\rm ZAMS} \gtrsim 17\,M_{\odot}$, apparently do not explode as Type II-P SNe, despite representing $\sim$15\% of RSGs under a Salpeter initial mass function \citep[IMF;][]{Smartt09, Smartt2015}. This is termed the RSG problem. One proposed explanation invokes observational biases due to a lack of archival IR imaging, which leads to underestimates of the dust content, luminosities, and initial masses of RSGs \citep{Beasor2025, Strotjohann24, Kilpatrick25}. NIR-MIR pre-explosion imaging of progenitors is therefore crucial to accurately characterize dusty RSGs \citep{Kilpatrick25}. Another explanation involves the direct collapse of massive RSGs into stellar-mass black holes, an event that manifests as a low-luminosity infrared transient rather than a SN \citep{Kochanek2014, Smartt2015, Sukhbold2020, Antoni2023, Tsuna2025, De2026}. While the significance of the RSG problem is below $2\sigma$, based on bias-corrected progenitor luminosity functions and comparisons to field RSGs, a high-mass RSG progenitor has yet to be detected \citep{Strotjohann24, Beasor2025}. A promising approach is to connect the field evolution and dust production of RSGs, as seen in resolved NIR populations, to inferences from their eventual SNe. 

Several studies in recent years have investigated the RSG populations of the Local Volume, including the Milky Way, the Small and Large Magellanic Clouds (SMC/LMC), M31, M33, and a handful of other galaxies \citep{Massey2009, Boyer11, Neugent20, Ren21, Wang2021, Massey2021, Sarbadhicary25, Bonanos2025, Li2025}. Conventionally, RSGs are identified in the NIR $J-K$ versus $K$ color-magnitude diagram (CMD) with linear cuts that isolate the brightest red stars forming a nearly vertical sequence, but blueward of the cooler Asymptotic Giant Branch (AGB) populations, which tend to be among the dustiest stars in a galaxy \citep{Massey2021}. Such selection techniques applied to M31 and M33 have yielded complete RSG catalogs down to $L\gtrsim10^4L_{\odot}$, below which contamination from AGB stars is difficult to remove \citep{Massey2021, Ren2021, Wang2021}. This AGB contamination can bias the RSG luminosity function and mass loss relations. Selection criteria have also been developed using {\it Spitzer Space Telescope} and {\it WISE} mid-infrared (MIR) data $(3.6\,{\rm\mu m},4.5\,{\rm\mu m})$, combined with optical data, to identify evolved dusty RSGs, some of which have undergone episodic mass loss \citep{Messineo2012, Britavskiy2015, Bonanos2024, deWit2024, deWit25}. Such dusty RSGs are of particular interest as stars in an unstable late stage immediately before core collapse. A primary drawback of using CMD selection criteria is quantifying contamination from AGB stars, which can be as high as $30\%$ at low luminosities ($\log(L/L_{\odot}) < 4.2$; \citealt{Boyer11}). Further, estimating physical parameters of RSGs from the CMD is empirical and poorly calibrated \citep{Levesque06,Massey2009}. Bolometric corrections for RSGs are large ($-4$ to $-1\,\text{mag}$) and temperature-dependent such that a $10\%$ error in $T_{\rm eff}$ leads to $0.3\,{\rm dex}$ differences in $\log(L/L_{\odot})$ derived from optical photometry \citep{Levesque06, Davies13}. A robust analysis of RSG evolution can instead be done by fitting their photometric Spectral Energy Distributions (SEDs) to infer their bolometric properties \citep{Beasor16, Beasor22, Wang23, Yang2023}. SED fitting will be fruitful in the NIR, where RSGs emit the bulk of their flux.

The {\it James Webb Space Telescope} ({\it JWST}) was designed in part for high-sensitivity imaging and spectroscopy in the IR \citep{Gardner06,Rigby23}, and NIRCam now delivers resolved stellar photometry in nearby galaxies at unprecedented depth \citep{Rieke23,Weisz24,Williams24,Correnti25}. \citet{Boyer2024} demonstrated the capabilities of deep {\it JWST} imaging by analyzing the stellar populations in the Wolf–Lundmark–Melotte galaxy and identified a significant, previously inaccessible, population of evolved, dusty stars, including RSGs. \citet{Sarbadhicary25} conducted a larger-scale analysis on 19 galaxies within 20\,Mpc, combining {\it JWST} and {\it Hubble Space Telescope} ({\it HST}) imaging, to identify $\sim97,000$ RSGs across these galaxies using the {\it HST} F814W and {\it JWST} F200W filters with CMD selection cuts. While physical parameters (e.g., bolometric luminosity; $L_{\rm bol}$, effective temperature; $T_{\rm eff}$) of RSGs are unavailable at this scale to date, these studies highlight improvements in crowding and sensitivity by {\it JWST} compared to existing facilities. {\it JWST} is also ideally suited to characterize the dustiest RSGs, such as the recently identified luminous RSG progenitor of SN\,2025pht, which are difficult to detect in the optical but are IR-bright \citep{Kilpatrick25, Beasor22, Verhoelst09}.

In this paper, we develop a uniform reduction and SED-modeling framework for archival NIRCam data, with the goal of identifying RSGs and constraining their physical properties across a multi-galaxy sample. Section \ref{sec:sample} presents the list of star-forming galaxies with extensive imaging by {\it JWST}, from which we identify RSGs. In Section \ref{sec:reduction}, we present {\tt jwst123}, a general-purpose reduction software for performing high-precision resolved stellar photometry on {\it JWST} imaging with {\tt DOLPHOT} following image alignment and mosaicking. In Section \ref{sec:models}, we develop radiative-transfer SED models for cool, luminous stars enshrouded by dust, such as RSGs. We present a scalable SED-fitting framework enabled by neural network posterior inference in Section \ref{sec:sbi}, applied to rapidly infer the physical properties of millions of luminous stars identified in our photometry. We present novel criteria for classification of stellar populations using the inferred physical parameters and our multi-galaxy RSG catalog in Section \ref{sec:cat}. In Section \ref{sec:discussion}, we present the first science results from our catalog, examining the distribution of RSG physical properties, their dependence on metallicity and star formation, and identifying a subsample of heavily dust-obscured RSGs. We summarize and conclude in Section \ref{sec:conclusion}. Future work will explore MIR and optical-UV datasets from {\it JWST}, {\it HST}, and the {\it Nancy Grace Roman Space Telescope (Roman)} to characterize these RSGs in more detail.

\section{Galaxy Sample Selection}
\label{sec:sample}
\afterpage{%
\startlongtable
\begin{deluxetable*}{cccccccc}
\tabletypesize{\small}
\tablecaption{Summary of galaxies analyzed, adopted properties and data attributes \label{tab:galaxy_sample}}
\tablehead{
\colhead{Galaxy} & \colhead{R.A.} & \colhead{Dec.} & \colhead{Distance} & \colhead{Metallicity} & \colhead{$\log_{10}{\rm SFR}^{\rm a}$} & \colhead{$N_{\rm filters}$} & \colhead{{\it JWST} Program IDs}\\
\nocolhead{} & \colhead{(h:m:s J2000)} & \colhead{(d:m:s J2000)} & \colhead{(Mpc)} & \colhead{($Z / Z_{\odot} $)} & \colhead{($ \log_{10}\,M_{\odot}/{\rm yr} $)} & \nocolhead{} & \colhead{(All GO)}
}
\startdata
NGC\,628 & 01:36:41.7936 & $+$15:47:01.2840 & $8.63 \pm 1.63^{\rm b}$ & $0.62 \pm 0.01^{\rm c}$ & $0.23$ & 8 & 1783,\,2107,\,2211,\,3990 \\
NGC\,1365 & 03:33:36.4080 & $-$36:08:24.6840 & $18.28 \pm 0.51^{\rm d}$ & $0.73 \pm 0.02^{\rm c}$ & $1.15$ & 8 & 1995,\,2107,\,5398 \\
NGC\,1637 & 04:41:28.1900 & $-$02:51:28.5001 & $12.02 \pm 0.39^{\rm e}$ & $1.29 \pm 0.7^{\rm f}$ & $-0.37$ & 8 & 3707,\,4793 \\
NGC\,3034 & 09:55:52.4299 & $+$69:40:46.9308 & $3.70 \pm 0.19^{\rm g}$ & $\til1$--$1.3^{\rm h}$ & $0.85$ & 8 & 1701,\,5145 \\
NGC\,4038 & 12:01:53.0136 & $-$18:52:03.4320 & $18.11 \pm 0.92^{\rm d}$ & $1.02 \pm 0.05^{\rm h}$ & $1.03$ & 6 & 1995,\,2581 \\
NGC\,4258 & 12:18:57.5040 & $+$47:18:14.2920 & $6.85 \pm 0.73^{\rm i}$ & $0.63 \pm 0.13^{\rm j}$ & $-0.03$ & 14 & 1685,\,1995,\,2080,\,2875 \\
NGC\,4449 & 12:28:11.1240 & $+$44:05:37.2120 & $4.02 \pm 0.59^{\rm b}$ & $0.34 \pm 0.03^{\rm c}$ & $-0.37$ & 6 & 1783 \\
NGC\,4485 & 12:30:30.9648 & $+$41:42:01.4040 & $8.75 \pm 1.41^{\rm b}$ & $0.25 \pm 0.03^{\rm k}$ & $-0.86$--$0.23$ & 6 & 1783 \\
NGC\,4536 & 12:34:27.0672 & $+$02:11:17.6640 & $15.21 \pm 0.35^{\rm d}$ & $0.62 \pm 0.13^{\rm l}$ & $0.47$ & 6 & 1995,\,3707 \\
NGC\,4548 & 12:35:26.4446 & $+$14:29:46.7591 & $15.00 \pm 0.35^{\rm m}$ & $\til1.45^{\rm n}$ & $-0.28$ & 8 & 3707,\,4793 \\
NGC\,5194 & 13:29:52.7112 & $+$47:11:42.7560 & $7.21 \pm 1.10^{\rm b}$ & $0.93 \pm 0.21^{\rm o}$ & $0.65$ & 17 & 1783,\,3435,\,3990 \\
NGC\,5236 & 13:37:00.9504 & $-$29:51:55.5120 & $4.61 \pm 0.28^{\rm e}$ & $0.74 \pm 0.01^{\rm c}$ & $0.62$ & 6 & 1783 \\
NGC\,5457 & 14:03:12.5448 & $+$54:20:56.2200 & $6.73 \pm 0.15^{\rm d}$ & $0.55 \pm 0.01^{\rm c}$ & $0.54$ & 12 & 1995,\,2452,\,3429,\,4087,\,5398 \\
NGC\,5643 & 14:32:40.7112 & $-$44:10:27.9480 & $12.47 \pm 0.44^{\rm p}$ & $0.62 \pm 0.01^{\rm q}$ & $0.33$ & 11 & 1685,\,1995,\,3707,\,4793 \\
NGC\,7320 & 22:36:03.3792 & $+$33:56:53.1960 & $18.11 \pm 4.50^{\rm r}$ & $0.49 \pm 0.11^{\rm h}$ & $-0.94^{\rm s}$ & 6 & 2732 \\
\enddata
\tablecomments{
The interacting galaxy pair NGC\,4485 and NGC\,4490 are both included in the NGC\,4485 field. The SFRs of -0.86 and 0.23 correspond to NGC\,4485 and NGC\,4490 respectively.\\
{\rm (a)} We use the uniformly derived SFRs from \citet{Leroy2019} for our galaxy sample, and adopt an uncertainty of $\pm 0.20$ as quantified in Table 4 of \citet{Leroy2019}. \\
{\it Distance and Metallicity References:}
{\rm (b)} \citet{Sabbi2018};
{\rm (c)} \citet{Galliano2021};
{\rm (d)} \citet{Riess2016};
{\rm (e)} \citet{Saha2006};
{\rm (f)} \citet{Wong2013};
{\rm (g)} \citet{Wagner2026};
{\rm (h)} \citet{DeVis2019};
{\rm (i)} \citet{Hoffmann2015};
{\rm (j)} \citet{Kudritzki2013};
{\rm (k)} \citet{Esposito2013};
{\rm (l)} \citet{Moreno-Raya2016};
{\rm (m)} \citet{Freedman2001};
{\rm (n)} \citet{Saha2006}, \citet{Sakai2004}; We note that a larger derived value of \til4.4 exists in \citet{Freedman2001}, but adopt the more recent \citet{Saha2006} measurement;
{\rm (o)} \citet{VanDyk2011};
{\rm (p)} \citet{Hoyt2021};
{\rm (q)} \citet{Pan2020};
{\rm (r)} \citet{Tully2016};
{\rm (s)} It is unclear whether this value applies specifically to the foreground galaxy of Stephan's quintet we analyze here.
}
\end{deluxetable*}
}

We use imaging data from the Near Infrared Camera (NIRCam) onboard {\it JWST}, spanning $0.6\,{\rm\mu m}$--$5.0\,{\rm\mu m}$. NIRCam performance and calibration context are described by \citet{Rieke23} and \citet{Rigby23}, and resolved-star programs (i.e., \citealt{Weisz24,Williams24,Correnti25}) illustrate the scientific reach of this mode. NIRCam uses a dichroic with a split at $2.3\,{\rm \mu m}$, delivering spatial resolution of $\til0.06''$ at the short wavelengths (SW) and $0.15''$ at the long wavelengths (LW; \citealt{Rieke23}). We use F200W as a benchmark for SW resolution and F444W for LW resolution, since these are the limiting values. The SW detectors enable a resolution of \til6\,pc or better at galaxies within \til20\,Mpc. The RSG SED peaks in the NIR at $\til 1.5$--$2\,\micronmath$ (Figure \ref{fig:rsg_sed}), making NIRCam ideally suited to capture the dominant flux from these sources. 

To leverage NIRCam's capabilities for resolved RSG studies, we curate a sample of nearby star-forming galaxies with archival observations meeting specific requirements on filter coverage. We assemble a set of galaxies that (1) are at distance $\lesssim20\,{\rm Mpc}$, (2) are star forming, with the star formation rate (SFR) $\gtrsim0.1\,{\rm M_{\odot}\,{\rm yr}^{-1}}$, (3) contain publicly available NIRCam observations before March 2026, including at least four unique filters for well-sampled stellar SEDs, (4) include F115W or bluer filters to constrain the Wien tail of RSG SEDs, and F335M or redder filters to characterize circumstellar dust emission, and (5) have NIRCam data covering at least $10\%$ of the on-sky area of the galaxy. We also include NGC\,1637 in our sample, despite its bluest filter being F150W, because this galaxy hosts the Type II SN\,2025pht, whose RSG progenitor was identified in NIRCam and MIRI imaging \citep{Kilpatrick25}. We anticipate a larger uncertainty on the effective temperature and dust optical depth for sources in NGC\,1637 due to the missing bluer filters. We query the Mikulski Archive for Space Telescopes (MAST)\footnote{\url{https://mast.stsci.edu}} for NIRCam images of all galaxies satisfying the above criteria, and NGC\,1637. We choose a final sample of 15 galaxies, with the interacting galaxy pair NGC\,4485/4490 counted as one. We summarize the galaxy sample and their nominal properties in Table \ref{tab:galaxy_sample}.

Our galaxy sample spans a wide range in host properties such as metallicity and SFR. Using global averages from the literature, we estimate the metallicity ranges from $Z/Z_{\odot}\sim0.25$--$1.45$. Metallicity estimates for NGC\,3034 and NGC\,4548 are uncertain, and we adopt a value of $Z/Z_\odot\til1.0$ for NGC\,3034 based on \citet{RodriguezMerino2011} and \citet{DeVis2019}, and $Z/Z_{\odot}\til1.45$ for NGC\,4548 based on \citet{Saha2006} and \citet{Sakai2004}. This extends the metallicity baseline relative to existing studies on Local Volume RSGs, which are much narrower and sparse in metallicity sampling. We also sample a range of star-forming environments, with SFR$\til\!0.2$--$11.2\,{M_{\odot}\,{\rm yr}^{-1}}$. This includes normal star-forming spirals and irregulars, starburst galaxies, and galaxy mergers, all conducive to producing core-collapse SNe \citep{Pessi2023}. We present the adopted metallicities and SFRs from the literature in Table \ref{tab:galaxy_sample}. 

\section{Data Reduction Pipeline: {\tt jwst123}}
\label{sec:reduction}

\begin{figure}[tp]
    \centering
    \includegraphics[width=0.8\columnwidth]{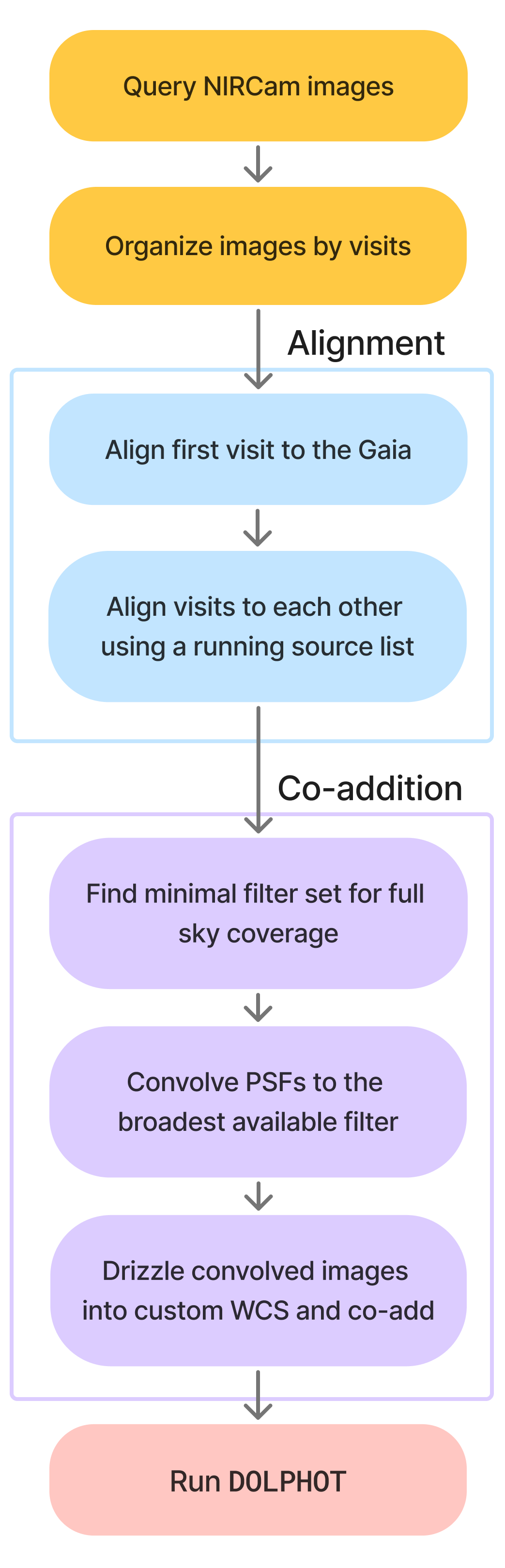}
    \caption{\texttt{jwst123} flowchart. {\tt jwst123} employs an alignment module using {\tt JHAT} to produce sub-pixel-aligned images, and a co-addition module to create a uniform single-filter reference image for {\tt DOLPHOT}. We use {\tt DOLPHOT} to produce high-precision photometry tables from crowded stellar fields.
    }\label{fig:jwst123_flowchart}
\end{figure}

We now describe {\tt jwst123}, a custom data reduction pipeline to obtain point-source photometry from NIRCam images. Although default data products created using the \jwstpipeline pipeline include photometry catalogs \citep{Bushouse25}, crowded stellar fields such as the ones we analyze require sophisticated algorithms to effectively deblend individual resolved stars from neighbors. To produce precise, confusion-limited photometry across a range of filters, PSF sizes, and depths, we use the photometry software {\tt DOLPHOT}, which has been widely used with {\it JWST} imaging \citep{Dolphin2000,Dolphin16,Weisz24,Riess23, Anand24, Lee25, Riess25}. To obtain optimal performance from {\tt DOLPHOT}, the input images for photometry must meet strict requirements regarding astrometric quality, depth, and spatial coverage. \texttt{DOLPHOT} works best when the images are astrometrically registered to each other with high precision \citep[$\lesssim 1$\,pix;][]{Dalcanton2012, Williams2014}. Additionally, \texttt{DOLPHOT} requires a reference image for fine astrometric alignment and source extraction \citep{Weisz24} that, ideally, covers the entire observational footprint of the input images. Neither requirement is necessarily satisfied by default \jwstpipeline pipeline data products. In what follows, we describe the astrometric alignment, creation of reference images that provide more optimal coverage of the observed footprint, and the photometric procedure within {\tt jwst123}. In \noindent Figure~\ref{fig:jwst123_flowchart}, we summarize the principal data-reduction stages from Level~2 products through alignment, coaddition, and \texttt{DOLPHOT}.

\subsection{Astrometric Alignment}
\label{ssec:align}

To consistently measure flux from the same source across all filters, we require the NIRCam images to be astrometrically aligned within a relative error of one pixel (30--60 mas). The archival NIRCam data we use here comprise several programs (Table \ref{tab:galaxy_sample}) that employed varied configurations including different filter combinations, pupils, position angles, and exposure times. Within {\tt jwst123}, we develop an alignment module (cyan box, Figure \ref{fig:jwst123_flowchart}) that produces sub-pixel-level relative alignment between such NIRCam images. We design {\tt jwst123} to work with Level-2 NIRCam images (``\texttt{*cal.fits}'' files), which contain calibrated integration-combined data for each exposure, and serve as the inputs to {\tt DOLPHOT}. We organize images of a given galaxy into spatially disjoint ``groups'', each pointed at a different region of the galaxy. Within each group, images are further organized into visits, which share the same pointing, using the {\tt VISIT\_ID} header keyword.

We use the {\it JWST}-{\it HST} Alignment Tool (\texttt{JHAT}; \citealt{Rest23}) for aligning images. Because JWST pointing and its PSF shape are exceptionally stable \citep{Rigby23}, NIRCam images with good overlap share $\gtrsim 1{,}000$ common sources and can be registered to each other with high precision. By contrast, a single image may contain fewer than six {\it Gaia} sources across its ${\sim}1\,\text{arcmin}^2$ field of view. We follow a multi-step process using {\tt JHAT} to ensure robust relative alignment between NIRCam images across a galaxy-wide dataset, followed by alignment to the {\it Gaia} astrometric frame. We detail the complete setup, including failure modes and their handling, in Appendix \ref{appendix_ssec:align}. Briefly, we begin with the visit that covers the largest on-sky area and align its images to within one pixel of each other. We mosaic these images into a larger footprint using the \jwstpipeline\ pipeline \citep{Bushouse25}, which contains enough {\it Gaia} sources for a robust solution, align that mosaic to {\it Gaia}, and propagate the solution back into the individual Level-2 images. We sequentially align the remaining visits to the first in order of spatial overlap, using a running catalog of common sources to limit error propagation.

To quantify the precision of the alignment solution, we use the RMS angular offset between the target and input positions of common sources, defined as 

\begin{equation}
    \sigma_{a} = \sqrt{\frac{1}{N} \sum_{i=1}^{N} d^2_{\theta_{i}, \theta}}\,,
\end{equation}
\noindent where $d_{\theta_{i}, \theta}$ is the angular distance for the $i^{\rm th}$ pair of cross-matched sources and $N$ is the total number of matches. Relative alignment between NIRCam images produced by {\tt jwst123} has a precision of $\sigma_a \approx 5$--$20\,$mas, while alignment to the {\it Gaia} frame is precise to $\sigma_a \approx 30\,$mas. We note that the absolute tie to {\it Gaia} does not affect {\tt DOLPHOT} photometry, which requires only that images be aligned to within $\til1$ pixel of each other.

\subsection{PSF-matched Image Coaddition}
\label{ssec:coadd}

In the co-addition module of {\tt jwst123} (lavender box, Figure \ref{fig:jwst123_flowchart}), we create a reference image required by \texttt{DOLPHOT} to perform fine alignment (within one pixel) and extract sources. {\tt DOLPHOT} can photometer only sources that fall inside the reference image, so that image must cover the full footprint of the Level-2 frames. The NIRCam module of \texttt{DOLPHOT} also requires that the reference image be in one of the NIRCam filters, which prevents arbitrary PSF homogenization across different filters (e.g., convolution to a common Gaussian PSF). When combining data across multiple programs, the likelihood of finding images in a single filter that cover the full footprint decreases rapidly. Hence, we create PSF matching kernels to convert sharper PSFs (e.g., F090W) to the broadest SW PSF available in a dataset (e.g., F200W). Simultaneous imaging in SW and LW bands ensures SW images alone can cover the full observed footprint. For a given dataset, we identify the minimal set of SW filters that cover $95\%$ of the footprint, convolve all images to the broadest PSF in this filter set, and coadd these images together to create the reference image. To keep {\tt DOLPHOT} runtime reasonable, we split this reference image such that each sub-image overlaps with fewer than 150 Level-2 images, and run {\tt DOLPHOT} independently for each subset. We find that the reference image typically achieves sky coverage close to 100\% of the footprint and improves signal-to-noise ratio (S/N) for most sources. We detail the image co-addition process in Appendix \ref{appendix_ssec:coadd}.

\subsection{\texttt{DOLPHOT} Photometry}
\label{ssec:DOLPHOT}

We use the NIRCam module of \texttt{DOLPHOT} to perform PSF photometry on Level-2 calibrated NIRCam images (\texttt{*cal.fits}, as recommended in \citealt{Weisz24}). We run pre-processing steps using \texttt{calcsky}, which calculates the sky background for each image, and \texttt{nircammask}, which masks saturated and bad pixels using the Data Quality (\texttt{DQ}) array. We follow the parameter configuration recommended in \citet{Weisz24}, with the following modifications: we allow the alignment module to calculate shifts and distortion corrections using a third-order polynomial (\texttt{Align=4}) and to incorporate rotations (\texttt{Rotate=1}), similar to \citet{Blanchard25}. Although our alignment procedure is robust, we adopt this configuration to ensure flexibility for occasional failures. Using the optimal image-splitting strategy described in Appendix \ref{appendix: jwst123}, we run \texttt{DOLPHOT} simultaneously on each image set across multiple cores. We emphasize that the input Level-2 images to \texttt{DOLPHOT} are not drizzled or co-added, to preserve flux calibration and PSF morphology. We set \texttt{NIRCAMvega=0} in {\tt DOLPHOT} to obtain photometric catalogs in the AB magnitude system, using NIRCam zero points\footnote{See the JWST NIRCam absolute flux calibration documentation: \url{https://jwst-docs.stsci.edu/jwst-near-infrared-camera/nircam-performance/nircam-absolute-flux-calibration-and-zeropoints}.}.

The initial photometric catalog produced by \texttt{DOLPHOT} contains a large number of contaminants and low-quality detections. We apply selection criteria following the recommendations in the \texttt{DOLPHOT} documentation\footnote{\url{https://DOLPHOT-jwst.readthedocs.io}} and \citet{Weisz24} to discard such sources. We filter the initial catalog using the following criteria aggregated across detections: (1) $S/N \geq 5$, (2) $\text{Sharpness}^2 \leq 0.04$, which measures how point-like a source is relative to the PSF, with zero being the ideal value, (3) $\text{Crowding} \leq 1.5\,\text{mag}$, which quantifies contamination from neighboring sources and (4) $\text{Type} \leq 2$ to exclude extended sources. We also correct the photometry for Milky Way extinction using the SFD dust map \citep{Schlegel98, Schlafly11} with $R_V = 3.1$.

Recent studies have found that running \texttt{DOLPHOT} with combined SW and LW imaging yields less clean PSF subtraction due to the coarser resolution of the LW images \citep{Riess23, Anand24, Lee25, Newman2024}. This produces a minor bias in photometry at the millimagnitude scale, which can be substantially reduced by employing warm-start mode, in which sources are identified in the higher-resolution SW images and subsequently photometered on the LW images. However, we choose to perform photometry using combined SW+LW imaging in each run to reliably identify highly obscured sources that are luminous in the LW filters but fade rapidly in the SW filters. Warm-start mode risks missing such sources, which are of particular interest as potential RSGs experiencing substantial mass loss. We emphasize that the bias introduced by this choice is well below the systematic uncertainty in SED modeling (0.05--0.1 mag; Section \ref{sec:models}). The photometric catalogs created by \texttt{DOLPHOT} contain $\til2$--6 million sources in each galaxy after quality cuts.

\section{SED Modeling}
\label{sec:models}
 
With fully reduced images, we now turn to modeling SEDs of luminous stars in our photometric catalogs. We aim to classify stellar populations using their inferred physical properties from fitting observed SEDs to our models. The primary contaminants in RSG catalogs are AGB stars, which represent evolved states of low-mass $(M_{\rm ZAMS} \lesssim 7\,M_{\odot})$ stars, and occupy similar regions in the CMD as RSGs due to their cool, dusty envelopes \citep{Boyer11, Britavskiy2015}. Luminous AGB stars and super-AGB stars show similar spectral features to RSGs in the optical, and overlap in the low end of the RSG luminosity function, from $-8.0 \lesssim M_{\rm bol} \lesssim -7.1$ \citep{vanLoon2005, Woods2011}.

For our sample, CMD-based selection is particularly limited because filter coverage is non-uniform: different parts of a galaxy may lack imaging in the bands (e.g., F115W or F200W) typically used to define selection cuts. Other filter combinations can increase AGB contamination and prevent cuts from transferring cleanly between CMDs while preserving purity and completeness. The empirical line cuts are themselves loosely defined, varying with the CMD used and with how populations cluster. We therefore treat CMD cuts only as a qualitative guide to isolate the RSG locus, and rely on multi-band SED fitting for classification using physical parameters. 

\begin{figure*}[tp]
\centering
\includegraphics[width=0.49\textwidth]{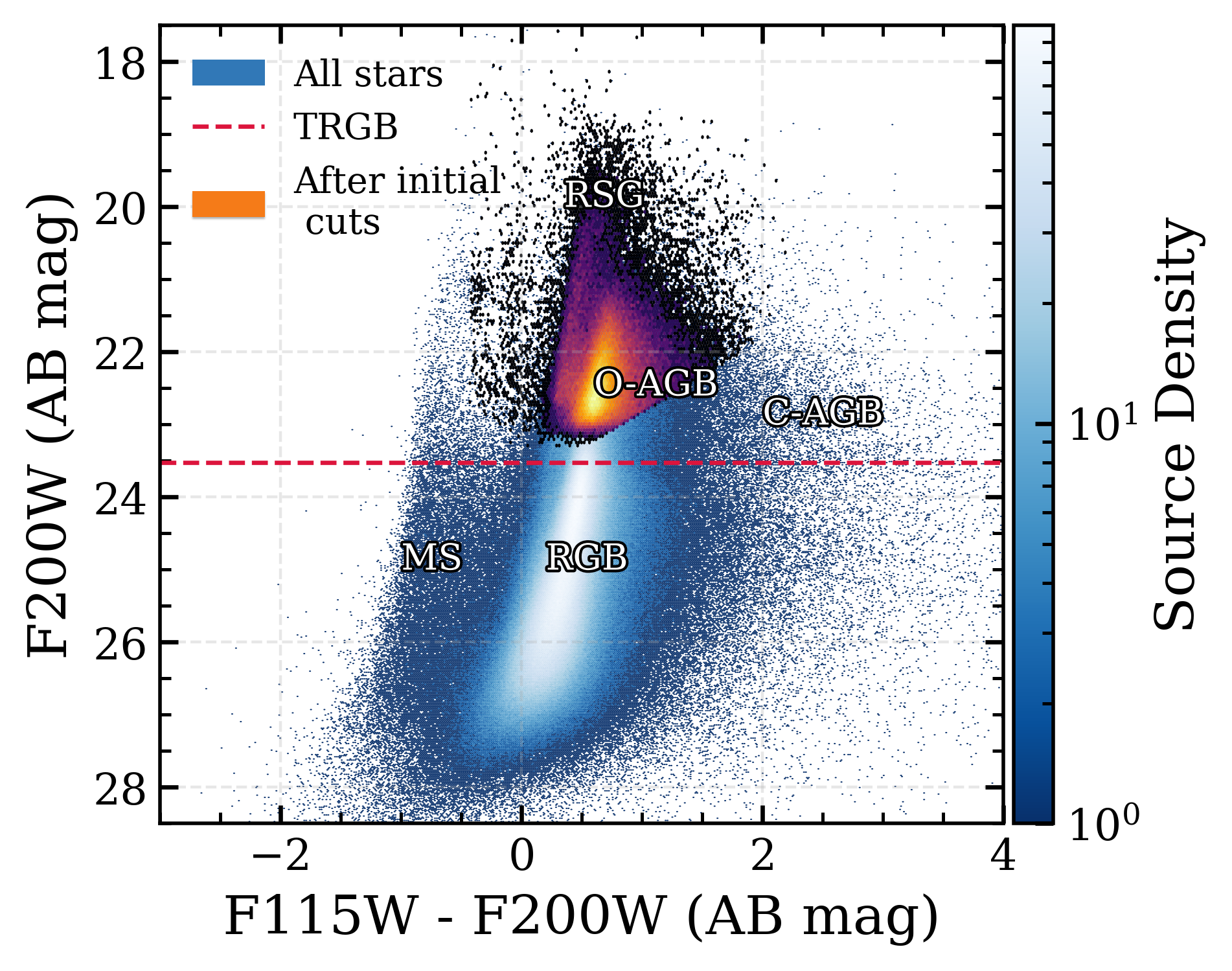}%
\hfill
\includegraphics[width=0.49\textwidth]{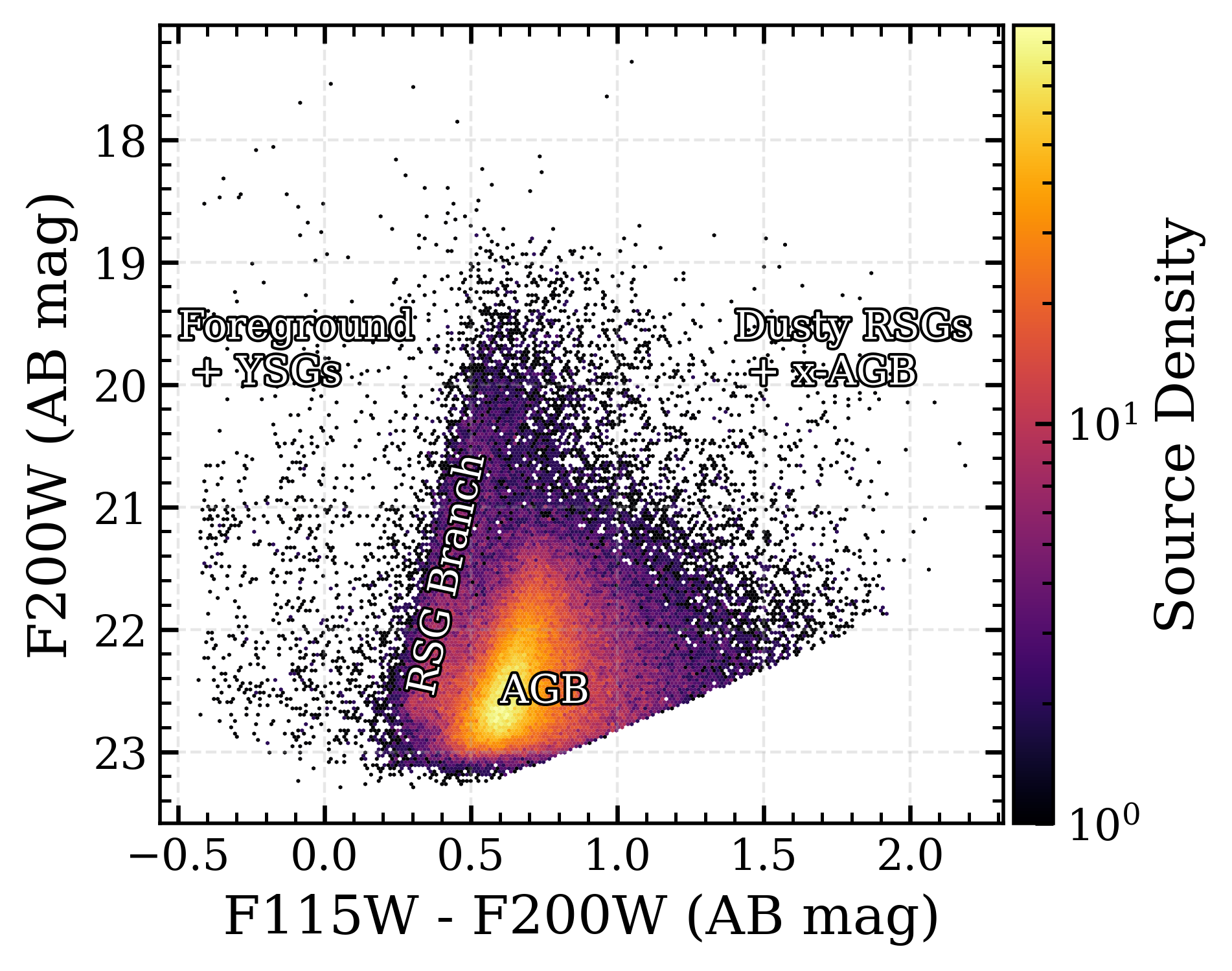}
\caption{Left: Color--magnitude diagram (CMD) with the stellar populations in NGC\,5236 ($d = 4.6\,{\rm Mpc}$), along with the candidate sources chosen for SED fitting, shown separately to highlight the effect of the coarse initial cuts applied (Section \ref{sec:models}). We select sources that are brighter than the tip of the red giant branch (red dashed line). Right: CMD of luminous stellar populations chosen for SED fitting. This includes RSGs, Oxygen-rich AGBs (O-AGBs), foreground stars, yellow supergiants, and extreme AGBs (x-AGBs).}
\label{fig:cmd_cuts}
\end{figure*}

To this end, we simulate synthetic photometry of RSGs in NIRCam filters using model atmospheric spectra of cool stars and add the effects of dusty CSM on the observed SED. We describe the SED modeling setup and initial cuts applied before parameter inference below. Figure~\ref{fig:cmd_cuts} illustrates the various stellar populations we aim to distinguish, before we discuss the SED forward model. Throughout the analyses detailed below, we use time-averaged SEDs for each source, as the sparse time coverage of the archival dataset is not ideal for variability analyses.

\subsection{SED model set-up with {\tt DUSTY}}

\begin{figure}[htp]
    \centering
    \includegraphics[width=0.46\textwidth]{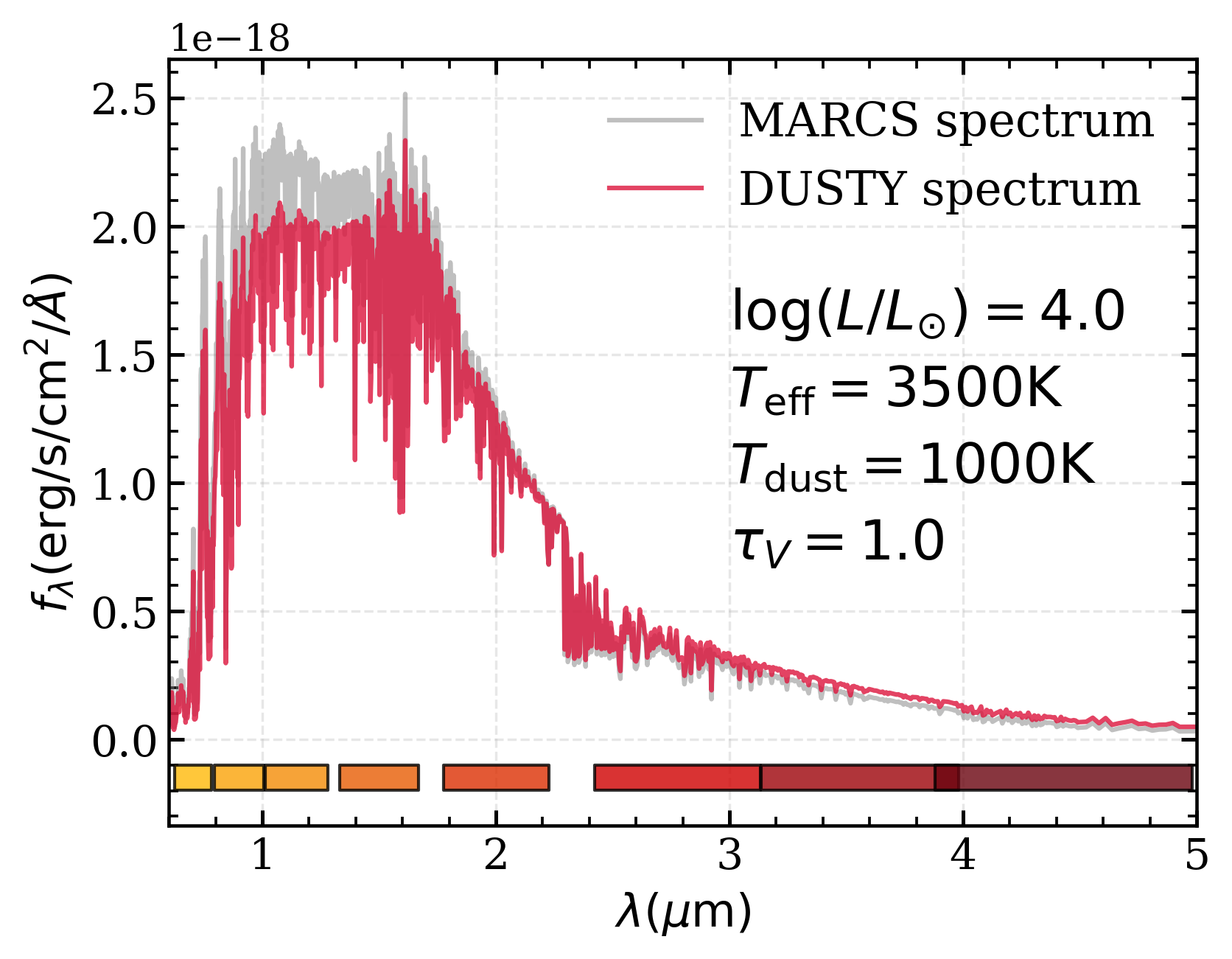}
    \caption{Synthetic RSG spectra modeled using \texttt{DUSTY} 1D-radiative transfer on MARCS atmospheric spectra of cool stars. We incorporate the effects of circumstellar dust (parameterized by $\tau_V$ and host galaxy extinction (parameterized by $A_V$ and $R_V$) on the model spectra and synthetic photometry. We show the MARCS spectrum in gray, and the {\tt DUSTY} spectrum in red. We compute SEDs in NIRCam filters, such as the wide filters, whose wavelength range is shown below the spectrum.
    }\label{fig:rsg_sed}
\end{figure}

We model the RSGs as a bare atmosphere described by $T_{\rm eff}$ and $\log(L/L_{\odot})$, surrounded by a spherically symmetric dust shell parameterized by $T_{\text{dust}}$ and $\tau_V$ (representing dust attenuation in the optical $V$-band). RSGs are enshrouded in dust due to extensive mass loss driven by stellar winds in their low surface gravity envelopes \citep{vanLoon2005, Chiavassa2011, vanLoon2025, Beasor22}. This dust obscures the stellar photosphere, so the stars are faint optically and relatively bright in the IR \citep{Lancon2007, Levesque2018}. We compute SED models using the 1D radiative-transfer software \texttt{DUSTY} (Figure \ref{fig:rsg_sed}; \citealt{Ivezic97, Ivezic1999}). \texttt{DUSTY} solves the radiative transfer equations for a radiation source interacting with either a spherical shell or a plane slab of dust, and provides flexibility in dust composition and density profile \citep{Ivezic1999}. {\tt DUSTY} has been widely applied to radiative transfer modeling of SN progenitors, dusty RSGs, and other stellar phenomena \citep[see e.g.,][]{Kochanek2012, Kilpatrick2018, Beasor2024, Kilpatrick25, Karambelkar2026}.

Since we do not know a priori whether a selected source is an RSG or belongs to another spectral or luminosity class, the model grid must be sufficiently general to fit a wide range of plausible candidates. This includes luminous AGB stars, which are redder and more variable than RSGs, as well as hotter blue-sqeuence stars \citep{Boyer11, Ren21}. The blue-sequence is likely a mix of yellow supergiants (YSGs), some blue supergiants (BSGs) and foreground contaminants \citep{Drout2012}. Hot blue stars and foreground contaminants have SEDs that differ strongly from those of RSGs and AGBs, which are themselves difficult to distinguish from one another. Hence, we generate models that describe RSGs and AGBs, and we discard other contaminants based on the resulting $\chi^2$ value of a model fit (Section \ref{ssec:chimin}). We use the same parameter set ($T_{\rm eff}, T_{\text{dust}}, L, \tau_V, R_V, A_V$) for both RSGs and AGBs, with the two classes occupying different regions of parameter space. AGB stars typically display more pronounced variability and reach lower luminosities and cooler effective temperatures than RSGs, though the two populations overlap substantially in both parameters \citep{Vassiliadis1993, Herwig05, Karakas2014, Paczynski1971, Sargent2011, Woods2011, Davies13}. 

In addition to the effects of circumstellar dust on the observed spectra, we must also account for absorption and scattering by line-of-sight interstellar dust from the host galaxy of the stars we model. Analyses of massive star progenitors of SNe use ${\rm Na}\,{\rm I}$ D absorption features to estimate the reddening due to the host environment \citep{Kilpatrick2018, Kilpatrick23, Kilpatrick2023b, JacobsonGalan2023}. In the absence of spectra, we fit for the host extinction, parameterized through $R_V$ and $A_V$, assuming the extinction law of \citet{Cardelli1989}. $A_V$ represents only the extinction due to the host, and $\tau_V$ accounts for CSM extinction. We constrain both quantities because they affect the SED differently, which helps quantify degeneracy between host and CSM dust contributions. Host extinction attenuates radiation disproportionately in bluer wavelengths, while photometry beyond \til2\,${\rm \mu m}$ is not affected significantly. On the other hand, CSM reprocesses the optical stellar light and emits radiation in the IR, altering the continuum. In parts of a galaxy with missing bluer filters (F090W, F115W), our inferred host extinction has larger uncertainties. 

\texttt{DUSTY} requires an input spectrum at $T_{\rm eff}$, corresponding to the energy source that heats the dust \citep{Ivezic1999}. We use synthetic atmospheric spectra of cool stars from the MARCS model grid \citep{Gustafsson08} as the input spectra; these models employ 1D LTE radiative transfer. We use the $1\,M_{\odot}$ model grid, as opposed to the $15\,M_{\odot}$ model grid generally used for supergiants, due to the finer $T_{\rm eff}$ grid spacing of $\Delta{T_{\rm eff}} = 100\,K$ and wider coverage from $2500\,K$ to $8000\,K$. This is necessary to model both RSGs and AGBs simultaneously and to reduce interpolation errors in our SED simulator, which is constructed by linearly interpolating the SEDs generated by \texttt{DUSTY} across the full parameter space. To validate that the use of the 1~$M_{\odot}$ grid does not introduce significant systematic uncertainties in parameter inference, we compare the results of using the 1~$M_{\odot}$ and 15~$M_{\odot}$ model grids in Appendix~\ref{appendix: marcs}, finding that the bias is well within the 1$\sigma$ error on $T_{\rm eff}$. As appropriate for RSGs, we use the models at surface gravity $\log\,g=0$, microturbulence $\xi = 4\,\text{km}\,\text{s}^{-1}$, solar-scaled alpha abundance, and spherical geometry as implemented in the MARCS grid \citep{Gustafsson08}. We note that down-sampling to broadband photometry largely erases fine spectral features. Thus, following common practice in RSG SED modeling \citep{Beasor16, Levesque2018, Britavskiy2019, Yang2023}, we fix surface gravity, microturbulence, and stellar mass, since these are strongly degenerate with one another and poorly constrained by broadband photometry \citep{Tabernero2018}. Consequently, $T_{\rm eff}$ is the only atmospheric parameter we vary in the MARCS grid. We use three sets of MARCS models, varying in metallicity to cover $\log(Z/Z_{\odot}) = -0.5$, $-0.25$, and $0.0$, corresponding to the global metallicity estimates of our galaxy sample (Table \ref{tab:galaxy_sample}), and create interpolated SED grids at each metallicity.

\startlongtable
\begin{deluxetable}{cc}
\tablecaption{Parameter grid of simulated \texttt{DUSTY} SEDs}
\tablehead{
\colhead{Parameter} &
\colhead{Grid Values}
}
\startdata
$T_{\rm eff}$ & $2500\,K - 4000\,K$ at $\Delta T_{\rm eff} = 100\,K$, \\     
& $4250\,K - 5000\,K$ at $\Delta T_{\rm eff} = 250\,K$ \\
$T_{\rm dust}$ & $200\,K, 500\,K, 800\,K, 1000\,K,$ \\
& $1200\,K, 1500\,K, 1800\,K$ \\
$\log(L/L_{\odot})$ & 3.0, 6.0\\
$\tau_V$ & $10^{-4}$, $0.1 - 1$ at $\Delta\tau_V=0.1$,\\
& $1.5 - 6$ at $\Delta\tau_V=0.5$\\
& $7 - 12$ at $\Delta\tau_V=1$\\
$R_V$ & $1$--$6$ at $\Delta{R_V}=1$\\
$A_V$ & $0 - 0.5$ at $\Delta{A_V}=0.1$\\
& $0.75 - 1$ at $\Delta{A_V}=0.25$\\
& $1 - 5$ at $\Delta{A_V}=1$\\
\enddata
\tablecomments{We only simulate SEDs at two luminosities ($\log(L/L_{\odot}) = 3.0$ and $6.0$) as we linearly interpolate this grid to create our model generator, and the SEDs scale to brighter magnitudes linearly with $\log L$.}\label{tab:dusty_grid}
\end{deluxetable}

We configure \texttt{DUSTY} to solve the radiative transfer equations for stellar radiation interacting with a spherically symmetric dust shell, with shell thickness $R_{\rm out}/R_{\rm in} = 2$, consistent with literature analyses \citep{Kilpatrick2018, Beasor22}. While the dust shell is not always spherically symmetric, it is a reasonable approximation (see \citet{Beasor22} for further discussion). We set the density profile to $\rho \propto r^{-2}$ as in \citet{Beasor22}. The composition of the circumstellar dust alters the SED morphology considerably. RSGs have been shown to have predominantly silicate-rich dust \citep{Verhoelst09, Goldman2017, Yang2023}. Direct measurements of dust grain sizes around individual RSGs, from optical polarimetric imaging and interferometry, find grains of \til0.5\,${\rm \mu m}$ \citep{Ohnaka08,Verhoelst09,Scicluna2015,Haubois2019}. These constraints come from a small number of extreme, nearby objects and need not be representative of the population. 

In this paper, we set the dust to be silicate-rich, following the \citet{Draine84} prescription, and adopt a modified MRN grain size distribution with ${a_{\rm min}}=0.005\,\micronmath$, ${a_{\rm max}}=1\,\micronmath$, and power law index for grain size distribution, $q=3.0$. While a small fraction $(\til10\%$) of RSGs may be consistent with carbon-rich dust models instead \citep{Wang2021}, we model all RSGs using a silicate dust shell, since we do not include MIR photometry, which is needed to constrain the dust composition. We use the silicate models to fit AGBs and hotter blue stars. Even though not all stars can be described using these models, we can reasonably characterize $T_{\rm eff}$ (and the temperature gradient across populations), $L$, and $\tau_V$, given the NIR data.

We simulate \texttt{DUSTY} SEDs according to the parameter grid listed in Table \ref{tab:dusty_grid} for each metallicity, resulting in three grids of $\til 400,000$ spectra each. \texttt{DUSTY} output spectra span a wavelength range from $0.01\,\micronmath$ to $3.6\,\mathrm{cm}$ to adequately capture the effects of the input stellar spectrum and calculate the bolometric luminosity. We simulate the spectra at a resolution of $R\til250$ in the $0.7$--$5\,\micronmath$ range, sufficient for synthetic photometry, and at a resolution of $R\til15$ elsewhere. For each output spectrum, we simulate synthetic photometry for each of the 27 NIRCam filters using \texttt{synphot} following standard synthetic-photometry practice for space-based instruments \citep{Bohlin14,Astropy18} to match our observed {\it JWST} photometry. We use only the broadband filters in the reduction and SED analysis described below (narrow-band synthetic photometry is not used for inference). 

We generate the grid at a fiducial distance of 10 Mpc ($\mu=30$\,mag) and rescale each SED to the distance of the target galaxy. To fit individual sources to the full range of model parameters ($T_{\rm eff}, T_{\text{dust}}, L, \tau_V, R_V, A_V$), we linearly interpolate the regular grid and use it as the forward model generator for our parameter inference setup.

\subsection{Initial cuts using $\chi^2$ minimization}
\label{ssec:chimin}

Before fitting SED models, we exclude the vast majority of sources after {\tt DOLPHOT} photometry and quality cuts that are not of interest as RSG candidates, being either main-sequence stars, RGB stars, or other low-mass analogs.
First, we exclude narrow-band photometry from SED fitting. This is because narrow-band flux can originate from nebular environments, potentially unassociated with the target source, and can contaminate the stellar flux. MARCS provides flux-sampled model atmospheres that reproduce the continuum well but need not reproduce individual line strengths accurately \citep{Gustafsson08}. Due to a combination of these effects, we find that a significant fraction of sources show deviations from the model predictions in the narrow filters, whereas the wide- and medium-band photometry fits the model predictions very well. We require a source to be detected in at least four filters, excluding narrow filters, to attempt SED fitting (Section \ref{sec:sample}). With fewer than four detections, the inference setup largely returns the prior, which is not useful for characterizing sources. We account for non-detections (due to faintness or missing coverage) during parameter inference.

\begin{figure}[tp]
    \centering
    \includegraphics[width=0.46\textwidth]{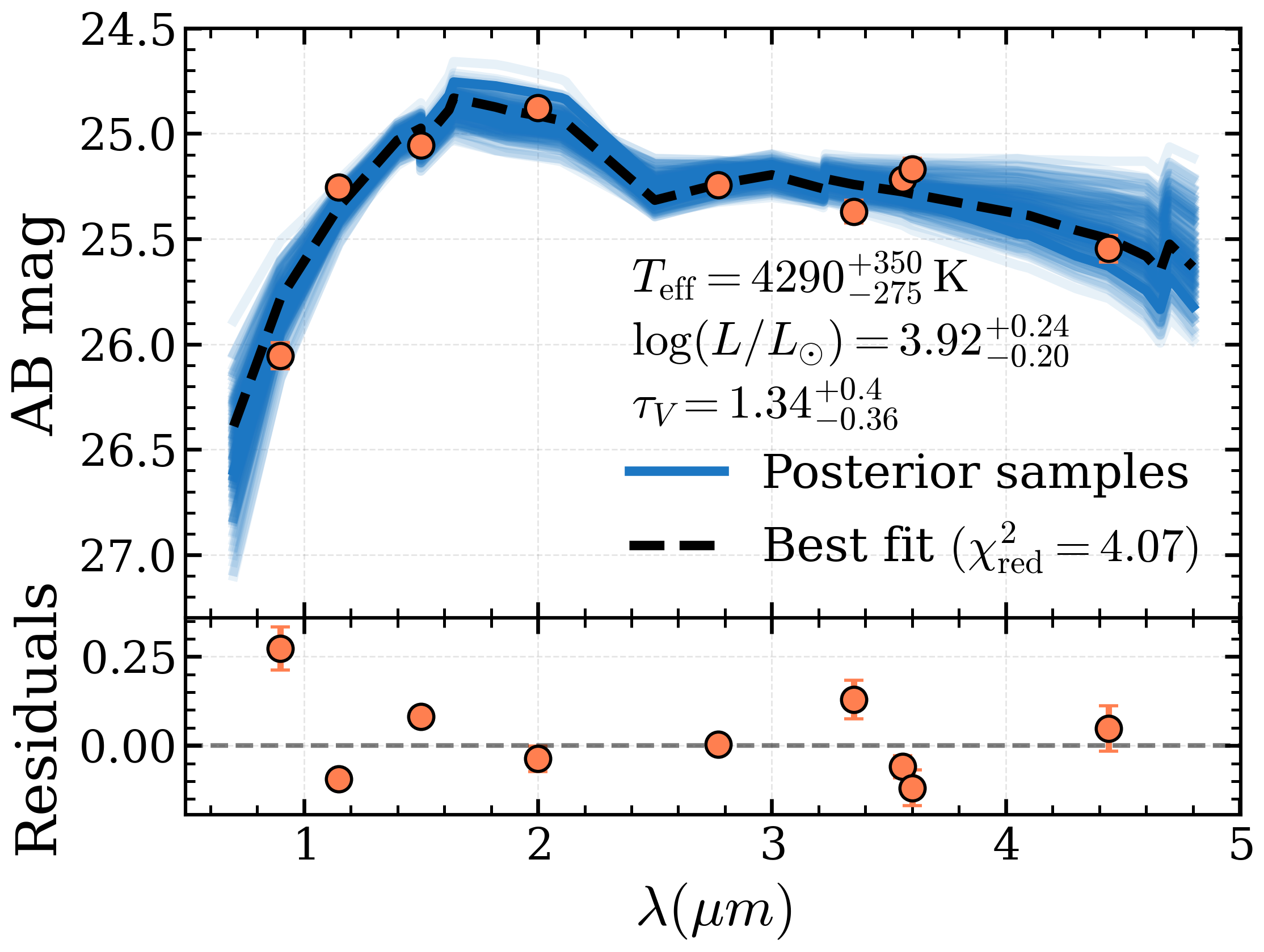}
    \caption{Example fit to the SED of a star in NGC\,5643 $(d=12.5\,\text{Mpc})$ using our \texttt{DUSTY} forward models. We use Markov chain Monte Carlo implemented using \texttt{emcee} \citep{ForemanMackey13} to fit our model to the observed SED, with 64 walkers running for 1000 steps, discarding the initial 75 steps of each walker as burn-in. Systematics from synthetic atmosphere models and interpolation errors contribute to the uncertainty in the fit, and we generally find well-calibrated estimates for $T_{\rm eff}, L$, and $\tau_V$.
    }\label{fig:mcmc_fit}
\end{figure}

Second, we impose a luminosity cut above the Tip of the Red Giant Branch (TRGB), defined independently for each galaxy and corresponding approximately to $\log (L/L_{\odot}) \gtrsim 3.5$. The TRGB has been well-calibrated in the $I$ band and is typically measured using {\it HST} F814W or {\it JWST} F090W \citep{Anand24, Newman2024}. We use TRGB values reported in the literature, typically measured using {\it HST} F814W or {\it JWST} F090W \citep{Riess23, Anand24, Newman2024}. The majority of candidates at this stage are discarded by this cut, leaving $\til 100{,}000$ plausible candidates per galaxy (Table \ref{tab:selection_cuts}; Figure \ref{fig:cmd_cuts}). We also impose that $\log (L/L_{\odot}) \lesssim 6$, which excludes a handful of bluer stars that are very likely foreground contaminants, with luminosity overestimated due to the assumed host distance.

The remaining candidates at this stage are primarily AGB stars, RSGs, and hotter stars. We generate a sparse grid of SEDs to run a coarse $\chi^2$ minimization procedure to discard implausible sources that cannot be reasonably described by our model. We generate a total of $\til300{,}000$ SED models, varying all parameters except $R_V$, which is fixed to 3.1. We calculate the reduced $\chi^2$ for each candidate against these SEDs and choose a best-fit model based on the minimum $\chi^2$, corresponding to a point estimate of the physical parameters. While the resulting SED fit is usually good, this method cannot account for degeneracies in the parameter space well and has an inherent trade-off between accuracy and runtime that scales poorly with larger grid sizes. Hence, we use the $\chi^2$ values to discard the bottom $25$th percentile of data that show very poor fits to the model SEDs and hence cannot be reasonably described by our models. This cut reduces SBI failures on pathological SEDs. We verify using CMDs that the vast majority of candidates rejected by this cut are low-luminosity AGB or main sequence stars. We retain a total of 1{,}531{,}629 candidates following the SED cuts (Table \ref{tab:selection_cuts}).

\section{Simulation-Based Inference}
\label{sec:sbi}

To rapidly derive physical parameters for the $\til1.5$ million candidate sources identified following initial cuts on the galaxy photometry, we fit SEDs using simulation-based inference (SBI) models for each galaxy. SBI is a machine learning method that maps a simulated training set of observables (SEDs in our case) to the probability density of the parameters of the physical model describing the data (e.g., $T_{\rm eff}$, $L$, and so on) without requiring an explicit likelihood. We train SBI models based on normalizing flows, a family of generative models that use bijective transformations to convert a simple base probability distribution, such as a uniform or Gaussian distribution, into the probability density estimate of the posterior, learned from the training samples \citep{Papamakarios21}. In particular, we use neural density estimators to learn the six-dimensional posterior density distribution and produce posteriors for new data in a process called Neural Posterior Estimation (NPE). The training is amortized, meaning that once an SBI model is trained on a set of simulated SED-parameter pairs, it can draw samples from the posterior density for any input observed SED that falls within the training sample prior range, without having to retrain the model. 

One major drawback of standard SBI models is the inability to infer posteriors for out-of-distribution (OOD) inputs. In our case, this can occur in several forms, by incorporating: (1) non-detections in some bluer filters when an RSG is heavily dust-enshrouded, (2) non-detections arising from missing filters in parts of a galaxy due to non-uniform filter coverage, or (3) larger uncertainties compared to the training set distribution. Such cases fall outside the support of the simulated training set, and SBI models will often produce unphysical parameter estimates or fail prior-based rejection sampling for OOD inputs, as the probability density shrinks outside the learned phase space. To circumvent these issues, we use \texttt{SBI++} \citep{Wang23}, which uses nearest-neighbor interpolation to fill in missing information and Monte Carlo to resample from OOD noise distributions (Section \ref{ssec:inference}). \texttt{SBI++} wraps around a baseline SBI model trained using simulated SEDs with the maximal filter set for a particular galaxy, with noise simulated using a model based on observations. The training distribution is used for the nearest-neighbor search, while the noise model is used to define input samples with OOD noise. \texttt{SBI++} provides a quick and elegant method for drawing meaningful posteriors from SEDs constructed from archival datasets such as the one used in this study, significantly improving inference completeness across galaxies and thereby yielding a more complete sample of RSGs. However, even with \texttt{SBI++}, we require a dedicated SBI model per galaxy dataset due to their unique filter combinations. 

\subsection{Training Dataset}

To create a training set for the SBI models, we simulate $300{,}000$ SED-parameter pairs in all 27 NIRCam filters using the interpolated SED simulator described in Section \ref{sec:models}, assuming a luminosity distance of $10\,{\rm Mpc}$ (including narrow-band synthetic photometry in the grid, which we omit from training inputs and inference for the reasons given in Section \ref{sec:models}). We generate SEDs in all filters to avoid simulating the training set repeatedly for each galaxy, and select the required filters before training. We generate three such grids, at metallicities of $\log(Z/Z_{\odot}) = 0, -0.25,$ and $-0.5,$, to account for different host environments. We select the appropriate metallicity grid before training a model for a specific galaxy, extract the filters present in the galaxy photometry, and scale the apparent magnitudes of the SEDs to the distance of the galaxy. 

\startlongtable
\begin{deluxetable}{cc} 
\tablecaption{Priors for physical parameters used in SBI \label{tab:sbi_prior}}
\tablewidth{\columnwidth}
\tablehead{
\colhead{Parameter} &
\colhead{Prior}
}
\startdata
$T_{\rm eff}$ & $\mathcal{U}[2500\,{\rm K},\,5000\,{\rm K}]$\\
$T_{\rm dust}$ & $\mathcal{U}[200\,{\rm K},\,1800\,{\rm K}]$\\
$\log(L/L_{\odot})$ & $\mathcal{U}[3.0,\,6.0]$\\
$\tau_V$ & TruncExp[rate=0.5,$10^{-4}$--12]\\
$R_V$ & $\mathcal{U}[2.0,\,6.0]$\\
$A_V$ & TruncExp[rate=0.5, 0--5]\\
\enddata
\tablecomments{TruncExp is a truncated exponential distribution used for $\tau_V$ and $A_V$ to ensure a hard cutoff for an exponential distribution at the lower and upper bounds. Beyond the bounds, these parameters are unphysical or outside training support.}
\end{deluxetable}

We randomly sample from the priors defined in Table \ref{tab:sbi_prior} to generate parameter values for simulating the SEDs. We use a truncated exponential prior for $\tau_V$ and $A_V$ to prevent the SBI model from overfitting individual SEDs and inferring skewed global extinction distributions across a galaxy, and to prevent density leakage into unphysical regions of parameter space, such as negative extinction values. Internally, the SBI model uses rejection sampling to discard posterior samples that fall outside the prior support. Similar to \citet{Wang23}, we generate $70\%$ of the training set using this ``mixed'' prior and simulate the remaining $30\%$ using a uniform prior across all parameters to ensure sufficient training samples in less dense regions of parameter space.

We include photometric uncertainties in each filter in the training set along with the SEDs. For real, noisy observations, this enables the model to calibrate the parameter uncertainties accordingly. As emphasized in \citet{Wang23} and \citet{Nugent26}, the noise model used to simulate the training set noise should be representative of the observed noise distribution of the real data on which we will run inference. In instances of missing observations, \texttt{SBI++} uses this noise model to generate the uncertainties for the model input. The noise model also needs to account for non-detections in real data, where the scaled $S/N$ relation in magnitude space below the $5\sigma$ limit yields impractically large uncertainties, potentially introducing confusion during training. In addition to the measured photometric uncertainties, we also account for systematic uncertainties in the SED models themselves by adding stochastic scatter to the simulated SEDs based on residuals of the $\chi^2$-minimized fit.

We create an empirical noise model for each galaxy using observed data. For each filter in the galaxy photometry, we bin the magnitude distribution such that each bin contains at least 100 observations. We model the uncertainties in each bin using a skewed Gaussian distribution (skew-normal), fitting for the mean, standard deviation, and skew. We resample from this distribution to assign uncertainties to all simulated training samples that fall within the magnitude bin. In instances where this fit fails, we assume a standard Gaussian distribution, i.e., with no skew. For training samples brighter than the brightest observed sources, we assign uncertainties sampled from a $\mathcal{N}(0.0, 0.02)$ distribution, as observed for the highest S/N sources. For samples fainter than the $5\sigma$ detection threshold in real data, we sample from the skew-normal distribution of the faintest magnitude bin. This faint component of the training set enables inference when \texttt{SBI++} uses nearest-neighbor interpolation to fill in values for non-detections. To prevent unrealistic uncertainties in the training set, we set the upper limit of the photometric error to $0.6$~mag, which is greater than the largest observed uncertainty in most cases. \texttt{SBI++} performs accurate parameter inference when the model encounters OOD noise realizations \citep{Wang23}. Throughout training and inference, we enforce a noise floor of $0.01$ mag to avoid overconfident posteriors. We add this noise floor in quadrature to the observed or simulated noise. We also modify the OOD noise identification step in \texttt{SBI++} to incorporate skew-normal error distributions instead of Gaussian errors. We calculate the $3\sigma$ uncertainties in different magnitude bins in each filter, and use these values as bounds within which uncertainties stay inside the training support. 

Finally, to represent systematic uncertainties in the SED model, arising from the synthetic spectra, choice of dust parameters, and interpolation, we use the point estimates derived using $\chi^2$ minimization as described in Section \ref{sec:models}. For each observed source, we compute the residuals with respect to the best-fit $\chi^2$-minimized SED. We use these residuals to build a probability distribution in each filter using Gaussian kernel density estimation (KDE) and resample from this distribution to generate noise added directly to the simulated SED magnitudes. We only apply this step to the simulated training set SEDs. We pass observed SEDs directly to the model with no modification except enforcing the noise floor.

\subsection{Training a Baseline SBI model}

We build a normalizing flow neural network model with the \texttt{sbi} Python package \citep{Greenberg19, Tejero-Cantero20}, implementing neural spline flows via the \texttt{nflows} framework \citep{Durkan19, Durkan20}. Neural spline flows are a class of normalizing flows that employ monotonic rational-quadratic spline transformations, providing flexible and expressive density estimation. We train 15 SBI models, corresponding to our galaxy sample. Even though some galaxies share the same combination of filters (e.g., NGC\,4449 and NGC\,4485), the noise properties of the photometry are different owing to different imaging configurations and exposure times. We include the interacting galaxy pair NGC4485/NGC4490 (Table \ref{tab:galaxy_sample}) in the same trained model as they have identical filter coverage.

The neural network of the SBI model is primarily controlled by three hyperparameters: hidden features (the width of the hidden layers), number of transforms (the number of successive transformations mapping the base Gaussian distribution to the posterior), and number of spline bins (the expressivity of the spline transforms used in neural spline flows). In addition, other hyperparameters, such as the learning rate of the neural net, input batch size, and the number of training epochs monitored for convergence (patience), also affect model performance. Tuning these hyperparameters is essential to constructing a high-performing and well-calibrated model. We use a $10\%$ validation split when training the model and train until validation loss fails to improve for $20-50$ consecutive epochs. 

We use a neural network with 10 hidden features, 5 transforms, and 10 spline bins as the default configuration to train models for any galaxy. We use a learning rate of $5\times10^{-4}$, batch size of 256, and patience of 50 epochs. Seven of our 15 models are trained using this configuration, with galaxies typically having photometry in six filters. For galaxies with a larger number of filters, we increase the number of hidden features to 25 from 10. For the most well-sampled photometry, with up to 14 filters of data, we use 45 hidden features, with an augmented training set of $500{,}000$ simulations created by resampling from our original training set and re-generating noise for duplicated samples to improve calibration.

\startlongtable
\begin{deluxetable}{lc}
\tablecaption{Hyperparameter tuning setup using \texttt{Optuna}\label{tab:optuna}}
\tablewidth{0pt}
\tablehead{
\colhead{Hyperparameter} & \colhead{Value}
}
\startdata
\multicolumn{2}{c}{\textit{Fixed Hyperparameters}} \\
Normalizing flow model & \texttt{nsf} \\
Training set size & $300,000$ \\
Maximum epochs & $400$ \\
Validation split & $10\%$ \\
\multicolumn{2}{c}{\textit{Tuned Hyperparameters}} \\
Learning rate & $[10^{-4},\,10^{-2}]$ \\
Hidden features & $[10,\,60]$ \\
Number of transforms & $[2,\,20]$ \\
Number of spline bins & $[5,\,30]$ \\
Batch size &  $[32,\,64,\,128,\,256,\,512]$ \\
Patience & $[20,\,50]$ \\
\enddata
\tablecomments{Tuned hyperparameters can have any value within the indicated range, except for batch size, which is discrete and selected from the list above.}
\end{deluxetable}

\begin{figure*}[tbp]
    \centering
    \includegraphics[width=0.99\textwidth]{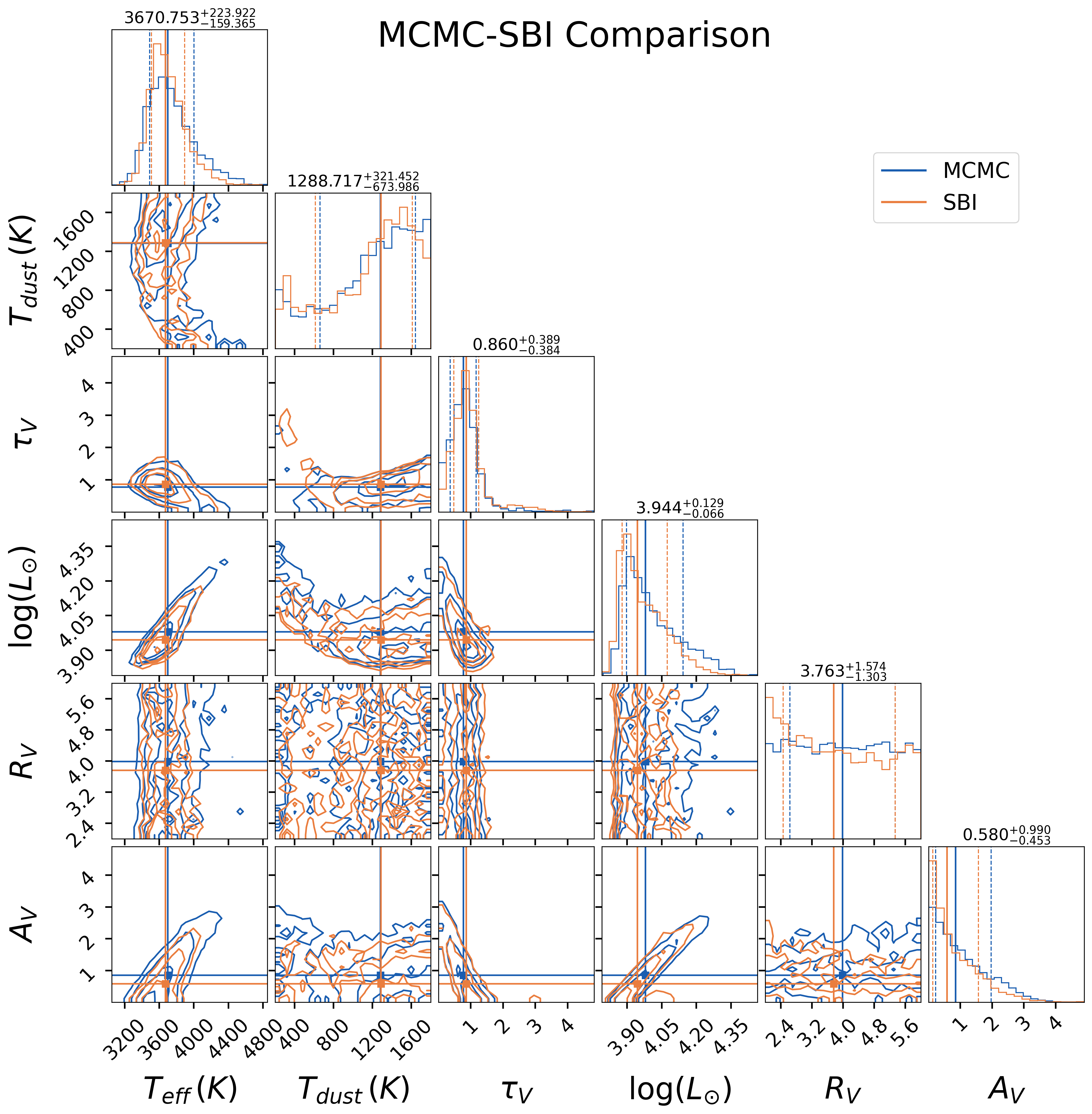}
    \caption{Comparison of posterior samples inferred by our SBI model for a star in NGC\,5194 $(d=7.2\,{\rm Mpc})$ to posteriors sampled using MCMC. We find excellent agreement between the marginalized point estimates across our parameter space. The SBI model also replicates the correlation across parameters as accurately as MCMC.
    }\label{fig:sbi_mcmc_compare}
\end{figure*}

Given the non-uniform set of filters and photometry across the galaxy sample, three of our models require more extensive hyperparameter tuning to achieve the same performance. To perform detailed hyperparameter tuning, we use the software \texttt{Optuna} \citep{Akiba19}, which implements optimization algorithms to search the parameter space efficiently and identify optimal hyperparameters given an objective function. We define a multi-objective setup, maximizing validation log probability on a withheld set and minimizing the Cr{\'a}mer-von Mises distance to the uniform CDF for SBC ranks. The hyperparameter space for \texttt{Optuna} is shown in Table \ref{tab:optuna}, and we run 50 trials per galaxy. We run hyperparameter tuning for three galaxies and indicate the best hyperparameters in brackets as hidden features, number of transforms, and number of spline bins, respectively: NGC\,4449 (12, 10, 24), NGC\,5236 (14, 6, 12), and NGC\,5457 (17, 5, 14).

\startlongtable
\begin{deluxetable*}{clccc} 
\tablecaption{Selection criteria for RSGs from resolved stellar population photometry\label{tab:selection_cuts}}
\tablehead{
\colhead{Step} &
\colhead{Selection cut / criteria} &
\colhead{Total candidates} &
\colhead{Surviving candidates} &
\colhead{\% Removed by cut}
}
\startdata
1 & Passes quality criteria on \texttt{DOLPHOT} photometry & 70{,}159{,}057 & 41{,}738{,}330 & $40.5\%$\\
2 & Detected in at least four filters & 41{,}738{,}330 & 22{,}289{,}663 & $46.6\%$ \\
3 & Source brighter than TRGB & 22{,}289{,}663 & 1{,}991{,}833 & $91.1\%$\\
4 & Coarse-grid-fit $\chi^2$ lower than $75^{\rm th}$ percentile & 1{,}991{,}833 & 1{,}531{,}629 & $23.1\%$\\
5 & Successful inference using SBI model & 1{,}531{,}629 & 1{,}507{,}852 & $1.6\%$ \\
6 & Passes KDE-based RSG selection cut $(P_{\rm RSG} > 0.7)$ & 1{,}507{,}852 & \nrsg  & $93.3\%$ \\
\enddata
\tablecomments{We describe {\tt DOLPHOT} quality cuts (step 1) in Section \ref{ssec:DOLPHOT}. We detail the SED selection cuts (steps 2-4) in Section \ref{sec:models} and SBI setup (step 5) in Section \ref{sec:sbi}. RSG selection criteria (step 6) follow from Section \ref{sec:cat}. The percentage removed is defined relative to the number of candidates entering that step.}
\end{deluxetable*}

\subsection{Inference using {\tt SBI++}}
\label{ssec:inference}

We infer physical parameters for all $\til1.5$ million candidates across 15 galaxies using {\tt SBI++}. We encounter four observational cases during parameter inference: (1) all filters present and in-distribution noise, handled by the baseline SBI model; (2)–(4) require SBI++ augmentation, as follows: (2) OOD noise, where \texttt{SBI++} does MC sampling within the noisy uncertainties and averages inference across multiple realizations of the SED with different noise, (3) missing observations in some filters (either non-detections or lack of coverage), where nearest neighbor search using the training set fills in the missing magnitudes, and we compute uncertainties using the noise model, and (4) both missing and noisy observations, where \texttt{SBI++} combines MC sampling and nearest neighbor interpolation. We follow the recommended setup from \citet{Wang23}, with 50 Monte Carlo (MC) samples for OOD observations, each with 50 posterior samples, for a total of 2500 posterior samples per object. We use an initial cut of $\chi^2\lesssim5$ to identify nearest neighbors, increasing this iteratively up to $\chi^2\lesssim500$ until at least 30 neighbors are found.

Using the posterior marginal distributions sampled for each parameter, we catalog their $16^{\rm th}$, $50^{\rm th}$, and $84^{\rm th}$ percentile values, along with the $\chi^2$ of the fit using the median parameter values. Inevitably, some sources in our candidate list fall outside the support of the training distribution, i.e., our SED models do not describe these objects. While most such cases are discarded by our $\chi^2$ minimization cut (step 4 of Table \ref{tab:selection_cuts}), a small fraction remains. In these cases, the SBI model fails rejection sampling for the posterior proposals it produces, and does not successfully run inference in a reasonable time (120s). To prevent this failure mode, we impose a maximum inference time of $10\,{\rm s}$ per object for the baseline SBI model and $60\,{\rm s}$ for the \texttt{SBI++} MC sampling.

On average, inference using our baseline SBI model takes $\til25\,{\rm ms}$ per source on an Apple M3 Pro chip. Parameter inference on the same machine using MCMC takes $\til30\,{\rm s}$ per source, for a factor of 1000 speed-up in inference time using SBI. Similar to \citet{Wang23} and \citet{Nugent26}, we find that \texttt{SBI++} performs robust posterior inference in cases where some observations are missing or noisy, as demonstrated in Figure \ref{fig:sbi_mcmc_compare}. We extensively test the uncertainty calibration and performance of each SBI model using several metrics, as presented in Appendix \ref{appendix: sbi}.

\section{The RSG Catalog}
\label{sec:cat}

We now present a novel method to classify RSGs, AGB stars, and hotter blue stars using their inferred physical parameters. CMD-based selection criteria isolate the RSG branch, typically using straight lines defining a red and blue edge. Due to increased AGB contamination in such samples, we use the CMD criteria only as a heuristic for stellar classes. We construct multi-dimensional probability distributions in $T_{\rm eff}$--$\log(L/L_{\odot})$--$\tau_V$ phase space for each stellar class, and use these distributions to assign each star a membership likelihood. In Section \ref{ssec:completeness}, we quantify the performance of the selection method by calculating the completeness and purity of a classified RSG sample using a simulated ground truth dataset.

\subsection{RSG Selection}

\begin{figure*}[tbp]
\centering
\includegraphics[width=\textwidth]{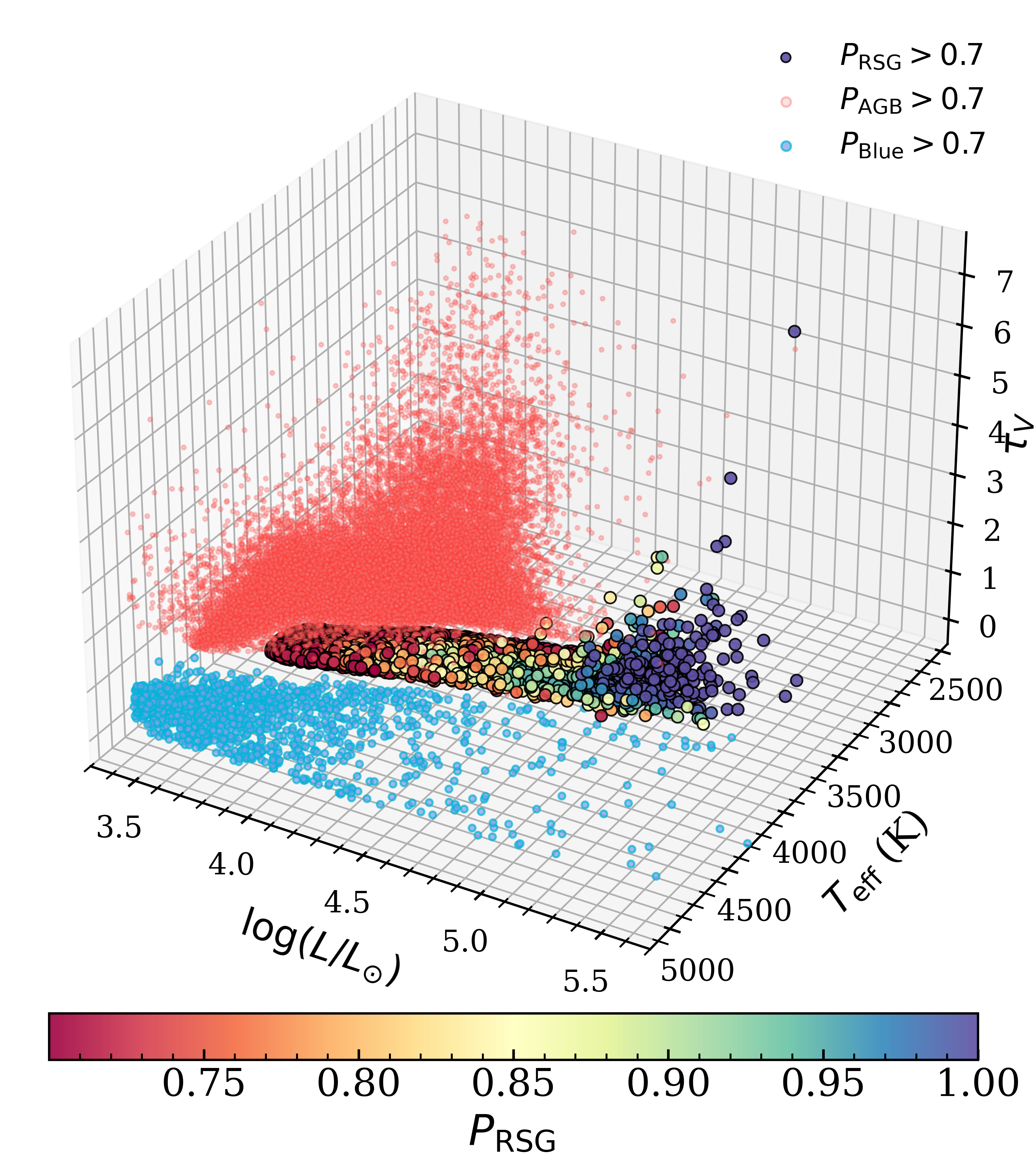}
\caption{Probabilistic classification of luminous stellar populations in NGC\,628 $(d = 8.6\,{\rm Mpc})$ based on the physical parameter phase space. We use the seed sample selected using a CMD in each galaxy to create a prior in the 3D phase space and build probability distributions for each stellar class. We use these distributions to assign a probability and class membership for each star in the galaxy, finding $5{,}024$ RSGs with $P_{\rm RSG} > 0.7$. We plot these RSGs colored by $P_{\rm RSG}$, as well as the stars classified as AGBs in coral, and blue-sequence stars colored in blue. We find that in the 3D parameter space, RSGs form a distinct locus from the other stellar populations, enabling us to form a clean sample of RSGs in each galaxy.
}\label{fig:kde_phase}
\end{figure*}

\begin{figure*}[tbp]
\centering
\includegraphics[width=\textwidth]{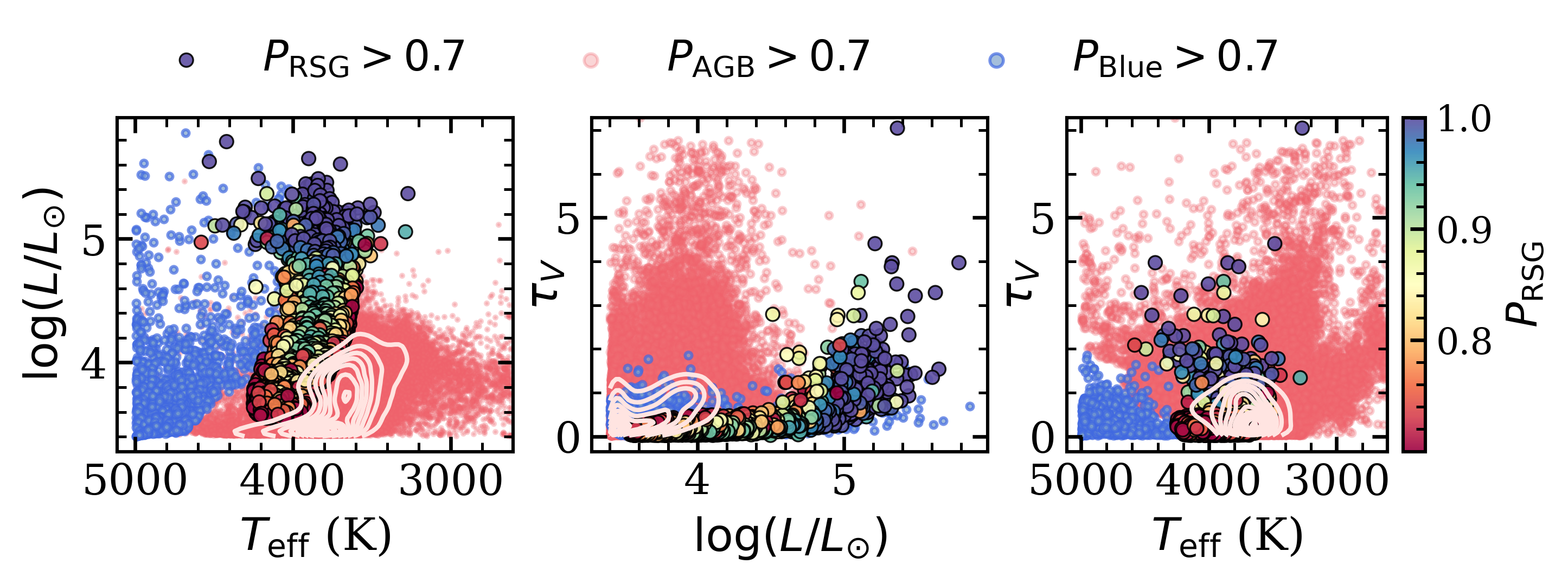}
\caption{2D projections of the physical parameter phase space shown in Figure \ref{fig:kde_phase}. We plot the RSGs colored by $P_{\rm RSG}$, AGB stars in coral, and blue-sequence stars in blue. Due to the large number of AGB stars, the red cloud is not representative of the AGB population density; the white contours trace the density of AGB stars in the respective projections. Any individual projection shows a large apparent overlap between AGBs and RSGs. Thus, the combined 3D phase space is necessary for separating RSGs from AGBs and other contaminants.
}\label{fig:kde_phase_proj}
\end{figure*}

\begin{figure}[tbp]
\centering
\includegraphics[width=0.4\textwidth]{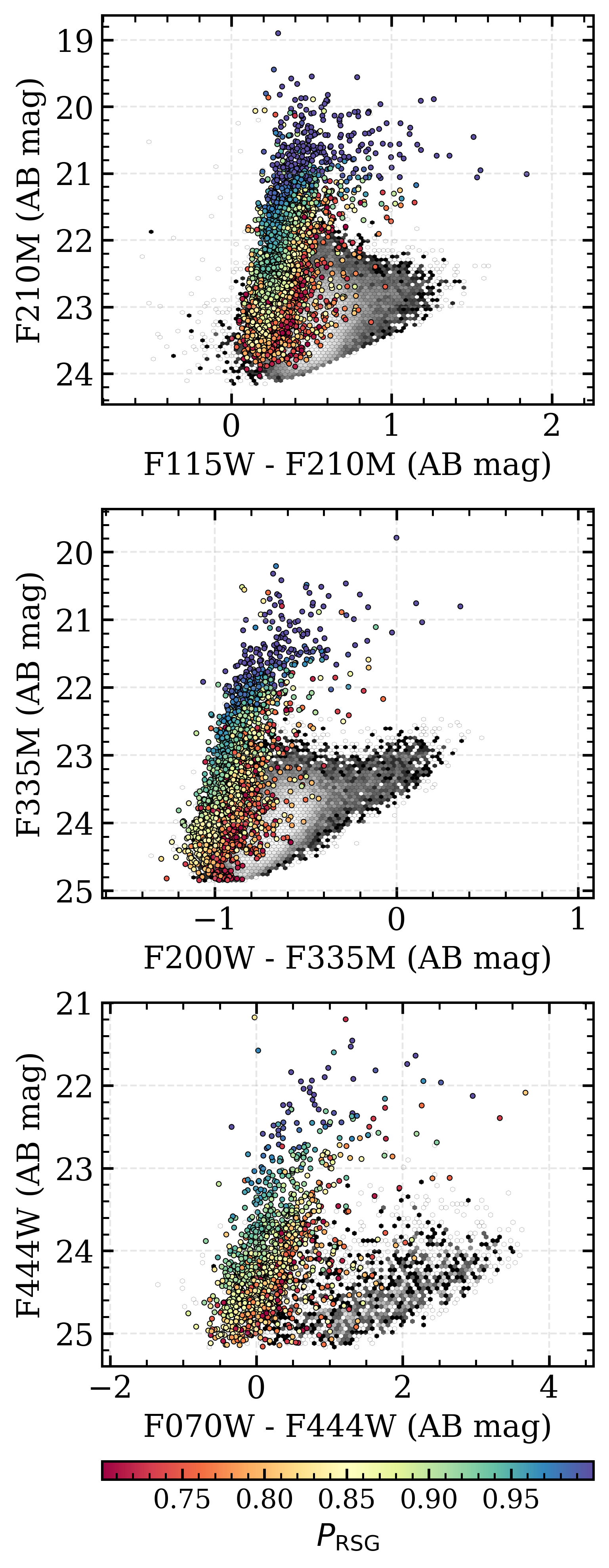}
\caption{RSGs selected using the KDEs plotted on example CMDs to show their locations. We show three different CMDs from NGC\,4258 ($d = 6.8\,{\rm Mpc}$), with the seed sample selected using the F115W-F210M CMD ({\it top panel}). The F200W-F335M CMD ({\it middle panel}) and F070W-F444W CMD ({\it bottom panel}) have only 54\% and 12\% of common sources with the F115W-F210M CMD ({\it top panel}), respectively. Despite the middle and bottom CMDs contributing little to the seed sample, we find that our selection method clearly isolates the RSG branch in each CMD, with a high assigned probability ($P_{\rm RSG} > 0.7$). We plot the RSGs colored by $P_{\rm RSG}$ and all sources in greyscale, colored by source number density.
}\label{fig:rsg_cmd}
\end{figure}

\begin{figure}[!h]
    \centering
    \includegraphics[width=0.45\textwidth]{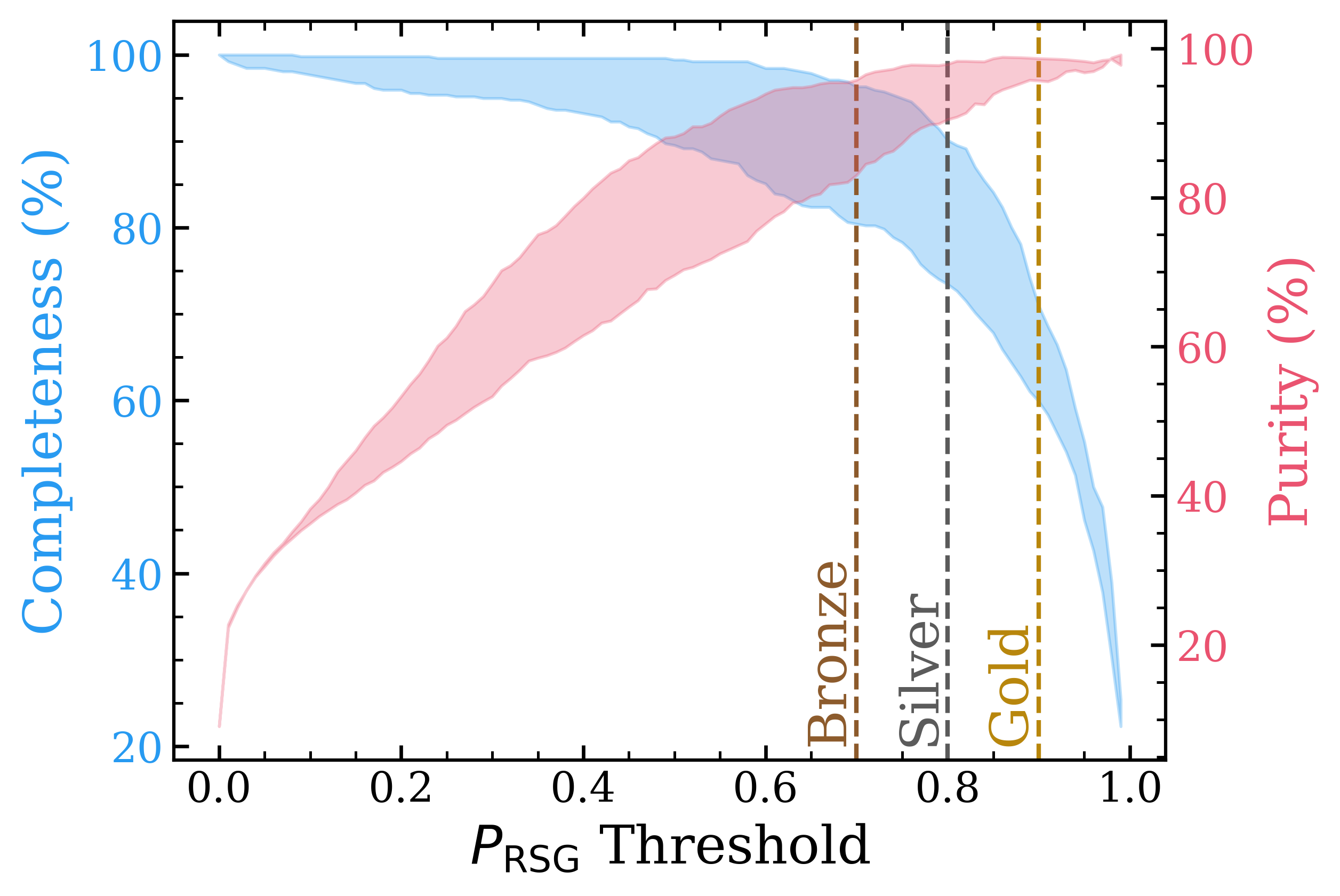}
    \caption{Completeness and purity of the KDE classifier using RSGs in NGC\,1365 $(d = 18.3\,{\rm Mpc})$ as a function of $P_{\rm RSG}$. We plot the lower and upper limits of the estimated completeness and purity functions in blue and red, respectively, and mark the bronze, silver, and gold cutoffs. We obtain good performance, with completeness and purity generally above $\sim80\%$ for the bronze sample. The gold sample is $>90\%$ pure as it is concentrated at higher luminosities than the silver and bronze tiers.
    }\label{fig:comp_purity}
\end{figure}

We aim to identify the RSG population of a galaxy empirically using the NIR photometry and inferred physical parameters of each star. Of the physical parameters, $T_{\rm eff}, \log(L/L_{\odot}), \tau_V$, and $T_{\rm dust}$ are intrinsic to a star, while $A_V$ and $R_V$ are properties of the host galaxy. For our sample, our modeling poorly constrains $T_{\rm dust}$ because we lack the needed MIR photometry (Figure \ref{fig:sbi_acc}). Hence, we restrict the stellar population classification to the $T_{\rm eff}, \log(L/L_{\odot}), \tau_V$ phase space, and define the region of this phase space that RSGs occupy. The luminous stellar populations we analyze form a nearly continuous density manifold in the 3D phase space, without obvious clustering into the constituent stellar classes. Thus, we use the CMD to pick a seed sample to serve as a prior for each stellar class \citep{Hirschauer20}, and inspect their location in parameter space. To form this seed sample, we use Gaussian Mixture Models to identify clusters in the CMD and refine their membership using simple physical parameter criteria. We detail the seed selection procedure in Appendix \ref{appendix: seed}.

Once we construct the seed sample, we use KDEs to build 3D probability distribution functions (PDFs) for each stellar class. We detail the KDE construction process and refinements to AGB contamination in the seed sample in Appendix \ref{appendix: seed}. In each galaxy, we evaluate the probability of all sources that pass our initial cuts (Table \ref{tab:selection_cuts}) using the stellar class-specific PDF. The seed sample requires detection in both filters used to construct the CMD. These filters may not cover all parts of the galaxy with NIRCam imaging, making it spatially incomplete. A fundamental advantage offered by the 3D phase space PDFs we construct is the ability to classify stars irrespective of their filter coverage, instead making use of their physical parameters. We use the KDEs to calculate the likelihood of each source being an RSG, AGB, or a blue-sequence star. This is filter-agnostic and includes all sources with derived parameters, beyond just the seed selection sample.

\begin{figure*}[t]
\centering
\includegraphics[width=\textwidth]{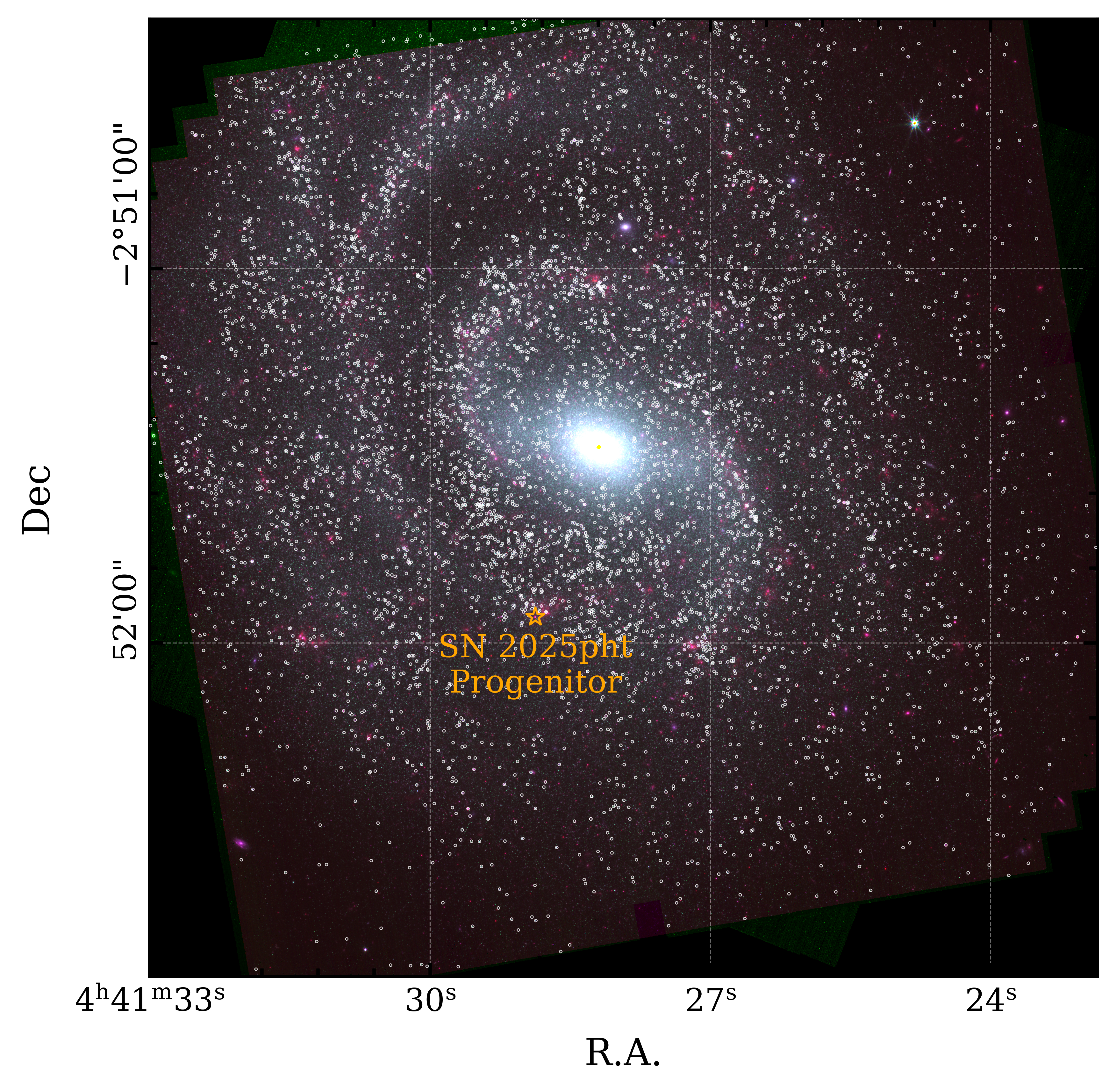}
\caption{Spatial distribution of RSGs ({\it white circles}) in NGC\,1637 $(d=12.0\,{\rm Mpc})$ selected such that $P_{\rm RSG} > 0.7$. Visually, we note that the RSGs closely track the star-forming regions of the galaxy, such as the spiral arms and clusters, as expected, since young RSGs do not have enough time to move far from their birth site. We also recover the progenitor star of recent SN\,2025pht in NGC\,1637 \citep{Kilpatrick25} independently from our analysis with $p_{\rm RSG} = 1.0$, marked using an orange star. The color image of the host galaxy is made using {\it JWST}/NIRCam observations ({\it JWST}-GO-3707, PI Leroy; {\it JWST}-GO-4793, PI Schinnerer) of NGC\,1637, using filters F444W, F277W, and F150W as RGB channels.
}\label{fig:ngc1367_rgb}
\end{figure*}

We normalize the likelihood such that the probability for each source ranges between zero and one, as 

\begin{equation}
P_{\rm RSG} = \frac{\mathcal{L}_{\rm RSG}}{\mathcal{L}_{\rm RSG} + \mathcal{L}_{\rm AGB} + \mathcal{L}_{\rm blue}}
\end{equation}

using the respective likelihoods of each class. We use $P_{\rm RSG}$ to select all sources with $P_{\rm RSG} > 0.7$ as RSGs and stratify the selected catalog into gold ($P_{\rm RSG} > 0.9$), silver ($0.9 > P_{\rm RSG} > 0.8$) and bronze tiers ($0.8 > P_{\rm RSG} > 0.7$). The probability assigned by our classification scheme is a relative likelihood that quantifies the degree of mixing between the stellar populations, that is, how likely a star is to be an RSG relative to the AGB and blue-sequence classes. We caution that these values are relative class weights, not absolute probabilities that a given star is an RSG. 

We show the 3D phase space occupied by the various stellar populations, with $P_{\rm RSG} > 0.7$, in Figure \ref{fig:kde_phase}, and their relevant 2D projections in Figure \ref{fig:kde_phase_proj}. Our new method using 3D phase space demonstrates that it is crucial to include all three parameters for RSG selection, as the RSG and AGB populations appear mixed in each 2D projection, but separate in the full 3D space. The parameters we use in our selection criteria reflect both the evolution of these stars on the HR diagram $(T_{\rm eff} - \log(L/L_{\odot})$ plane$)$, and their mass-loss relation $(\log(L/L_{\odot}) - \tau_V$ plane$)$, leading to cleaner separation. This is particularly relevant at low luminosities $(\log(L/L_{\odot}) \lesssim 4.2)$, where $\tau_V$ is needed to distinguish RSGs from AGBs. 

We overlay the selected RSGs on several CMDs in Figure \ref{fig:rsg_cmd} to assess whether the seed selection using a particular CMD yields meaningful results throughout the galaxy. Clustering in multi-dimensional space is difficult due to increased data sparsity and non-intuitive distance metrics \citep[the ``curse of dimensionality'';][]{Michel05}. We verify that our selection criteria are not biased or overfitted to represent only stars selected in the seed sample, and that they can recover the RSG locus across multiple CMDs. We also find excellent spatial coincidence of RSGs with the star-forming regions and spiral structure (if present) of their host galaxies, shown as an example for NGC\,1637 in Figure \ref{fig:ngc1367_rgb}. We also note that upon lowering the probability criteria (e.g., $P_{\rm RSG} > 0.5$), we find more stars diverge from the star-forming loci, indicating an increased fraction of AGB stars, which are distributed throughout the galaxy since they are older \citep{Lee25}.

\subsection{Completeness and Purity}
\label{ssec:completeness}

We now assess how well the instrument- and filter-agnostic RSG selection method performs in terms of completeness and purity. Since the underlying physical parameter distributions of RSGs are not established independently from this paper, it is not possible to evaluate the true completeness and purity of the sample we present. However, we can quantify the completeness and purity of sources classified as RSGs, as a function of $P_{\rm RSG}$, if the KDEs encounter a statistically similar distribution of stellar parameters. This is a useful benchmark of the KDE models, for example, when applied to future {\it JWST} or {\it Roman} observations, where we expect a similar distribution of inferred parameters.

Calculating the completeness and purity of an RSG sample for each galaxy requires a ground-truth sample to compare against the KDE-predicted classification. To create this ground-truth sample, we generate a set of RSG and AGB physical parameters by resampling from a GMM fit to high-probability ($p>0.8$) RSGs and AGBs, respectively. The temperature scales of RSGs vary across galaxies, such that sampling from the global distribution of the full RSG sample is no longer representative of any individual galaxy (Figure \ref{fig:rsg_metallicity_dependence}). However, we expect the RSG luminosity function to show similarities across galaxies, with roughly similar lower and upper bounds. The observed differences between the luminosity functions are likely due to selection effects limiting RSG recovery at lower luminosities due to increased AGB contamination. To account for this, we use the combined luminosity function of RSGs across all galaxies to sample RSGs below the faint limit of each galaxy's luminosity distribution, extending to the global luminosity minimum ($\log(L/L_{\odot}) \sim 3.55$). We construct the truth sample using 500 RSGs and 5000 AGBs per galaxy with resampled physical parameters, roughly tracking the $\sim 10:1$ ratio of AGBs to RSGs.

We test the KDE performance on three different datasets: (1) we use the ground-truth parameters to generate stellar SEDs using the {\tt DUSTY} models (Section \ref{sec:models}), and re-infer physical parameters using an appropriate SBI model (Section \ref{sec:sbi}), which is then passed into the KDE, (2) we build a KDE using only half the data available for a galaxy, and assign $P_{\rm RSG}$ to the left-out sources, which we compare to results from using the full dataset, and (3) we pass the ground-truth parameter set as-is to the KDE model. The first method tests classification performance under the assumption that the RSG parameter distributions we infer from the SBI models reasonably represent the true parameters of RSGs. This is a lower-bound estimate of completeness and purity since the SBI model adds an intrinsic scatter to the true parameters (Figure \ref{fig:sbi_acc}). The second test is leave-out validation, as a proxy for cases when new data becomes available. The final test represents the upper bound of KDE performance, where it encounters a dataset very similar to one it was trained on. We present the results of tests (1) and (3) in Table \ref{tab:comp_purity}. We find that leave-out validation recovers RSGs very well, with only 1--10$\%$ of the sample lost when only half the data is used to build the KDEs.

From Table \ref{tab:comp_purity}, we find that the completeness of the RSG sample typically ranges between $\sim65\%$--$80\%$ at the lower limit for the bronze sample, and $\sim80\%$--$95\%$ at the upper limit, indicating good recovery of the RSG population. Similarly, we find a purity of $\sim80\%$--$95\%$ at the lower limit and $>95\%$ at the upper limit for the bronze sample, with the purity improving in the silver and gold tiers. We show the full completeness and purity curves, taking NGC\,1365 as an example, in Figure \ref{fig:comp_purity}. For science applications that require a pure and complete sample of RSGs, we find that the silver tier offers high purity ($>90\%$) and is reasonably complete ($60$--$90\%$). We conclude that the KDE selection method will deliver a clean sample of RSGs when working with datasets similar to the one we analyze in this paper. We show example rows from the RSG catalog in Table \ref{tab:catalog_example}.

In starburst galaxies such as NGC\,3034 or NGC\,4038, the heavy imbalance between the number of AGB stars and RSGs can make the RSG seed locus difficult to isolate, leading to lower completeness values than the other galaxies. More sophisticated selection techniques will be necessary to accurately identify RSGs in galaxies which display significantly different CMD morphologies than typical star-forming spirals, and will be explored in future work.  These photometric classifications can also be tested with multi-epoch \textit{Roman} and \textit{JWST} observations. RSGs and AGBs differ in pulsation amplitude and occupy partially distinct period--luminosity sequences \citep{Conroy18}. This provides an independent handle on their classifications, which can be combined with our framework to understand stellar variability across spectral types.

\startlongtable
\begin{deluxetable*}{lccc|ccc|ccc|ccc|c}
\tablecaption{Completeness and Purity Statistics Across the Galaxy Sample \label{tab:comp_purity}}
\tablewidth{0pt}
\tablehead{
\colhead{} & 
\multicolumn{3}{c}{Lower-limit Completeness} & 
\multicolumn{3}{c}{Lower-limit Purity} &
\multicolumn{3}{c}{Upper-limit Completeness} &
\multicolumn{3}{c}{Upper-limit Purity} \\
\cline{2-4} \cline{5-7} \cline{8-10} \cline{11-14}
\colhead{Galaxy} & 
\colhead{Bronze} & 
\colhead{Silver} & 
\colhead{Gold} &
\colhead{Bronze} & 
\colhead{Silver} & 
\colhead{Gold} & 
\colhead{Bronze} & 
\colhead{Silver} & 
\colhead{Gold} &
\colhead{Bronze} & 
\colhead{Silver} & 
\colhead{Gold} &
\colhead{$N_{\rm RSG}$}
}
\startdata
NGC\,628 & 0.82 & 0.65 & 0.29 & 0.84 & 0.91 & 0.95 & 0.98 & 0.90 & 0.42 & 0.97 & 0.99 & 1.00 & $5{,}024$ \\
NGC\,1365 & 0.77 & 0.69 & 0.54 & 0.87 & 0.92 & 0.97 & 0.97 & 0.91 & 0.71 & 0.96 & 0.98 & 0.99 & $2{,}724$ \\
NGC\,1637 & 0.63 & 0.47 & 0.28 & 0.79 & 0.93 & 0.99 & 0.98 & 0.93 & 0.44 & 0.98 & 0.99 & 1.00 & $4{,}565$ \\
NGC\,3034 & 0.54 & 0.51 & 0.49 & 0.95 & 0.97 & 0.98 & 0.61 & 0.59 & 0.54 & 0.98 & 0.99 & 1.00 & $2{,}682$ \\
NGC\,4038 & 0.65 & 0.59 & 0.48 & 0.77 & 0.84 & 0.90 & 0.75 & 0.70 & 0.59 & 0.97 & 0.98 & 0.99 & $21{,}638$ \\
NGC\,4258 & 0.79 & 0.64 & 0.35 & 0.93 & 0.96 & 0.99 & 0.97 & 0.90 & 0.42 & 0.98 & 0.99 & 0.99 & $6{,}609$ \\
NGC\,4449 & 0.70 & 0.60 & 0.41 & 0.77 & 0.84 & 0.91 & 0.98 & 0.95 & 0.70 & 0.92 & 0.94 & 0.98 & $2{,}951$ \\
NGC\,4485 & 0.66 & 0.48 & 0.27 & 0.76 & 0.88 & 0.94 & 0.95 & 0.83 & 0.34 & 0.95 & 0.98 & 0.99 & $9{,}916$ \\
NGC\,4536 & 0.64 & 0.60 & 0.48 & 0.94 & 0.98 & 0.99 & 0.72 & 0.69 & 0.56 & 0.99 & 0.99 & 0.99 & $3{,}184$ \\
NGC\,4548 & 0.63 & 0.43 & 0.05 & 0.94 & 0.97 & 0.96 & 0.96 & 0.88 & 0.08 & 0.98 & 0.99 & 0.95 & $1{,}288$ \\
NGC\,5194 & 0.69 & 0.44 & 0.25 & 0.88 & 0.93 & 0.96 & 0.99 & 0.91 & 0.30 & 0.96 & 0.98 & 0.97 & $15{,}508$ \\
NGC\,5236 & 0.79 & 0.67 & 0.29 & 0.77 & 0.89 & 0.94 & 0.97 & 0.91 & 0.48 & 0.95 & 0.97 & 0.97 & $11{,}147$ \\
NGC\,5457 & 0.78 & 0.65 & 0.38 & 0.93 & 0.97 & 1.00 & 0.94 & 0.87 & 0.51 & 0.97 & 0.97 & 0.99 & $8{,}055$ \\
NGC\,5643 & 0.79 & 0.68 & 0.46 & 0.95 & 0.98 & 1.00 & 0.98 & 0.93 & 0.58 & 1.00 & 1.00 & 1.00 & $5{,}087$ \\
NGC\,7320 & 0.85 & 0.73 & 0.51 & 0.72 & 0.86 & 0.92 & 0.85 & 0.75 & 0.35 & 0.97 & 0.99 & 1.00 & $841$ \\
\hline
& & & & & & & & & & & & \textbf{Total} & $\nrsg$ \\
\enddata
\tablecomments{Completeness and purity values for different probability thresholds. The lower limit values correspond to the fully simulated SED inputs with SBI re-inferred parameters passed to the KDE, while the upper limit values correspond to GMM resampled values passed as-is to the KDE}
\end{deluxetable*}

\begin{deluxetable*}{lllrrrrcccr}[!t]
\tablecaption{Example catalog rows of the RSG catalog\label{tab:catalog_example}}
\tablewidth{0pt}
\tabletypesize{\scriptsize}
\tablehead{
\colhead{RSG{$\_$}ID} &
\colhead{RA (h:m:s)} &
\colhead{Dec (d:m:s)} &
\colhead{SNR} &
\colhead{$T_{\rm eff,med}$} &
\colhead{$T_{\rm eff,1\sigma_-}$} &
\colhead{$T_{\rm eff,1\sigma_+}$} &
\colhead{$\log(L/L_{\odot})_{\rm med}$} &
\colhead{$\log(L/L_{\odot})_{1\sigma_-}$} &
\colhead{$\log(L/L_{\odot})_{1\sigma_+}$} &
\colhead{$P_{\rm RSG}$} \\
}
\startdata
ngc4258{$\_$}1{$\_$}0{$\_$}613.0 & 12:18:55.19 & +47:16:50.88 & 1285.10 & 2803.36 & 163.42 & 583.88 & 5.17 & 0.05 & 0.11 & 1.00 \\
ngc4038{$\_$}0{$\_$}1{$\_$}582.0 & 12:01:55.61 & -18:52:44.06 & 848.80 & 3780.43 & 404.26 & 575.32 & 5.63 & 0.14 & 0.19 & 1.00 \\
ngc4548{$\_$}0{$\_$}0{$\_$}684.0 & 12:35:21.79 & +14:29:37.95 & 776.50 & 3620.12 & 233.34 & 662.26 & 5.02 & 0.16 & 0.44 & 1.00 \\
ngc3034{$\_$}0{$\_$}7{$\_$}1406.0 & 09:55:52.34 & +69:40:46.92 & 1042.40 & 3472.65 & 532.82 & 586.81 & 5.57 & 0.18 & 0.19 & 1.00 \\
ngc1365{$\_$}0{$\_$}1{$\_$}616.0 & 03:33:43.58 & -36:08:49.15 & 471.20 & 3916.98 & 351.63 & 344.85 & 5.39 & 0.26 & 0.18 & 1.00 \\
ngc4038{$\_$}0{$\_$}1{$\_$}1469.0 & 12:01:54.73 & -18:52:50.06 & 621.40 & 3656.44 & 410.86 & 428.66 & 5.53 & 0.18 & 0.20 & 1.00 \\
ngc3034{$\_$}0{$\_$}7{$\_$}95.0 & 09:55:55.31 & +69:40:52.29 & 4329.70 & 3014.45 & 277.02 & 582.45 & 5.25 & 0.11 & 0.19 & 1.00 \\
ngc3034{$\_$}0{$\_$}7{$\_$}85.0 & 09:55:52.82 & +69:40:44.49 & 2974.20 & 3132.34 & 195.06 & 224.12 & 5.36 & 0.15 & 0.06 & 1.00 \\
ngc3034{$\_$}0{$\_$}7{$\_$}73.0 & 09:55:53.55 & +69:40:48.10 & 3196.80 & 3079.90 & 246.59 & 311.91 & 5.38 & 0.17 & 0.20 & 1.00 \\
\enddata
\tablecomments{The full catalog of \til1.5 million sources with physical parameters, and probabilistic classification is available as a machine-readable table on Zenodo. We sort the candidates above in decreasing order of $P_{\rm RSG}$. The physical parameter values represent the median, and $1\sigma$ lower and upper error bars.}
\end{deluxetable*}

\section{Discussion}
\label{sec:discussion}

In this Section, we use our catalog of extragalactic RSGs to summarize their temperature, luminosity, and mass-loss rate distributions. We also perform a preliminary analysis of correlations between the RSG physical properties and their environmental properties such as metallicity or star formation rate. We investigate the dust-enshrouded RSGs as candidate late evolutionary stages approaching core collapse, and highlight the power of our catalog in identifying short-lived phases of RSGs. 

\subsection{Temperature Scale of RSGs}

\begin{figure*}[htp]
\centering
\includegraphics[width=0.32\textwidth]{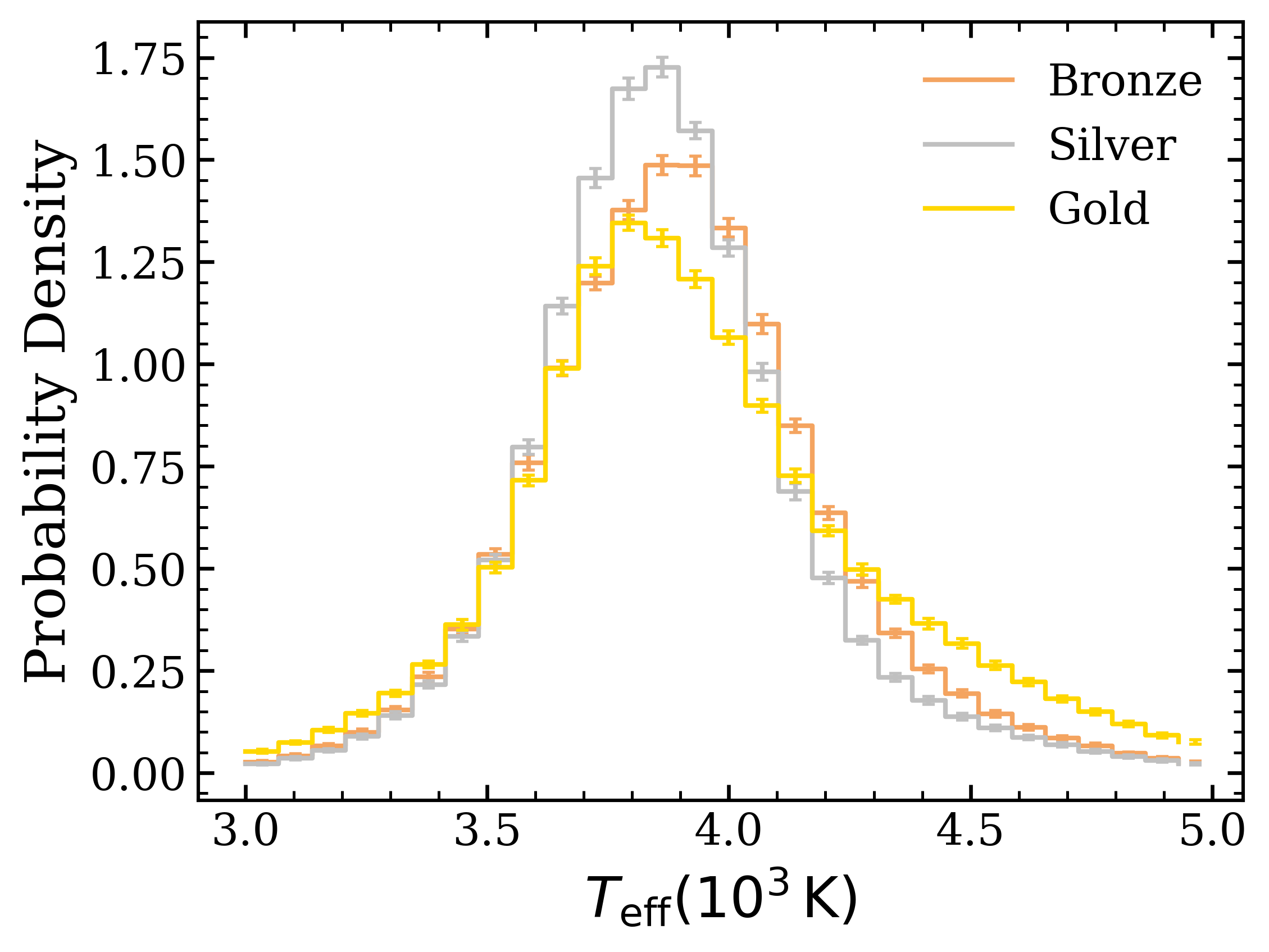}
\includegraphics[width=0.32\textwidth]{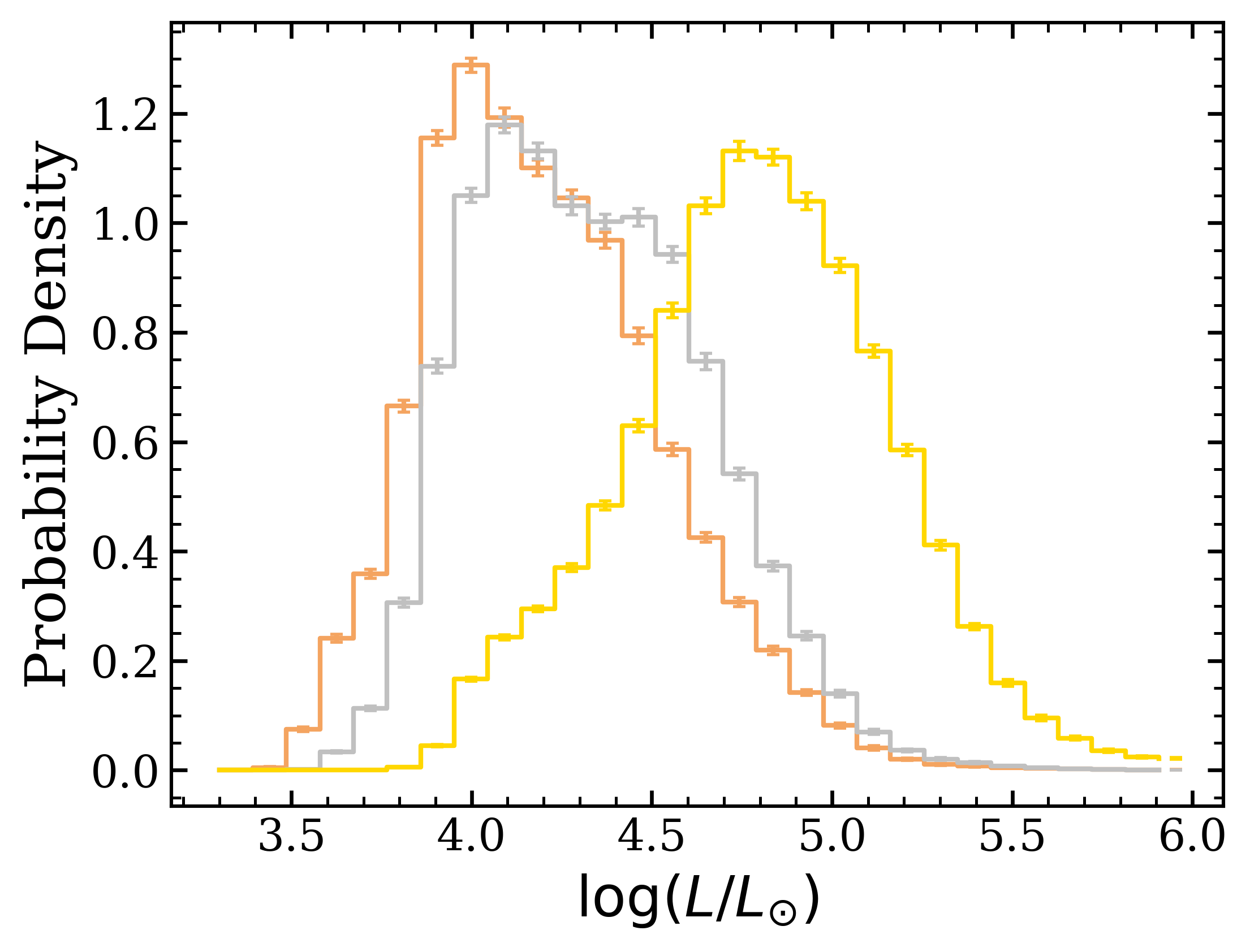}
\includegraphics[width=0.32\textwidth]{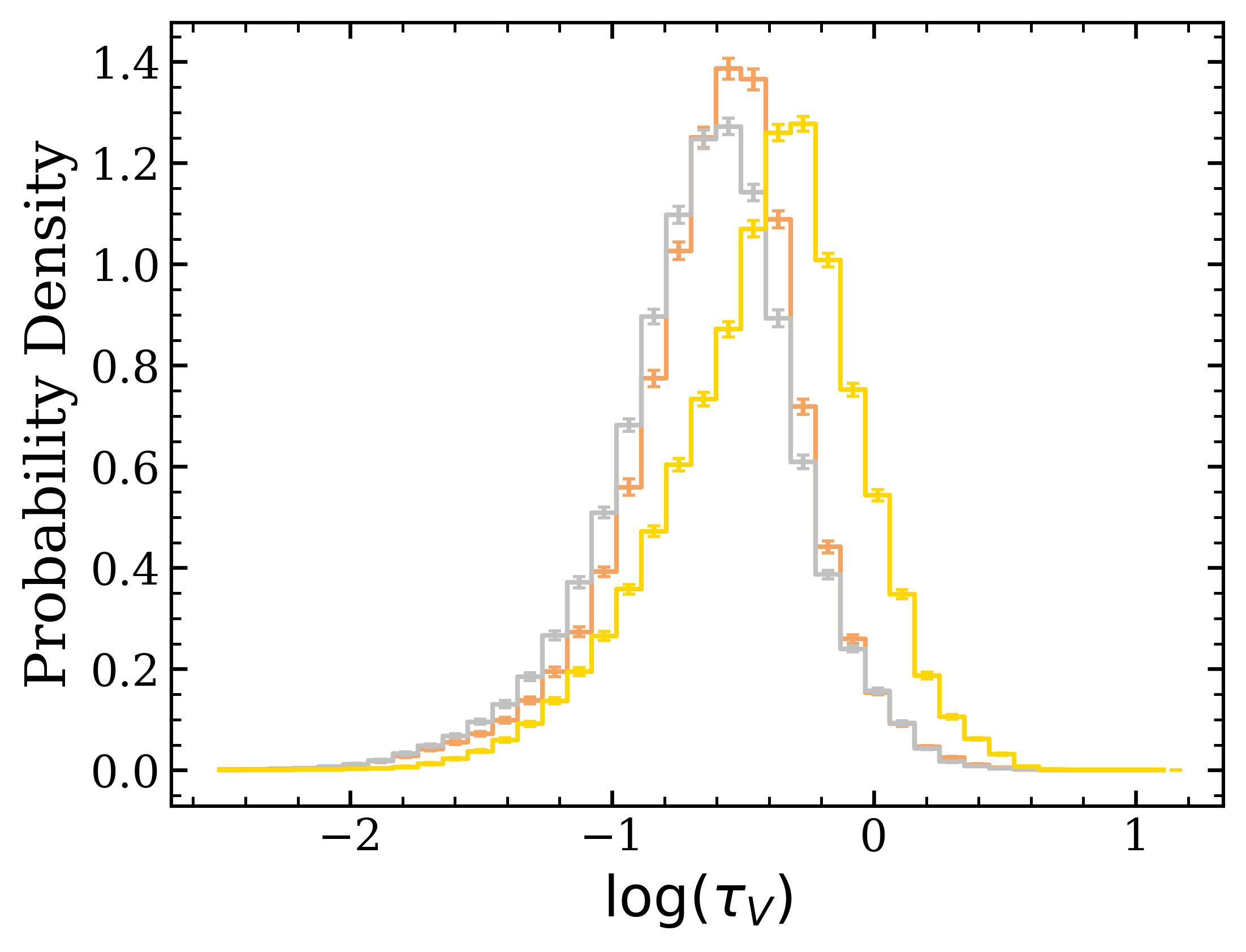}
\caption{Effective temperature, bolometric luminosity, and dust optical depth distributions of the RSG catalog. We show histograms of the median physical parameters with $1\sigma$ resampled errors. We split the catalog into bronze $(P_{\rm RSG} > 0.7)$, silver $(P_{\rm RSG} > 0.8)$, and gold tiers $(P_{\rm RSG} > 0.9)$, with each tier containing a progressively cleaner sample of RSGs, at the cost of completeness (Table \ref{tab:comp_purity}). We find that our inferred parameters generally follow expected theoretical ranges, and the gold sample contains the more luminous RSGs, which can be selected with high confidence due to low AGB contamination. 
}\label{fig:param_dist}
\end{figure*}

We compare our inferred RSG $T_{\rm eff}$ to MIST single-star evolution tracks \citep{Choi16, Dotter26}, and test the metallicity dependence of the RSG temperature scale predicted by stellar structure models, using the global metallicity measurements of our host galaxies. RSGs follow nearly vertical tracks on the HR diagram, occupying a narrow range of $T_{\rm eff}$ set by the Hayashi limit, which is predicted to shift to cooler temperatures at higher metallicity \citep{Hayashi61}. This dependence has been difficult to quantify empirically, primarily due to statistical uncertainty from small sample sizes in both the number of RSGs (\til100 per galaxy) and the range of host environments to date \citep{Dorda2016, Tabernero2018}. Our homogeneously derived $T_{\rm eff}$ values across 15 galaxies spanning $Z/Z_{\odot}\sim0.25$--$1.45$ (Table \ref{tab:galaxy_sample}) let us test this prediction on a significantly larger scale than was possible before. Anchoring the $T_{\rm eff}$-$Z$ dependence is important for stellar structure predictions of Type II-P progenitors, their hydrogen envelope masses before core collapse \citep{Hiramatsu2021SP}, shock breakout signatures \citep{Chun18, Goldberg2022}, and bolometric corrections \citep{Davies13}.

\begin{figure*}[t]
\centering
\includegraphics[width=\textwidth,height=0.9\textheight,keepaspectratio]{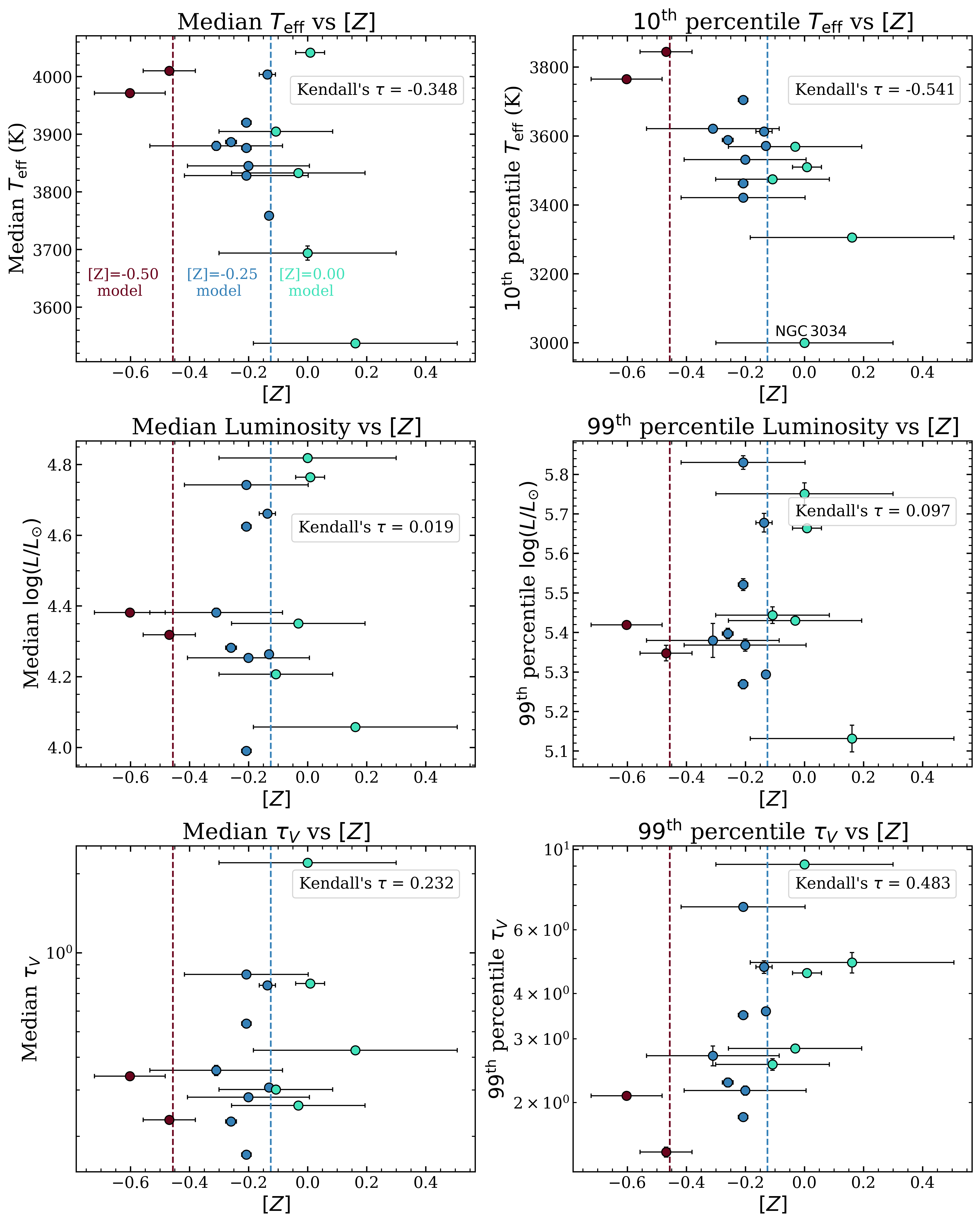}
\caption{Variation of RSG physical parameters with metallicity across our multi-galaxy sample, sampling metallicity from $[Z] \sim -0.6$ to $+0.2$. We color the points by the specific discrete MARCS grid used to fit the SEDs: $[Z]=-0.50$ in red, $[Z]=-0.25$ in blue, and $[Z]=0.00$ in green. We quantify the correlation between quantities in each panel using Kendall's $\tau$ statistic and find that RSG effective temperatures decrease with increasing metallicity. In contrast, the bolometric luminosity and the upper luminosity limit show no strong correlation with metallicity. The upper end of the $\tau_V$ distribution appears to positively correlate with metallicity, while the median optical depth is independent of metallicity.
}\label{fig:rsg_metallicity_dependence} 
\end{figure*}

We present the overall temperature distribution of our RSG catalog in Figure \ref{fig:param_dist}. To account for uncertainties in the overall parameter distributions, we draw 50 samples for each source using the median, $16^{\rm th}$ and $84^{\rm th}$ percentiles of the inferred parameters and construct the distributions. We observe that RSG temperatures follow similar ranges between the gold, silver, and bronze tiers, with the gold sample containing the coolest RSGs. As expected from literature RSG population studies, we find that the temperature spread of RSGs within a galaxy is typically narrow (median of $1\,\sigma$ spread $\Delta T$ across galaxies is $\pm 150\,{\rm K}$; \citealt{Davies13, Chun21}). We observe that the median temperature of the RSG population varies among galaxies and is indeed influenced by metallicity (Figure \ref{fig:rsg_metallicity_dependence}), which increases the standard deviation of RSG temperatures in Figure \ref{fig:param_dist} to $\sim230\,$K. The global RSG $T_{\rm eff}$ distribution of our catalog peaks at 3{,}888~K (between spectral types K7 and M0), and spans the range from 3{,}000--5{,}000~K (spectral types K3--M6). The cool tail of our inferred $T_{\rm eff}$ spans the range 2{,}800--3{,}600~K (spectral types M2--M8) and also varies with metallicity (Figure \ref{fig:rsg_metallicity_dependence}). The true underlying $T_{\rm eff}$ distribution is likely narrower, and we observe an inflated dispersion due to modeling uncertainties (Figure \ref{fig:sbi_acc}). 

We find a statistically significant anti-correlation: the $10^{\rm th}$ percentile of $T_{\rm eff}$ per galaxy, used here as a proxy for the Hayashi limit, decreases with increasing metallicity ({\it top right panel;} Figure \ref{fig:rsg_metallicity_dependence}). The quantitative relationship between the Hayashi limit/cool $T_{\rm eff}$ RSGs and metallicity has not been established, but we find an approximately linear trend between the two quantities. We use the $10^{\rm th}$ percentile value to avoid bias from some anomalously cool RSGs in the sample ({\it see below}). To quantify the strength of the correlation, we use Kendall's $\tau$ statistic, which is a non-parametric test of correlation between two variables. We obtain $\tau=-0.54$ with a p-value of $p=0.006$, indicating a strong correlation between the cool end of RSG temperatures and $Z$. We find a weaker correlation between the population median $T_{\rm eff}$ and $Z$, with $\tau=-0.39$, corresponding to a p-value of $p=0.07$ ({\it top left panel;} Figure \ref{fig:rsg_metallicity_dependence}). As found in \citet{Tabernero2018}, comparing RSG $T_{\rm eff}$ across different metallicities shows significant variations. For example, the dusty, starburst galaxy NGC\,3034 hosts one of the coolest populations of RSGs in our catalog, with the $10^{\rm th}$ percentile $T_{\rm eff}$ of $3{,}165\,{\rm K}$, significantly offset from the other galaxies ({\it top right panel;} Figure \ref{fig:rsg_metallicity_dependence}).

\begin{figure}[htp]
\centering
\includegraphics[width=0.48\textwidth]{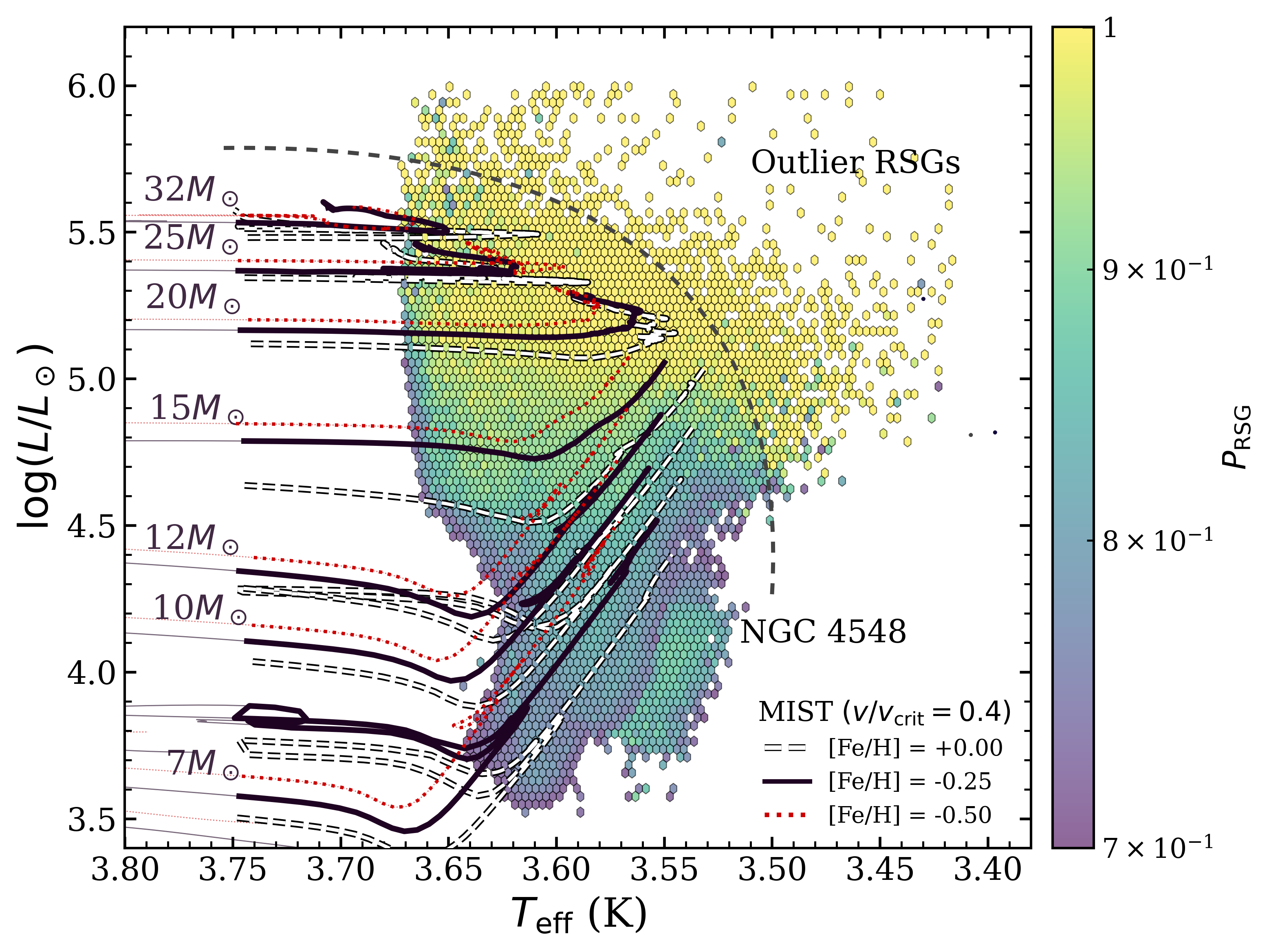}
\caption{HR diagram of RSGs ($P_{\rm RSG} > 0.7$) overlaid with MIST evolutionary tracks of stars with ZAMS masses ranging from $7\,{\rm M}_{\odot}-32\,{\rm M}_{\odot}$, with $v/v_{\rm crit} = 0.4$. We plot tracks with the three metallicity bins we adopt in our analysis, namely $[Z] = 0.0, -0.25$ and $-0.5$. Crucially, we note that our selection method is completely empirical, classifying RSGs solely based on CMD population density and their 3D parameter phase space. Yet, we find that our RSG sample shows excellent agreement with theoretical stellar tracks. A subpopulation of cool and/or luminous RSGs occupies a unique region of the HR diagram, veering off the predicted evolutionary tracks, rightward of the grey dotted curve.}\label{fig:hr_mist} 
\end{figure}

Compared to theoretical single-star evolution tracks from MIST \citep[v2.5;][]{Choi16, Dotter26}, we find excellent agreement between our derived $T_{\rm eff}$ and $\log(L/L_{\odot})$ values and their expected position on the HR diagram. We use the solar-scaled models, with ZAMS masses ranging from $7-32\,{\rm M}_{\odot}$ for RSGs, and $v/v_{\rm crit} = 0.4$. We highlight the He-burning stages of these evolutionary tracks in Figure \ref{fig:hr_mist}. Interestingly, we observe that about $0.67\%$ of the RSG population (682 RSGs) deviates from the evolutionary tracks. We select this sample as RSGs having $T_{\rm eff} < 3000$\,K or $\log(L/L_{\odot}) > 5.55$ as the evolutionary tracks do not predict such RSGs (Figure \ref{fig:hr_mist}). Such cool RSGs, which are not consistent with theoretical tracks, have been found in very small numbers to date (\til10), which we expand by an order of magnitude using our catalog \citep{Beasor16, Dorda2016, Tabernero2018}. This subpopulation of cool, luminous RSGs is also dusty, is likely experiencing enhanced mass loss and envelope inflation, and is discussed in Section \ref{ssec:rsg_progenitors}. 

\subsection{Luminosity Function of RSGs and its Host Dependence}

\begin{figure}[htp]
\centering
\includegraphics[width=0.46\textwidth]{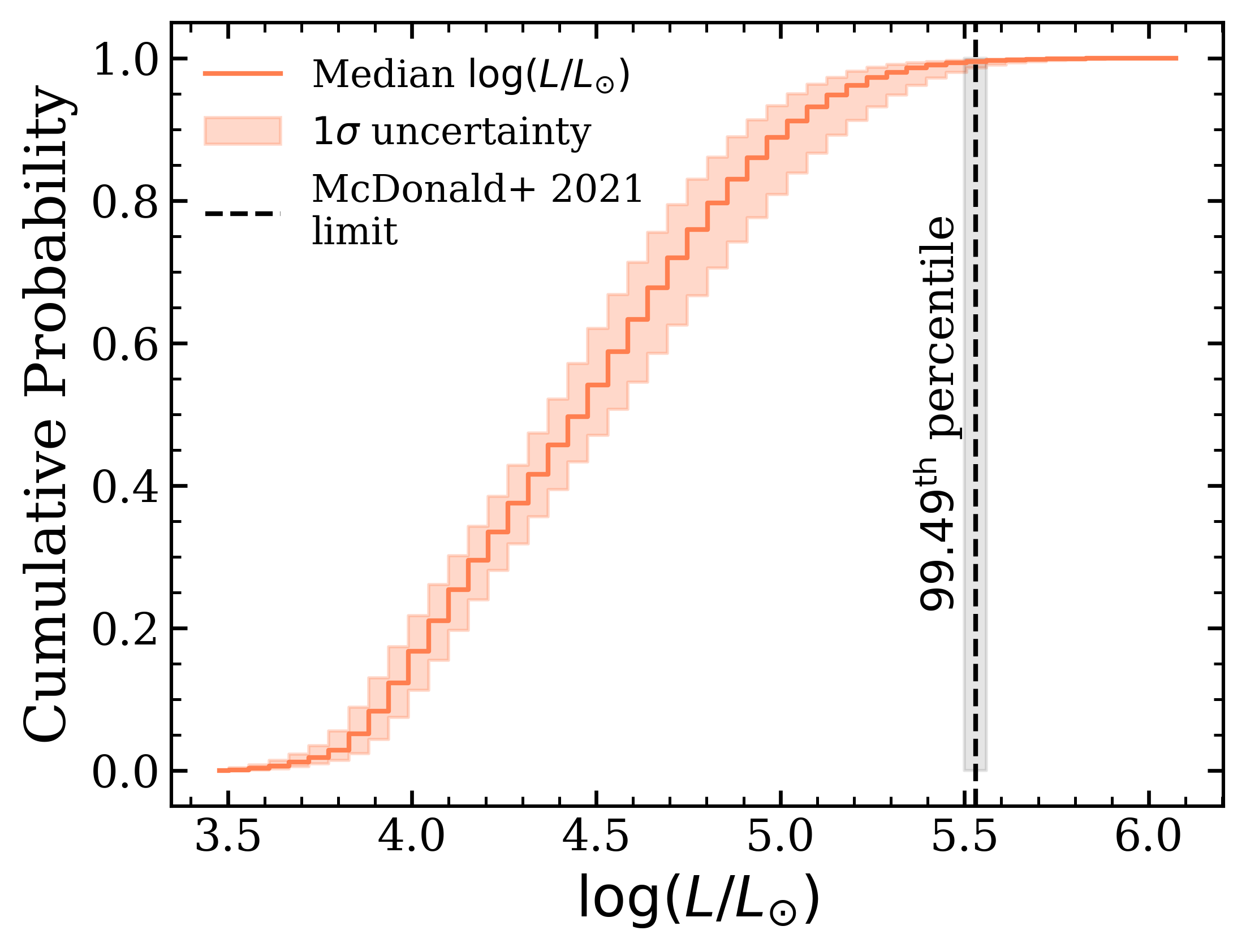}
\caption{Cumulative distribution function of RSG luminosity, combined across all our galaxies ($P_{\rm RSG} > 0.7$). We find that the empirical HD limit at $\log(L/L_{\odot}) = 5.53\pm 0.03$ for M31 \citep{McDonald22} falls at the $99.49^{\rm th}$ percentile of our sample, i.e., only $0.59\%$ of RSGs in our sample are more luminous. The nature of the most luminous RSGs in our catalog is unclear and requires further observations to confirm the classification.}\label{fig:hd_cumulative}
\end{figure} 

Our catalog is well-suited to characterize the luminosity function of RSGs, as the $1{-}5\,\mu{\rm m}$ SEDs capture the bulk of their emitted flux. Here, we analyze the complete sample with $P_{\rm RSG} > 0.7$ across all galaxies to construct the overall luminosity function in Figure \ref{fig:hd_cumulative}. We recover the RSG luminosity function down to $\log(L/L_{\odot}) \approx 3.5$, probing a phase space that is generally inaccessible due to contamination from AGB stars, which are far more common at these luminosities and occupy similar regions of the CMD. The faint RSGs and luminous AGBs are of interest as potential progenitors of the proposed electron capture supernova scenario \citep{Nomoto1984, Moriya2014, Hiramatsu2021, Limongi2024, Rose2025}.

RSGs have been observed to have an empirical maximum luminosity limit, often called the ``Humphreys-Davidson'' (HD) limit \citep{Humphreys79, Massey2003}. The HD limit is thought to be an imprint of intense mass-loss in massive RSGs ($M_{\rm ZAMS} \approx 20$--$30\,M_{\odot}$), causing them to shed a significant part of their envelopes and evolve into hotter blue stars (YSGs/BSGs). Hence, the HD limit sets the maximum mass for a star to undergo an RSG phase, and potentially explode as a Type II-P SN. The influence of metallicity on radiative winds in RSGs is theoretically expected to drive increased mass loss at higher $Z$, with stronger winds driving the increased-opacity material outward \citep{Vink2001, Maeder2003}. At the population level, this should manifest as a lower HD limit at high $Z$, and a larger fraction of stars lose their hydrogen envelopes and evolve blueward \citep{Cheng26}. Observationally, RSG populations of nearby galaxies such as LMC, SMC, and M31 find the HD limit to be consistent with each other within $\sim0.1\,$dex, and independent of metallicity, in tension with theoretical predictions \citep{McDonald22, Davies2018}. 

At the luminous RSG tail of our catalog, we find that the HD limit of $\log(L/L_{\odot}) = 5.53 \pm 0.03$, inferred in \citet{McDonald22} for RSGs in M31, falls at the $99.49^{\rm th}$ percentile of our sample, indicating good agreement with previous analyses (Figure \ref{fig:hd_cumulative}). A subpopulation of RSGs shows luminosities above the expected HD limit. The nature of these stars is unclear. We hypothesize that this may be due to stars briefly crossing the HD limit to evolve to hotter stellar classes, pulsations in terminal RSGs \citep{Laplace26}, or due to line-of-sight effects with non-spherical CSM geometry around these stars, leading to an overestimate of their true luminosity. Additional observations are needed to confirm whether these stars are RSGs and to examine their evolution in more detail.

\begin{figure*}[tbp]
\centering
\includegraphics[width=0.95\textwidth]{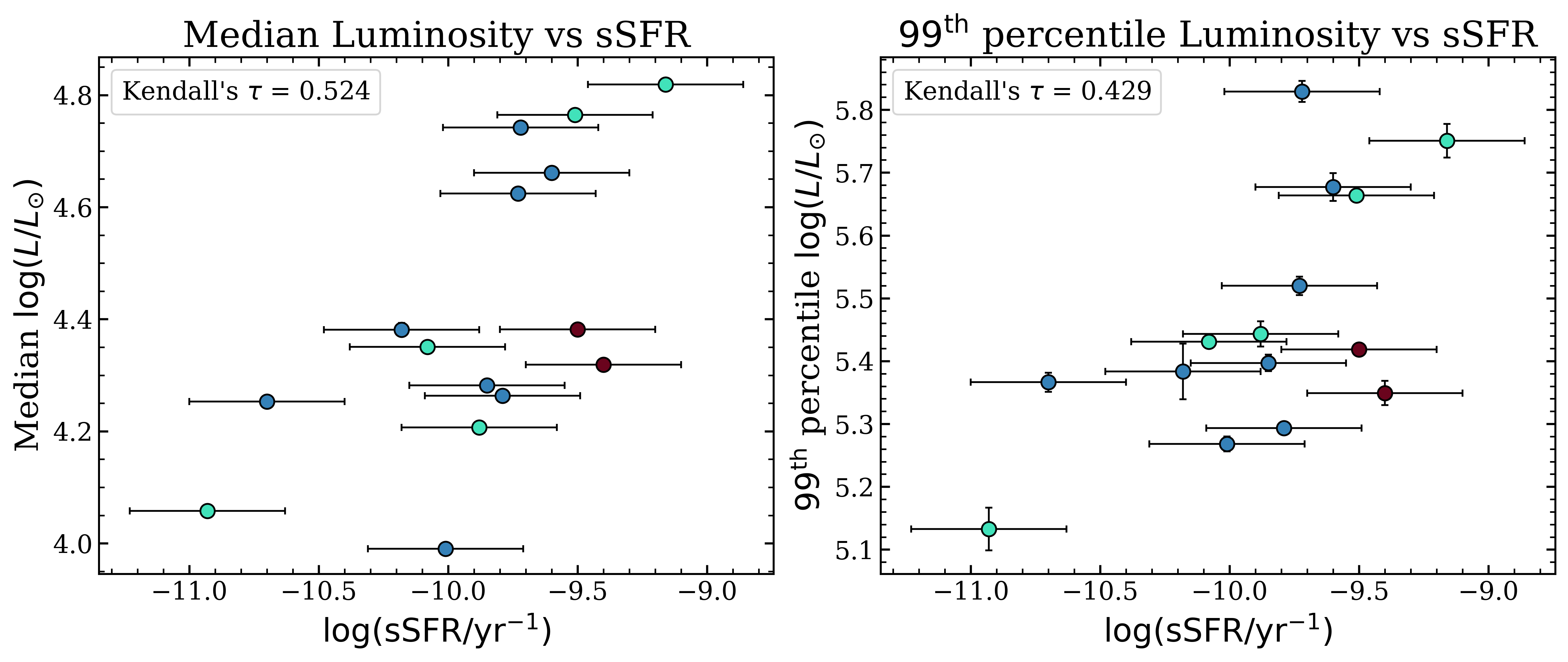}
\caption{Impact of host galaxy specific star formation rate (sSFR) on the observed luminosities of their RSG populations. We find a statistically significant correlation between both the median population luminosities and the $99^{\rm th}$ percentile population luminosities and sSFR. We hypothesize that this may be due to an increased number of massive RSGs in high sSFR galaxies, leading to a larger number of stars near/crossing the HD limit. }\label{fig:rsg_sfr_dependence} 
\end{figure*}

Consistent with \citet{McDonald22}, we do not find evidence for the evolution of the HD limit with metallicity ({\it middle right panel;} Figure \ref{fig:rsg_metallicity_dependence}). We obtain Kendall's $\tau=0.1$, which is not statistically significant, with a p-value of $p=0.48$. However, the upper luminosity limit of RSGs shows significant galaxy-to-galaxy variations, spanning $\log(L/L_{\odot})\sim5.1$--$5.8$. Here, we define the HD limit as the $99^{\rm th}$ percentile of the RSG luminosity function to avoid bias by the overluminous outlier sample. The median luminosity of RSGs in a galaxy, which is potentially affected by selection effects stemming from AGB contamination at low luminosities, also does not follow a clear trend with metallicity ({\it middle left panel;} Figure \ref{fig:rsg_metallicity_dependence}). Once again, Kendall's $\tau = 0.02$ and is not statistically significant ($p=0.92$). Notably, we find that the median luminosities of high SFR galaxies show an interesting trend, clustering at high values ($\log{(L/L_{\odot})}>4.6$; Figure \ref{fig:rsg_sfr_dependence}). High SFR galaxies appear to form an increased number of luminous RSGs, which is also reflected in their HD limit.

The SFR dependence of RSG population luminosities has implications for the IMF of the stellar populations. Upon inspecting CMDs across our galaxies, we note that starburst galaxies host a much larger number of very luminous and red stars, above the nominal AGB maximum luminosity of $\log(L/L_{\odot}) = 5.0$. For instance, the starburst interacting galaxy pair NGC\,4038/4039 \citep[SFR $8.954^{+0.422}_{-0.788}\,M_{\odot}\,{\rm yr^{-1}}$;][]{Lanz2018} contains the largest number of RSGs in the catalog with $\log(L/L_{\odot}) > 5.0$. Motivated by this, we inspect how the SFR impacts stellar parameters and luminosity in particular.

Figure \ref{fig:rsg_sfr_dependence} shows how the median and $99^{\rm th}$ percentile luminosity vary across galaxies as a function of the specific star formation rate (sSFR). We use sSFR instead of SFR to normalize over the wide range of stellar masses in our galaxy sample. We find a weak positive correlation with sSFR for both quantities (Kendall's $\tau=0.52$ and 0.43, respectively), i.e., the luminosity increases with sSFR. Both results are statistically significant, with $p=0.008$ and $0.03$, respectively. This result is unexpected. If all the galaxies in our sample follow the same IMF, we expect the median luminosity of RSGs to remain unaffected by sSFR. We find that our proxy for the HD limit at the $99{\rm th}$ percentile also increases with SFR, and this correlation becomes stronger when we change the representative value from the $99{\rm th}$ to the $99.9{\rm th}$ percentile. The correlation between the HD limit and sSFR cannot be easily explained by AGB contamination, which should not depend on global galaxy properties and have luminosities well below the RSG HD limit. In principle, different modeling choices between galaxies could be the reason for this trend. We rule out this explanation, as it should lead to a stronger correlation with metallicity, where the models differ in their physics, instead of SFR, which is not explicitly modeled.

In the RSG phase, they can briefly ($\lesssim5\%$ of the RSG lifetime) cross the HD limit beyond which they may evolve blueward into YSGs/BSGs with hotter temperatures \citep{Higgins2020}. High sSFR galaxies can form an increased number of high-mass stars, leading to an abundance of massive star populations and clusters. Thus, with a large enough sample of RSGs, we observe these short phases in their evolution where they appear to be highly luminous, which is also consistent with only $0.5\%$ of our sample lying in this phase space. Alternatively, variability in evolved RSGs can also cause brief excursions into the higher luminosity tail, contributing a small fraction of luminous RSGs per galaxy \citep{Laplace26}. This would be particularly relevant for the cool, luminous RSGs ($T_{\rm eff} \lesssim 3000\,{\rm K}; \log(L/L_{\odot} \gtrsim 5.55$), since we expect stars crossing the HD limit to evolve to hotter temperatures ($T_{\rm eff} \gtrsim 4000\,{\rm K}$). While these hypotheses may explain the high luminosity end of the sSFR-$\log(L/L_{\odot})$ relation (Figure \ref{fig:rsg_sfr_dependence}), it is unclear whether the number of massive RSGs formed in high sSFR environments is sufficient to increase median population luminosities. 

Further investigation into how star formation impacts RSGs will be necessary,  as we observe that sites of local star formation alter the nature of the RSG population. The global SFR estimates we use in this analysis (Table \ref{tab:galaxy_sample}) use empirically calibrated UV and MIR tracers of the last $\sim100\,$Myr of star formation \citep{Leroy2019}, and could potentially mask stronger local effects. Integral Field Unit observations of these galaxies would enable a finer measurement of their spatially resolved star formation rate, and assess how star formation affects RSGs in more detail. For instance, NGC\,4548 has a circumnuclear starburst ring, traced by RSGs with $\log(L/L_{\odot}) > 5.5$. For a short-lived phase crossing the HD limit, this may be possible if these RSGs have co-evolved to high luminosities simultaneously. Alternatively, binary mass transfer and stellar mergers can be responsible for at least a fraction of the luminous RSGs we observe. Disentangling the multitude of evolutionary channels and physical processes contributing to the RSG population properties we observe will require a multi-wavelength dataset comprising the UV-MIR, tracing emission from companions, dust, and surrounding stellar populations.  

\subsection{RSG Dust Optical Depth and Mass-Loss Rates}
\label{ssec:rsg_metallicity}

The rate at which RSGs lose mass and the dust mass ejected throughout their lifetimes have far-reaching consequences for the nature of the eventual SN, evolution into broader classes of SN progenitors, and the resulting compact object post-explosion \citep{Merritt2025, vanLoon2025, Zapartas2025}. Theoretical analyses of RSG mass loss and insights from Type II-P SNe point to low-mass RSGs experiencing insufficient mass loss to shed their hydrogen envelopes, while high-mass RSGs $(M>12\,M_{\odot})$ experience episodic mass loss, evolving into dusty stars \citep{Fuller2024, JacobsonGalan2025}. Understanding mass loss in RSGs across evolutionary stages, and especially in their final decades to centuries, is necessary to connect massive stellar evolution to their SN outcomes. The environmental dependence of RSG mass-loss rates has also been studied only in a small sample of galaxies to date \citep{Antoniadis2025}. Here, we examine the fraction of dusty RSGs in our catalog, how dust optical depths depend on metallicity, and compare our RSGs with literature prescriptions for mass-loss rates.

The dust optical depth of RSGs in our catalog follows an approximately log-normal distribution, with a median of $\tau_V = 0.32$ (Figure \ref{fig:param_dist}). $16.5\%$ of the sample have $\tau_V > 1$, which are moderately dust enshrouded, while $1{,}155$ RSGs, i.e., $1.1\%$ of the sample, are heavily dust-enshrouded with $\tau_V > 4$. These $\tau_V$ values represent the $84^{\rm th}$ and $99^{\rm th}$ percentile of the catalog respectively. We note that the value of dust optical depth inferred depends on the dust composition and geometry assumed (Section \ref{sec:models}). Circumstellar dust produced by stellar winds in the RSG phase at very high mass loss rates could potentially strip the hydrogen envelopes of these stars, ultimately resulting in a hydrogen-poor SN. However, recent work has suggested that wind-driven mass loss alone cannot strip hydrogen envelopes for most RSGs \citep{Beasor22}. Following the definition of a dust-enshrouded RSG (DE-RSG) from \citet{Beasor22} as $\dot{M}>10^{-4}\,{M_{\odot}\,{\rm yr^{-1}}}$ ({\it see below for converting $\tau_V$ to $\dot{M}$}), we find 131 RSGs can be classified as DE-RSGs. This corresponds to $0.13\%$ of the full catalog, and agrees with the conclusion of \citet{Beasor22} that single-star evolution channels leading to extensive mass loss apply to a very small fraction of the RSG population.

Inspecting the metallicity dependence of RSG dust optical depths, we find an interesting pattern: the dustiest RSGs in our catalog are located in galaxies with high metallicity ({\it lower right panel;} Figure \ref{fig:rsg_metallicity_dependence}). However, the median dust optical depth of RSGs does not show a significant correlation with host galaxy metallicity ({\it lower left panel;} Figure \ref{fig:rsg_metallicity_dependence}). Once again, we use Kendall's $\tau$ test to assess the significance of the correlation. We find statistically significant correlation for $99^{\rm th}$ $\tau_V$ against Z, with $\tau = 0.48\,(p=0.01)$, while median $\tau_V$ does not show a significant correlation ($\tau = 0.23;\,p=0.23$). We also find that the correlation becomes systematically stronger as we increase the percentile from $50$ to $99.9$. It is puzzling that we find the dustiest RSGs to be disproportionately affected by mass loss, while the broader population does not appear to follow this trend. 

Together, our observations indicate that there could exist a turnover point in dust produced/mass lost by RSGs, beyond which the metallicity of the host environment has a measurable impact at the population level. Hence, the dustiest RSGs preferentially occur in high-metallicity galaxies. We note that due to this metallicity dependence, the fraction of DE-RSGs increases from zero to $\sim0.3\%$ as we move to higher-$Z$ galaxies, but this value is still smaller than the $3\%$ value of \citet{Beasor22}. We anticipate this to be due to different RSG selection criteria. While a very small fraction of the RSG population, if these DE-RSGs lose enough mass to shed their envelopes, it is consistent with an increased fraction of stripped envelope SNe in higher $Z$ environments \citep{Pessi2023}. We caution here that the metallicity values we use are global averages, with large scatter. Future analyses will incorporate local galaxy properties and can help us understand the impact of host environment on RSG evolution in better detail. Further, observations of RSGs with MIRI would help constrain the MIR excess due to dust emission and improve inferred values of $\tau_V$ and the mass loss rate.

\begin{figure}[htp]
    \centering
    \includegraphics[width=0.47\textwidth]{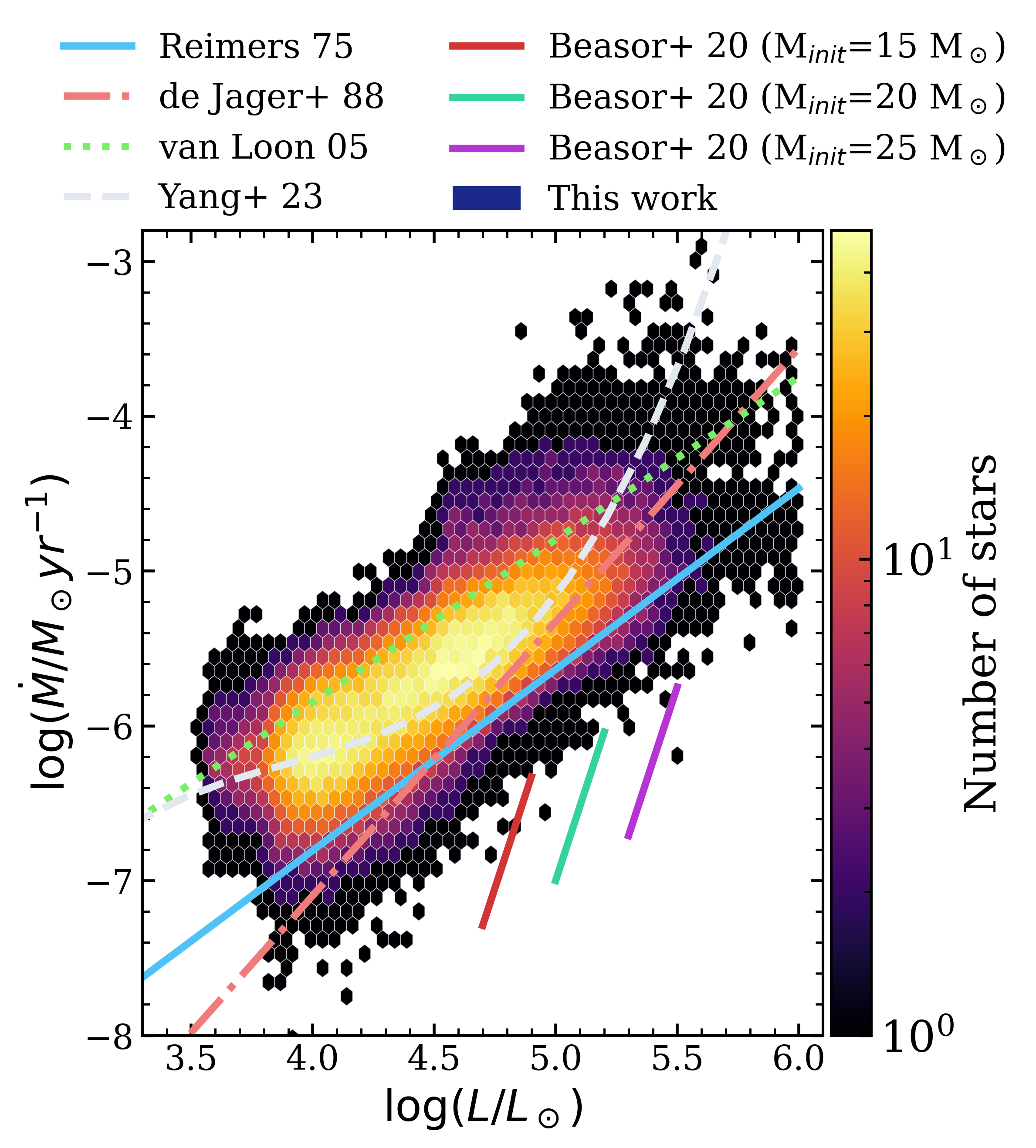}
    \caption{Mass-loss rates of RSGs in our catalog compared with several literature prescriptions of the $\dot{M} - L$ relation. We find a large intrinsic scatter in derived $\dot{M}$ values, and note that a broader range of literature prescriptions could be consistent. We caution that the numerical value of $\dot{M}$ is sensitive to assumed dust parameters. For instance, the \citet{Beasor2020} relation has a consistent slope with our inferred values, but is offset vertically, likely due to modeling choices. 
    }\label{fig:mdot_l}
\end{figure}

The mass-loss rates of RSGs in the Local Volume ($ d\lesssim3\,{\rm Mpc} $) have been studied extensively across a wide range of wavelengths, spanning the optical to mid-infrared \citep{Beasor2020, Yang2023, Antoniadis2024}.  Systematic differences in modeling and parameter estimation of RSG  properties and their mass-loss rates lead to a large scatter (\til2 dex) between different prescriptions of how the mass-loss rate varies with luminosity -- the RSG mass-loss relation. This relation traces the RSG mass-loss across the evolutionary sequence, since luminous RSGs are more evolved. Here, we compare the mass-loss rates of the RSGs in our sample to several existing literature prescriptions. We emphasize that this is not a detailed comparison between different mass-loss relations to identify the true underlying relation, but rather a qualitative comparison to highlight the science enabled by our large multi-galaxy RSG catalog. 

We convert $\tau_V$ to the mass-loss rate using Equation 1 of  \citet{Beasor2020}:
\begin{equation}
\label{equation:mdot}
    \dot{M} = \frac{16\pi}{3}\frac{R_{\rm in}\tau_V\rho_d a v_{\infty}}{Q_V}r_{gd}
\end{equation}
where $R_{\rm in}$ is the inner radius of the dust shell calculated using the inferred dust temperature and luminosity, $\tau_V$ is the inferred dust optical depth, $\rho_d$ is the bulk density of dust grains, assumed to be $3.5\,{\rm g}\,{\rm cm^{-3}}$ for \citet{Draine84} astronomical silicates, $a/Q_V$ is the grain size to extinction efficiency ratio, set to $0.1\,{\rm \mu m}$ (following the composition detailed in Section \ref{sec:models}), $v_{\infty}$ is the steady-state wind velocity, assumed to be $25\pm10\,{\rm km}/{\rm s}$ consistent with Galactic RSG measurements. $r_{\rm gd}$ is the gas-to-dust ratio, which is a metallicity-dependent quantity that varies across galaxies, and we assume it to be $300\pm100$. From our photometric dataset, it is difficult to constrain the specific dust parameters such as dust grain size, composition, or wind velocity. Therefore, assumptions and uncertainties in these parameters lead to a large systematic scatter in the inferred mass-loss rate. Nevertheless, the relative distribution of RSG mass-loss rates is set by $\tau_V$, which sets the ``slope'' of the mass-loss relation, while the numerical value of the mass-loss rate can be altered by changing any of the unconstrained variables in Equation \ref{equation:mdot}.

We plot the mass-loss rate of RSGs in all galaxies excluding NGC\,3034 in Figure \ref{fig:mdot_l}, with a set of commonly used literature relations taken from \citet{Yang2023}. We plot the relations as is, but compare only the slopes and not the numerical values. We exclude NGC\,3034 in this comparison due to the extreme dust obscuration in the stellar populations of this galaxy, with RSGs largely occupying the otherwise empty region $\log(L/L_{\odot})\gtrsim5.0$ and $\log(\dot{M}/M_{\odot}\,{\rm yr}^{-1})\gtrsim-5$ in Figure \ref{fig:mdot_l}, skewing the overall distribution. We adopt a similar steady-wind $r^{-2}$ density profile for our {\tt DUSTY} models as several of the literature studies we compare to. Among RSGs in other galaxies, we find that the slope of the $\dot{M}-L$ relation is consistent with the \citet{Reimers1975} and \citet{vanLoon2005} relations. We do not observe a strong spike in the slope when $\log(L/L_{\odot})\gtrsim4.5$, for the overall sample, as found by \citet{Yang2023} and \citet{Antoniadis2024} for SMC and LMC RSGs, respectively. 

We note that our derived $\dot{M}$ values are an order of magnitude larger numerically than \citet{Beasor2020}, and caveat that the absolute value of our $\dot{M}$ should not be taken at face value due to reasons mentioned above. If our values were offset to lower $\dot{M}$, say due to different grain size composition, the \citet{Beasor2020} relation would overlap reasonably well with the $M_{\rm ZAMS}>15\,M_{\odot}$ RSGs in our catalog. Moreover, \citet{Beasor2020} argues against deriving the RSG mass-loss relation using all field RSGs, and instead using clusters of co-eval RSGs. While this analysis is beyond the scope of this paper, this could be why the deviation from \citet{Beasor2020} observed in Figure \ref{fig:mdot_l} may be expected.

Supplementing our NIR SEDs with optical photometry from the {\it Hubble Space Telescope} and MIR photometry from {\it JWST}/MIRI can help reduce the systematic errors in our inferred mass-loss rate. Near- and mid-infrared spectroscopy of RSGs in our catalog by {\it JWST} can significantly aid in inferring their mass-loss rates to better precision, and constrain dust parameters leading to the large error floor in Figure \ref{fig:mdot_l}.

\subsection{Identifying Potential Supernovae Progenitors Before Explosion}
\label{ssec:rsg_progenitors}

Pre-explosion imaging of RSG SN progenitors remains challenging but has succeeded in a small number of cases \citep{Smartt09}; an RSG catalog with homogeneously inferred luminosities and dust optical depths narrows the pool of high-probability targets in galaxies where follow-up is feasible. Analysis of RSG progenitors of nearby Type CSM-interacting IIP SNe (see SN\,2023ixf \citealt{Kilpatrick23}; SN\,2024ggi \citealt{Xiang24}; SN\,2025pht \citealt{Kilpatrick25}) and flash spectroscopy of the SNe suggest enhanced mass loss in the final years preceding explosion ($\til 10^{-3}\,{M_{\odot}}{\rm yr^{-1}}$). The physical mechanism(s) behind the elevated terminal mass loss and fraction of RSGs undergoing such mass loss is not well understood. In a single case to date \citep[SN\,2020tlf;][]{Jacobson-Galan22}, mass loss from an RSG was high enough $(\dot{M}\til10^{-2}M_{\odot}\,{\rm yr^{-1}})$ to produce precursor emission characterized by an inflated envelope $(R\til1000R_{\odot})$ and high luminosity $(L\til10^6L_{\odot})$ before a normal Type II-P/L SN. The confined circumstellar material around terminal RSGs makes them very dusty ($\tau_V > 1$). Here, we expand the sample of luminous, heavily dust-enshrouded RSGs using our catalog, lying in a similar part of parameter space to SN progenitors from the literature.

\begin{figure}[htp]
\centering
\includegraphics[width=0.47\textwidth]{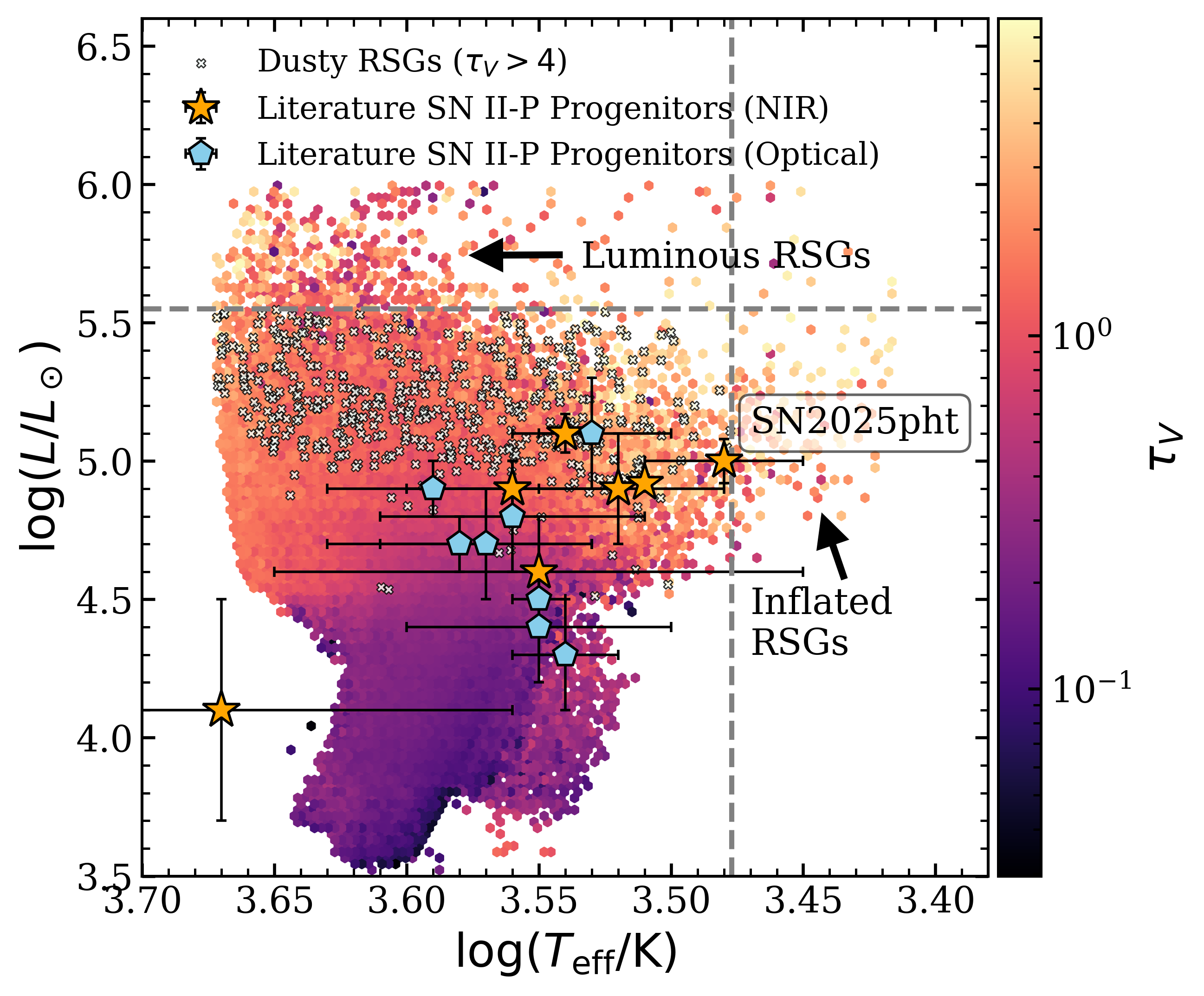}
\caption{Literature sample of Type IIP progenitors plotted on the HR diagram with our RSG catalog. We plot literature progenitors with at least two NIR detections ($>1\,\mu{\rm m}$) as orange stars, and those with optical detections as blue pentagons. We mark the three subpopulations of outlier RSGs with the luminous sample ($\log(L/L_{\odot}) > 5.55$), cool/inflated sample ($T_{\rm eff} < 3000\,{\rm K}$), and dusty sample ($\tau_V > 4$). We find that progenitors of recent CSM-interacting SNe such as SN\,2023ixf, SN\,2024ggi, or SN\,2025pht are generally luminous ($\log(L/L_{\odot})>4.5$ and cool ($\log(T_{\rm eff}/{\rm K}) < 3.55$), blending with our inflated RSG population.}\label{fig:progenitor_hrd}
\end{figure}

To this end, we select a subsample of ``outlier'' RSGs in our catalog (Figure \ref{fig:hr_mist}) that show elevated dust optical depth ($\tau_V > 4$), or anomalously cool effective temperatures ($T_{\rm eff} < 3000\,{\rm K}$), or are very luminous ($\log(L/L_{\odot}) > 5.55$). We recover 1{,}669 RSGs (163 cool, 531 overluminous, and 1{,}155 dusty RSGs)in this phase space, representing $1.8\%$ of the total catalog. While not all of this sample is consistent with progenitor RSGs (particularly the overluminous RSGs; \citealt{Smartt09}), we capture a broad range of interesting outliers to understand their evolutionary pathways. We plot a sample of 15 RSG progenitors of Type IIP SNe reported in the literature against our RSG catalog (\citealt{VanDyk2025} and references therein), and mark the outlier sample in Figure \ref{fig:progenitor_hrd}. We find that progenitors of recent CSM-interacting Type II-P SNe (SN\,2023ixf, SN\,2024ggi and SN\,2025pht), characterized using NIR data, are consistent with the extended, cool, luminous sample of RSGs in our catalog ({\it right of the vertical grey dashed line;} Figure \ref{fig:progenitor_hrd}). We also plot several progenitor detections of non-CSM-interacting Type II SNe, which occupy a broad range of luminosities, but concentrate on the cooler end of RSG temperatures. Promisingly, the progenitor star of SN\,2025pht \citep{Kilpatrick25} in NGC\,1637 is independently classified as an RSG in our catalog (Figure \ref{fig:ngc1367_rgb}), with $P_{\rm RSG} = 1.0$, high inferred luminosity, and dust optical depth. 

\begin{figure}[tp]
\centering
\includegraphics[width=0.45\textwidth]{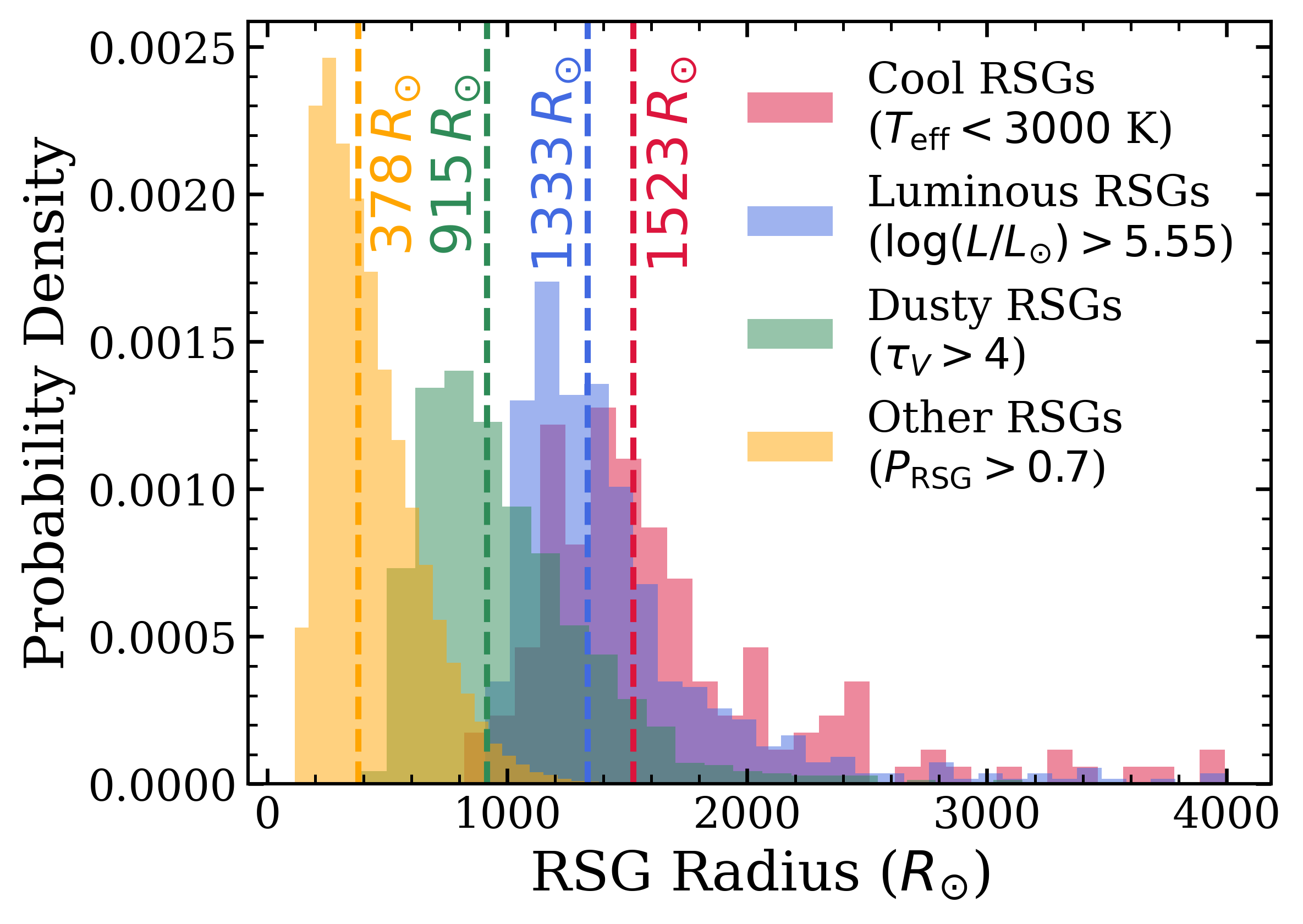}
\caption{Blackbody radii of our outlier RSG population compared to the distribution of all other RSGs in the catalog $(P_{\rm RSG} > 0.7)$. We plot the outlier sample of 1{,}669 RSGs split into three subclasses: cool RSGs  $(T_{\rm eff} < 3000\,{\rm K})$ with 163 stars, overluminous RSGs $(\log(L/L_{\odot}) > 5.55)$ with 531 stars, dusty RSGs with 1{,}155 stars. We observe that the cool and dusty RSGs, which are similar to SN progenitors, have systematically higher blackbody radii than the broader sample, with median values of $>1200\,R_{\odot}$ and $378\,R_{\odot}$, respectively.}\label{fig:rsg_radii}
\end{figure}

The outlier RSGs are systematically extended relative to the broader catalog (median $378\,R_{\odot}$), with median blackbody radii ($R_{\rm BB}$) of $915\,R_{\odot}$ for the dusty subsample, $1{,}333\,R_{\odot}$ for the luminous subsample, and $1{,}523\,R_{\odot}$ for the cool subsample (Figure~\ref{fig:rsg_radii}). These values are consistent with the progenitor radii inferred from early-time light curves of nearby Type~II SNe \citep{Irani2024}, indicating that our outlier population occupies the same physical regime as SN progenitors. We compute $R_{\rm BB}$ directly from $T_{\rm eff}$ and $\log(L/L_{\odot})$, which we fit jointly with $\tau_V$. Conventional definitions of RSG radii, such as at $\tau_V=2/3$, are instead non-trivial to compute for dusty RSGs, which form molecules over a range of radii \citep{Davies13}. Given that several RSGs are also known to have asymmetric, clumpy dust \citep{Smith2001, Scicluna2015, Montarges21}, we treat our radius values as a qualitative diagnostic. The dusty and cool subpopulations thus offer a direct pathway to systematically probe RSGs in their final decades to centuries.

\section{Conclusion}
\label{sec:conclusion}

In this paper, we present a catalog of \nrsg extragalactic RSGs and their inferred physical parameters, representing the largest population analysis to date. We identify and characterize RSGs using a novel data reduction and parameter inference workflow on archival imaging of nearby galaxies $(d\lesssim20\,{\rm Mpc})$ by {\it JWST}/NIRCam. Below, we summarize the workflow and first results from our catalog.

\begin{enumerate}
    \item We present \texttt{jwst123}, a general-purpose data reduction pipeline for high-precision resolved stellar photometry on archival-scale NIRCam datasets. Our sequential alignment procedure using {\tt JHAT} produces NIRCam images aligned to the Gaia frame, as well as relative error within one pixel between images. We develop a PSF-matched image coaddition process to homogenize PSFs across filters to produce a single-filter reference image for {\tt DOLPHOT} covering the full sky footprint of the observations.
    \item We apply {\tt jwst123} to 15 galaxies within $d \lesssim 20\,\text{Mpc}$ with extensive NIRCam imaging, producing photometric catalogs of $\sim$2--6 million sources per galaxy. Our galaxy sample spans metallicities $Z/Z_\odot\sim0.25$--$1.45$ and SFRs $\til 0.1$--$14.1\,M_\odot\,\text{yr}^{-1}$, making it well suited to studying RSGs across evolutionary stages and environments.
    \item With the rich dataset of stellar SEDs from NIRCam photometry spanning 0.6--5\,${\rm \mu m}$, we infer the physical parameters ($T_{\rm eff},T_{\rm dust},\tau_V,\log(L/L_{\odot})$,\\$R_V,A_V$) of luminous stellar populations including RSGs and AGBs using MARCS+{\tt DUSTY} forward models. At the scale of our dataset (\til1.5 million luminous stars across 15 galaxies), we use SBI to perform rapid likelihood-implicit posterior inference using neural networks, at 25ms per object.
    \item We present a novel probabilistic classification scheme to identify RSGs using the 3D $T_{\rm eff}$--$\log(L/L_\odot)$--$\tau_V$ phase space, with the regions occupied by various stellar populations seeded by GMM-based CMD selection. This method is filter-agnostic, recovers the RSG locus across multiple CMDs, and separates RSGs from AGBs at low luminosities ($\log(L/L_\odot) \lesssim 4.2$) where 2D projections overlap. We apply this scheme across galaxies to produce a catalog with \nrsg RSGs, the largest sample to date with measured physical parameters by an order of magnitude. Our sample shows good agreement with MIST evolutionary tracks for ZAMS masses of 7--32 $M_\odot$, despite an entirely empirical selection.
    \item Across our sample, we find RSG temperatures to decrease with metallicity, while the median luminosities and dust optical depths remain uncorrelated with metallicity. However, we find that the dustiest RSGs prefer high-metallicity galaxies. If such RSGs shed their envelopes through mass loss, this would lead to a higher occurrence of stripped-envelope SNe in higher metallicities, agreeing with SN environment studies \citep{Pessi2023}. We also find the median and maximum luminosities of RSGs increase with sSFR, possibly due to efficient sampling of the high-mass end of the IMF and brief phases of RSGs crossing the HD limit.
    \item We identify a subsample of 1{,}669 outlier RSGs selected on $\tau_V > 4$, $T_\text{eff} < 3000\,\text{K}$, or $\log(L/L_\odot) > 5.55$, representing $1.8\%$ of the catalog, with inflated blackbody radii (median 1300 $R_\odot$ versus 382 $R_\odot$ for the parent sample). A subset of these outliers occupy the same regions of the HR diagram as Type II-P SN progenitors identified from pre-explosion imaging.  We independently recover the RSG progenitor of SN\,2025pht in NGC\,1637 \citep{Kilpatrick25} with $P_\text{RSG} = 1.00$, high $\log(L/L_{\odot})$, and high $\tau_V$, validating our selection against a known progenitor. The other RSGs represent an interesting sample exhibiting unique evolution, whose identification is enabled by the size of our catalog and will require further follow-up for interpretation.
\end{enumerate}

Since our classification operates on inferred physical parameters rather than filter-specific colors, the resulting per-class probability densities can be applied directly to resolved stellar populations observed with different instruments and filter sets. This is particularly relevant for {\it Roman}, whose wide-field NIR imaging will resolve stellar populations across a far larger number of galaxies than are currently accessible. Further, the Vera C. Rubin Observatory LSST will produce multi-epoch optical lightcurves, which can be combined with our selection to identify variable and dust-obscured RSGs and their explosive endpoints. A coherent parameter-inference framework spanning these facilities would allow the KDEs derived here to seed RSG selection in surveys with no direct CMD analog, and the amortized nature of our SBI models makes scaling to the resulting source counts computationally tractable.

We note that individual sites of higher and lower metallicity, as well as star formation, exist across our galaxies, which we do not account for in this paper. A significant fraction of our galaxy sample contains supplementary archival data from the Multi Unit Spectroscopic Explorer (MUSE), enabling spatially resolved metallicity gradient and SFR measurements. Further, all galaxies in our sample also include complementary imaging in the optical (and some in ultraviolet) by {\it HST}, and in the MIR by {\it JWST}/Mid-Infrared Instrument (MIRI), which are reserved for future work. Supplementing our NIR SEDs with HST optical photometry would extend wavelength coverage into the regime where foreground dwarfs and blue contaminants separate most cleanly. Similarly, \textit{JWST}/MIRI photometry would constrain the circumstellar dust emission more precisely than NIR data, directly improving estimates of $T_\text{dust}$, dust composition, and the mass-loss rates that currently carry $\sim$2 dex of systematic uncertainty. Near- and mid-infrared spectroscopy of catalog members by {\it JWST} would independently calibrate the $T_\text{eff}$ and $\tau_V$ scales against which our photometric inferences are benchmarked, and would test the silicate-dust and spherical-shell assumptions built into our forward model. Detailed dust models with varying shell configurations and density profiles can also be tested against NIR-MIR data. Multi-epoch imaging from \textit{JWST} and {\it Roman} would characterize the RSG variability that our single-epoch SEDs average over, and would enable the AGB--RSG separation based on $T_\text{eff}$--$L$ evolution.

\section*{Acknowledgments}

We thank Andrew Dolphin, Phelipe Darc, and Alex Gagliano for helpful discussions regarding photometry and simulation based inference. C.D.K. gratefully acknowledges support from the NSF through AST-2432037, the HST Guest Observer Program through HST-SNAP-17070 and HST-GO-17706, and from JWST Archival Research through JWST-AR-6241 and JWST-AR-5441. W.F. gratefully acknowledges support by National Science Foundation under grant Nos. AST-2206494, AST-2308182, AST-2432037, and CAREER grant No. AST-2047919, the David and Lucile Packard Foundation, and the Research Corporation for Science Advancement through Cottrell Scholar Award \#28284. D.H. is supported by STScI grants HST-GO-17770.002, JWST-GO-12468.001, and JWST-GO-09964.001. M.R.D. acknowledges support from NSERC through grant RGPIN-2025-06224, the Canada Research Chairs Program (CRC-2023-00127), the Ontario ERA program (ER22-17-164) and the Dunlap Institute at
the University of Toronto. W.J.-G.\ is supported by NASA through Hubble Fellowship grant HSTHF2-51558.001-A awarded by the Space Telescope Science Institute, which is operated for NASA by the Association of Universities for Research in Astronomy, Inc., under contract NAS5-26555.

This work is based on observations made with the NASA/ESA/CSA James Webb Space Telescope. The data were obtained from the Mikulski Archive for Space Telescopes at the Space Telescope Science Institute, which is operated by the Association of Universities for Research in Astronomy, Inc., under NASA contract NAS5-03127 for JWST. These observations are associated with the following Cycle 1--3 programs and PIs: GO 1685, Adam Riess; GO 1701, Alberto Bolatto; GO 1783, Angela Adamo; GO 1995, Wendy Freedman; GO 2080, Jason Glenn; GO 2107, Janice Lee; GO 2211, David Trilling; GO 2452, JD Smith; GO 2581, Rupali Chandar; GO 2732 Klaus Pontoppidan; GO 2875, Adam Riess; GO 3429, Christopher Clark; GO 3435, Karin Sandstrom; GO 3707, Adam Leroy; GO 3990, Takahiro Morishita; GO 4087, Caroline Huang; GO 4793, Eva Schinnerer; GO 5145, Adam Smercina; GO 5398, Jeyhan Kartaltepe.

Data availability: Public JWST/NIRCam archival imaging analyzed here can be downloaded via MAST. All observations we use are available under this DOI: \href{https://doi.org/10.17909/cr4p-mw12}{10.17909/cr4p-mw12}. The full catalog of luminous sources with inferred parameters and photometry is available on Zenodo with DOI: \href{https://doi.org/10.5281/zenodo.22788058}{10.5281/zenodo.22788058}.

\software{DOLPHOT \citep{Dolphin2000, Dolphin16},
    JHAT \citep{Rest23},
    JWST Calibration Pipeline \citep{Bushouse25},
    Astropy \citep{Astropy18, Astropy2022},
    photutils \citep{Bradley25},
    reproject \citep{Robitaille2024},
    DUSTY \citep{Ivezic97, Ivezic1999},
    MARCS \citep{Gustafsson08},
    MIST \citep{Choi16, Dotter26},
    sbi \citep{Tejero-Cantero20},
    Optuna \citep{Akiba19},
    NumPy \citep{Harris2020},
    SciPy \citep{Virtanen2020},
    OverCite \citep{Shariat2026},
    {\tt jwst123}\footnote{\url{https://github.com/aswinsuresh24/jwst123}},
    {\tt rsg\_phot}\footnote{\url{https://github.com/aswinsuresh24/rsg_phot}}
}

\facility{JWST (NIRCam)}

\appendix

\section{Data Processing with {\tt jwst123}}
\label{appendix: jwst123}

We provide additional details of the methods implemented in {\tt jwst123} here.

\subsection{Image Alignment}
\label{appendix_ssec:align}

We process images between different galaxies independently. For a given galaxy, we use Level-2 images (``\texttt{*cal.fits}'' files) organized into groups and visits (Section \ref{ssec:align}. By design, a single NIRCam exposure produces eight SW and two LW Level-2 images \citep{Rieke23}. These images, being within the same pointing, maintain nearly identical astrometric accuracy \citep{Williams24}.
 
Since images across different groups have no spatial overlap, we perform self-consistent relative alignment independently between visits within each group before aligning them to the {\it Gaia} astrometric frame. We use the catalog of sources in {\it Gaia} DR3 to provide the position and proper motion reference \citep{GaiaDR3}. However, the number of {\it Gaia} sources contained in the $\til1\,\text{arcmin}^2$ field of view of a single NIRCam detector can be small $(\lesssim6)$, which hinders deriving a robust alignment solution. For this reason, PHANGS performs absolute alignment to bright HST sources instead \citep{Williams24}. We leverage the excellent relative alignment between visits to construct a larger alignment mosaic, thereby increasing the number of available {\it Gaia} sources.

The \jwstpipeline\ pipeline \citep{Bushouse25} uses \texttt{TweakReg} to align NIRCam images, following the drizzle-based mosaic philosophy developed for {\it HST} \citep{FruchterHook02}. We use the {\it JWST}-{\it HST} Alignment Tool (\texttt{JHAT}; \citealt{Rest23}) for image alignment. {\tt JHAT} internally uses {\tt TweakReg} for alignment but optimizes the source and reference catalogs to pick the best available matches between the images to be aligned. {\tt JHAT} modifies the world coordinate system (WCS) of a NIRCam image by applying a linear shift and rotation to minimize coordinate offsets between the source catalog and reference catalog. Compared to \texttt{TweakReg}, \texttt{JHAT} offers greater user-defined control over alignment and generally produces better results by carefully selecting the best set of matches between photometric catalogs and pre-computing optimal rotation offsets. \texttt{JHAT} can also align images to an external frame, such as to {\it Gaia} astrometric calibrators, by replacing the photometric catalog of the reference image with the proper-motion-corrected positions of {\it Gaia} sources in the image. 

Due to {\it JWST}'s exceptionally stable imaging \citep{Rigby23}, we find that aligning NIRCam images relative to each other using \texttt{JHAT} achieves alignment precision of \til5--20 mas, because NIRCam images with sufficient overlap share a large number $(\gtrsim 1{,}000)$ of common sources. This advantage does not extend to aligning individual NIRCam frames to \textit{Gaia}, which can contain fewer than 3 sources per frame in some instances. Consequently, we employ a systematic approach that prioritizes relative alignment between NIRCam images within a group before attempting alignment to \textit{Gaia}.

We compute the sky coverage for each visit using the \texttt{S\_REGION} header keyword, which defines the observed region. We begin with the visit covering the largest on-sky area within a group, excluding narrow-band filter observations. We align all images within this first visit to the pointing with the highest exposure time, and we drizzle the SW images into a Level-3 (\texttt{*i2d.fits}) mosaic using the \jwstpipeline pipeline, skipping the \texttt{tweakreg} step since images are already aligned to each other. We select the SW channel for its finer native resolution of $0.031''$, which enables more accurate source catalogs for alignment. We then repeat this process for the visit with the maximal area overlap with the first visit. The resulting drizzled mosaic from the second visit is aligned to the first visit's mosaic, and all Level-2 images in the second visit are subsequently re-aligned to this new mosaic, which inherits the astrometric frame of the first visit. This iterative process continues through remaining visits, but sequential alignment can propagate astrometric uncertainties through each step, potentially resulting in the final visit having non-negligible WCS shifts relative to the first visit. To mitigate this uncertainty propagation, we maintain a running source catalog that we update after each alignment iteration. New sources are added only if they lie outside the spatial coverage of the previous iteration's list, to prevent source duplication. This unified source list is passed directly to \texttt{JHAT} as the reference catalog. By the end of this loop, all images in the group are sequentially aligned relative to each other, proceeding in order of area overlap with the spatial coverage of the common source list.

We align only the Level-3 mosaic from the first visit to {\it Gaia}, and this alignment is propagated to all images when they are aligned sequentially to the first visit.  The larger sky coverage of the Level-3 mosaic mitigates most instances where few {\it Gaia} stars are available for alignment, but a few mosaics, often in the outskirts of galaxies, can still face this limitation. In such cases, we quantify the alignment uncertainty of the mosaic with respect to {\it Gaia} sources and propagate it to the individual NIRCam frames. We emphasize that astrometric calibration to the {\it Gaia} frame does not affect the photometry catalogs output by \texttt{DOLPHOT}, which requires only the relative alignment between images to be within \til1 pixel for optimal results.

An infrequent failure mode in the alignment process occurs when images exhibit substantial WCS shifts, either relative to other NIRCam visits or to the {\it Gaia} frame. These shifts can occur, for example, due to guide star acquisition failure, especially in crowded fields or extended guide targets in certain filters \citep{Williams24}. Our sequential alignment process is especially sensitive to large alignment offsets that occur early in the loop, as these can cause a significant number of images to exceed the required alignment error threshold of 1 pixel. For each image alignment step performed with \texttt{JHAT}, we require $\sigma_a < 0.031''$ (i.e., alignment accurate to within one pixel for SW images) or $\sigma_a < 0.063''$ for LW images. If an alignment solution produces dispersion greater than one pixel, we retry alignment using \texttt{JHAT} with an expanded search radius of $2''$. We also lower the source detection threshold from $5\sigma$ to $3\sigma$, as alignment occasionally fails due to insufficient matched sources. We align most images within the one-pixel threshold following one of these configurations. For images that still fail (e.g., if the alignment shift is much larger than $2''$), we run a coarse grid search to calculate a best-guess alignment offset. In case of large offsets, \texttt{JHAT} works best when provided with a rough initial estimate of the shift. To obtain this estimate, we compute the alignment dispersion at each point on a grid spanning a search radius of 50 pixels and select the grid point with the minimum dispersion. 

\subsection{Image Coaddition}
\label{appendix_ssec:coadd}

Since NIRCam simultaneously images in the SW and LW filters, each LW image is accompanied by higher-resolution SW images of the same sky region. Therefore, we create reference images using only SW filters without sacrificing coverage. We identify the minimal set of SW filters whose co-added images cover $>95\%$ of the total footprint. Narrow-band imaging is excluded at this step because shallow exposures or sparse source populations can introduce downstream photometry issues. If a single filter covers the full footprint, all images in that filter are drizzled using the default \jwstpipeline pipeline configuration, skipping the \texttt{TweakReg} step. Often, two or three filters are required. In these cases, we select the SW filter with the longest wavelength (e.g., F200W) as the target filter for PSF conversion, since the FWHM of NIRCam PSFs increases with wavelength \citep{Rieke23}.

We use simulated PSF models from \texttt{STPSF} \citep{Perrin14} that include detector-sampling and distortion effects. Using these models, we perform PSF matching and convolution to effectively degrade the sharper images to match the target PSF. We construct a convolution kernel from the source PSF to the target PSF using \texttt{photutils} \citep{Bradley25}, applying a split cosine bell window with $\alpha=1.3$ and $\beta=1.5$ to taper high-frequency noise \citep{Matsuura24, Mowla24}. PSF growth curves show that deviation from the target PSF after convolution remains below a relative error of 0.040\%. We note that more sophisticated methods for PSF convolution exist, such as in \citet{Aniano11}. However, our implementation is sufficient for inter-conversion between SW PSFs at identical pixel scales.

Before drizzling to create the Level-3 mosaic, we apply this matching kernel to the Level-2 images to avoid rotating the simulated PSF to different position angles. We also apply the convolution procedure to the \texttt{`ERR'} frames. We then drizzle these PSF-homogenized images using the \jwstpipeline pipeline to create mosaics in all required filters. We drizzle the images onto a fixed-pixel grid defined by a precomputed WCS (\textit{see below}) that accounts for the overall footprint orientation and filter-dependent mosaic areas. We coadd the resulting mosaics as an inverse-variance-weighted sum, with inverse-variance calculated as the \texttt{WHT} data frame divided by the mosaic exposure time.  We compute the coadd for both the \texttt{`SCI'} and \texttt{`ERR'} data frames, and update the header keywords for exposure time and photometric calibration for the coadded image appropriately. In addition to expanded sky coverage, the co-added image provides an enhanced S/N ratio (defined here as the ratio of measured flux to its photometric uncertainty) for most sources.

Due to computational and memory constraints, a single \texttt{DOLPHOT} run can require up to $\til 4$ days for 100 images, and more than 14 days for 250 images on a single core of an Intel Xeon Platinum 1.9\,GHz processor with 4\,TB of RAM. This constraint sets an upper limit on the number of images that can be simultaneously processed in a single run. We find that \texttt{DOLPHOT} photometry yields the optimal depth from a given set of images when all images containing a source are processed together in the same run, regardless of filter, as this incorporates information from each detection and enhances S/N. Hence, we optimize the computational time of \texttt{DOLPHOT} while still extracting the most information from the input images by picking an optimal set of images for each run. We set a maximum number of images that can be concurrently processed in a single run to be 150. 

However, under this constraint, it is often not possible to process every image containing a given source while simultaneously satisfying this condition for all sources, unless we process some images in multiple runs. To minimize runtime, we take advantage of the fact that \texttt{DOLPHOT} performs photometry only on sources within the sky footprint of the reference image. We drizzle multiple reference images with a minimum necessary overlap of 15 pixels to recover sources truncated at image edges. For images within a group, we rotate the overall sky coverage polygon using the \texttt{reproject} package (Astropy-affiliated imaging utilities; \citealt{Astropy18}) to minimize the area of the resulting rectangular bounding box. This rotated polygon defines a WCS that describes the target coordinate frame for the reference images.


\section{MARCS Stellar Spectra Models}
\label{appendix: marcs}

We construct the forward models used to infer physical parameters of luminous stellar populations in this paper using MARCS stellar atmosphere grids, surrounded by a spherical dust shell processed by {\tt DUSTY}. The MARCS theoretical spectra are calculated using 1D LTE radiative transfer at discrete effective temperatures, surface gravities, metallicities, and stellar masses \citep{Gustafsson08}. We use linear interpolation to produce a grid of models over the range of physical parameters we require (Section \ref{sec:models}). Therefore, the underlying model grids need sufficiently good sampling in parameter space for the interpolated models to have good accuracy. Part of the accuracy constraint is relaxed due to the effects of the dust shell, and down-sampling to broadband photometry with fewer features than a spectrum. However, deviations of up to \til0.3\,mag can occur due to poor grid sampling for interpolation.

\begin{figure*}[!htp]
\centering
\includegraphics[width=0.95\textwidth]{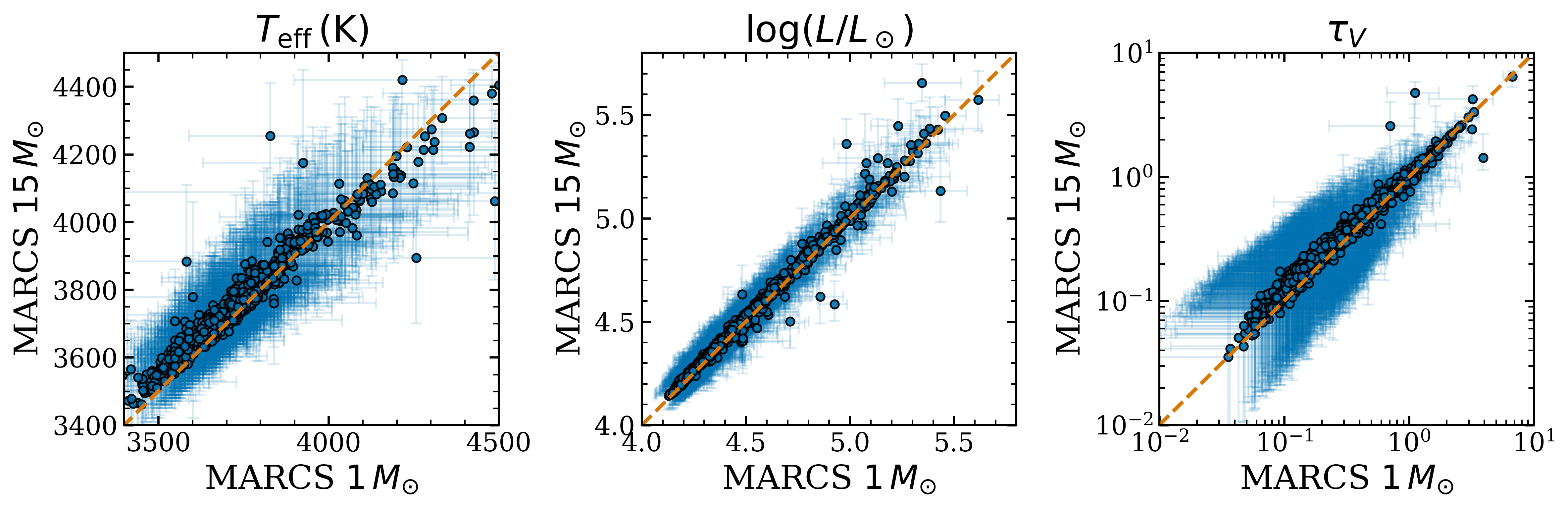}
\caption{Comparison of physical parameters derived using MARCS $1\,M_{\odot}$ and $15\,M_{\odot}$ models.
}\label{fig:marcs_comparison}
\end{figure*}

\begin{figure*}[!htp]
\centering
\includegraphics[width=0.95\textwidth]{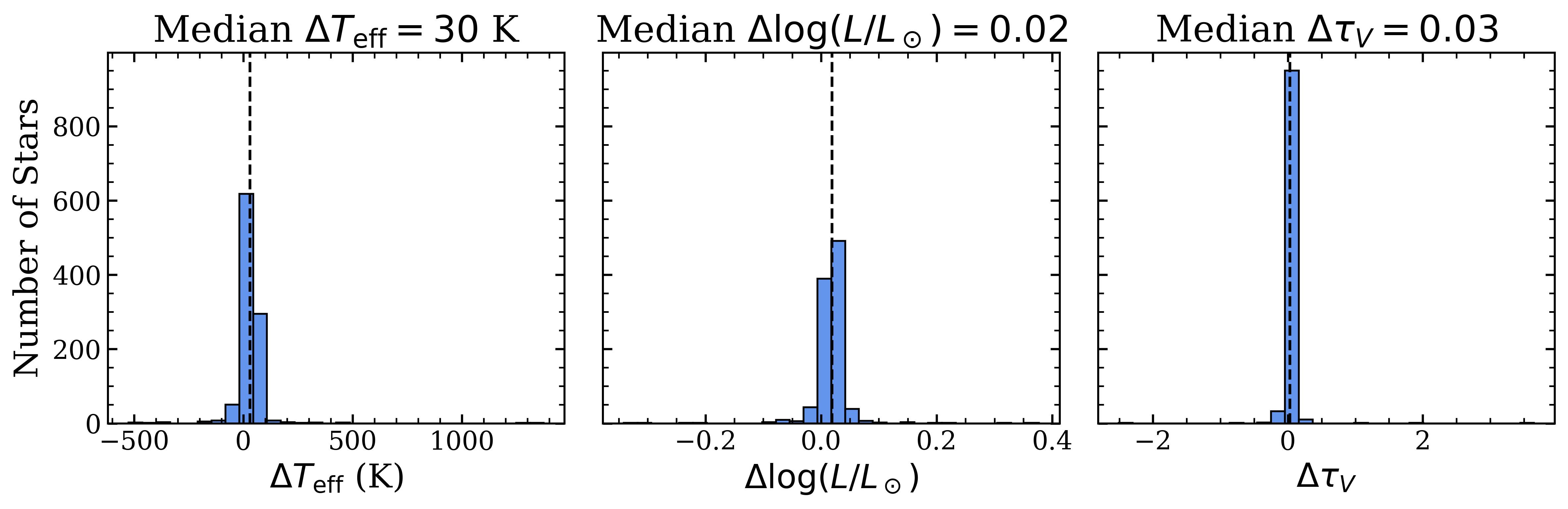}
\caption{Difference in median of derived parameters using MARCS $1\,M_{\odot}$ and $15\,M_{\odot}$ models.
}\label{fig:marcs_comparison_median}
\end{figure*}

Hence, we choose the $1\,M_{\odot}$ grid of MARCS spectra to model RSGs despite the availability of a $15\,M_{\odot}$ grid, as the $1\,M_{\odot}$ grid has much finer $T_{\rm eff}$ grid spacing at 100\,K, as compared to 250\,K for $15\,M{\odot}$. Further, the $1\,M_{\odot}$ models also span a wider range in effective temperature (2500\,K--8000\,K), making it suitable for both RSGs and AGBs. We avoid mixing models across different stellar masses to simplify model fitting. Here, we quantify the impact of model choice on the derived physical parameters. We fit both model grids (1 and 15\,$M_{\odot}$) to a sample of 1{,}000 randomly chosen stars in NGC\,5194 using MCMC. We run 64 walkers for 350 steps each, which we find to be sufficient for convergence. We treat the inferred values for the $15\,M_{\odot}$ model as ground truth for RSGs, and compare results from the $1\,M_{\odot}$ model. We compare $T_{\rm eff}$, $\log(L/L_{\odot}$ and $\tau_V$ from both models in Figure \ref{fig:marcs_comparison}, and their median estimates in Figure \ref{fig:marcs_comparison_median}, since these parameters affect the stellar classification method directly (Section \ref{sec:sample}).

We observe that the choice of a specific MARCS grid has a minimal impact on our derived parameters. While there is a weak preference for higher $T_{\rm eff}$ when using the $15\,M_{\odot}$ grid, the median difference in $T_{\rm eff}$ is \til30\,K between the two grids, which is well within our uncertainties. Further, the restricted $T_{\rm eff}$ range of the $15\,M_{\odot}$ model (3400--4500\,K; appropriate for RSGs) biases estimates for cooler AGB stars, making it unsuitable for our analysis. For all three parameters, we note that the derived parameters are consistent within error bars. $\log(L/L_{\odot})$ and $\tau_V$ are consistent with both models, with minimal difference in their median values. Therefore, we use the $1\,M_{\odot}$ grid for SED modeling.

\section{SBI Model Calibration Diagnostics}
\label{appendix: sbi}

Assessing the performance of SBI models using calibration tests and accuracy diagnostics is crucial to establishing the reliability of the model output and flagging potential failure modes when used on real observations. We verify our model performance using SBC, Tests of Accuracy using Random Points (TARP), and accuracy on a withheld test set, which the model does not access during training. Model performance in SBC and TARP tests is necessary for the posterior uncertainties to be reasonable. Testing the accuracy of the marginal parameter estimates assesses how well the model recovers the true parameters of the underlying system.

SBC assesses whether, when true parameters $\theta^*$ are drawn from the prior $\pi(\theta)$ and synthetic data are generated from the model, the resulting ensemble of posteriors is statistically consistent with those assumptions \citep{talts20}. SBC uses this principle to compute the ranks of an ensemble of posteriors $\{\theta_1, \theta_2, \ldots, \theta_N\}$ relative to the true parameters drawn from the prior. We calculate ranks as in \citet{talts20},

\begin{equation}
    r(\{f(\theta_1), \ldots, f(\theta_N)\}, f(\theta^{*})) = \sum_{n=1}^{N} \mathbf{1}\left[f(\theta_n) < f(\theta^*)\right]
\end{equation}
\noindent These ranks range from 0 to the number of posterior samples for each observation. An ideally calibrated model will follow the uniform distribution in its ranks. An overconfident model (underestimated variance) shows dispersed U-shaped ranks, while an underconfident model (overestimated variance) shows centrally peaked $\cap$-shaped ranks. Skewed rank distributions indicate biases in the model inference.

\begin{figure*}[tbp]
\centering
\includegraphics[width=0.49\textwidth]{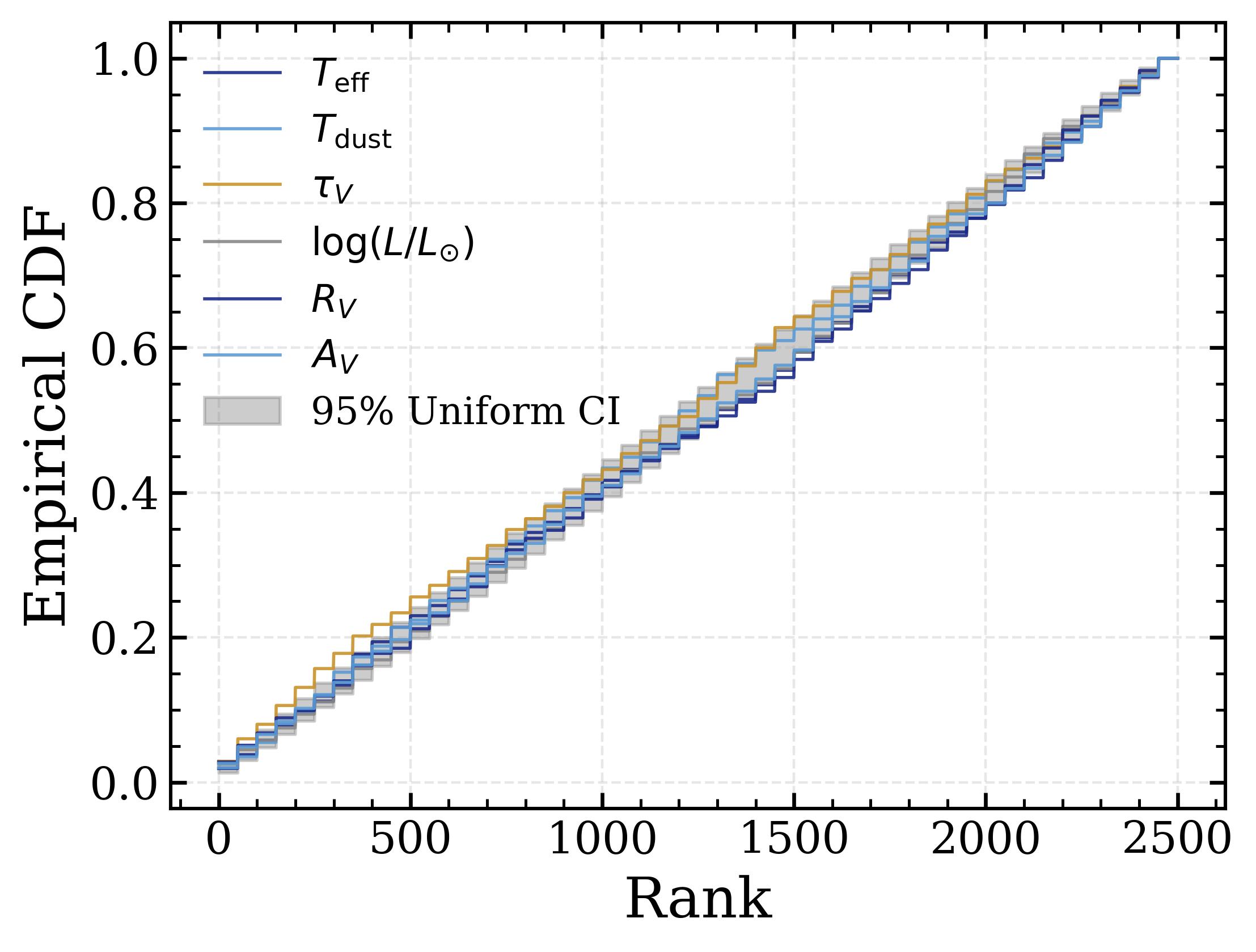}
\includegraphics[width=0.49\textwidth]{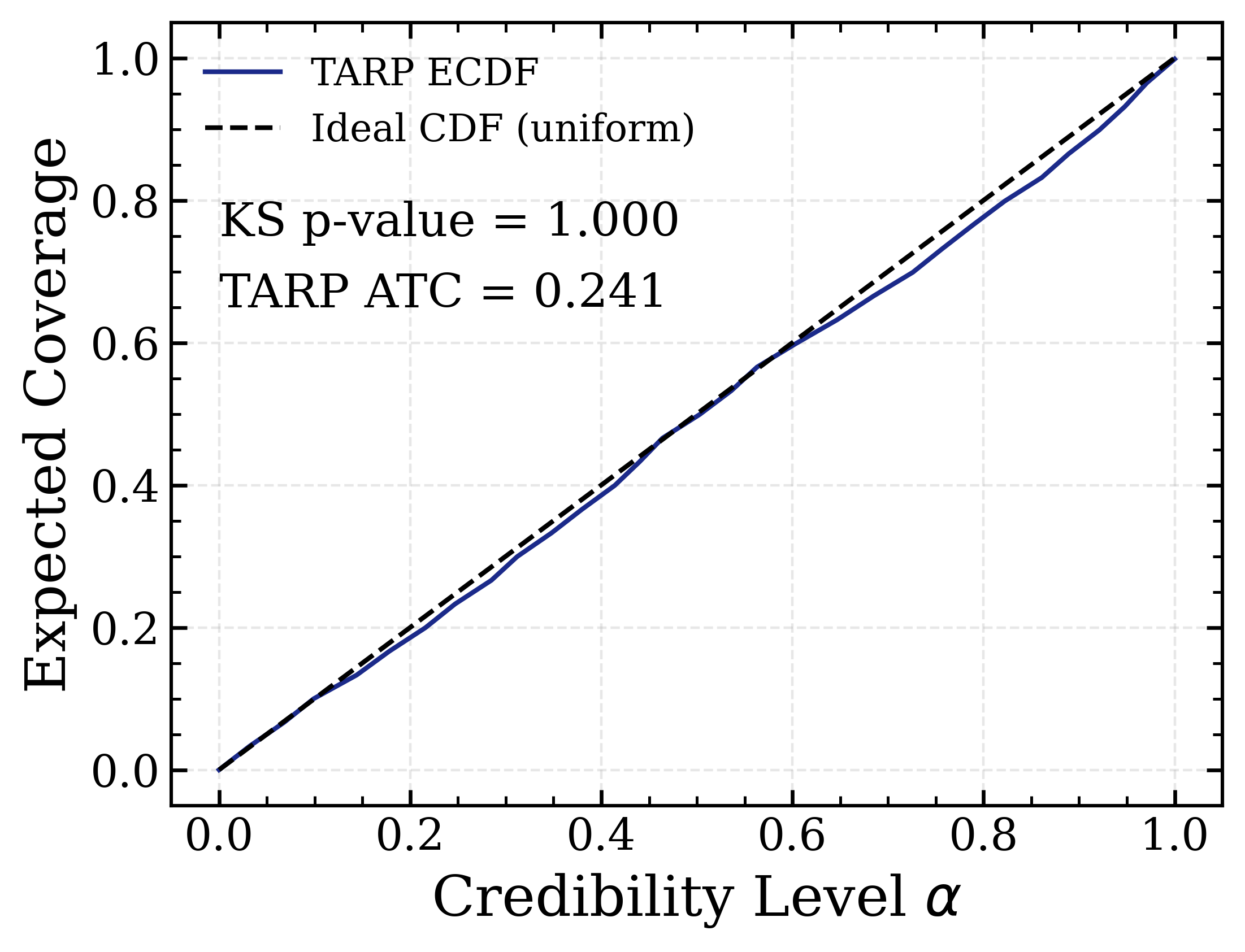}
\caption{Left: Representative example of Simulation-Based Calibration (SBC) using the galaxy NGC\,4258 ($d = 6.8\,{\rm Mpc}$) for our trained SBI model, evaluated using a withheld simulated test set of 1000 SEDs, with 2500 posterior draws per sample. Uniform ranks in SBC indicate that the uncertainties in the SBI model are well calibrated, neither overconfident nor underconfident. Right: Tests of Accuracy with Random Points (TARP) for the same SBI model. Ideal performance in TARP is a necessary and sufficient condition for model calibration.
}\label{fig:sbc_tarp}
\end{figure*}

\begin{figure*}[tbp]
    \centering
    \includegraphics[width=\textwidth]{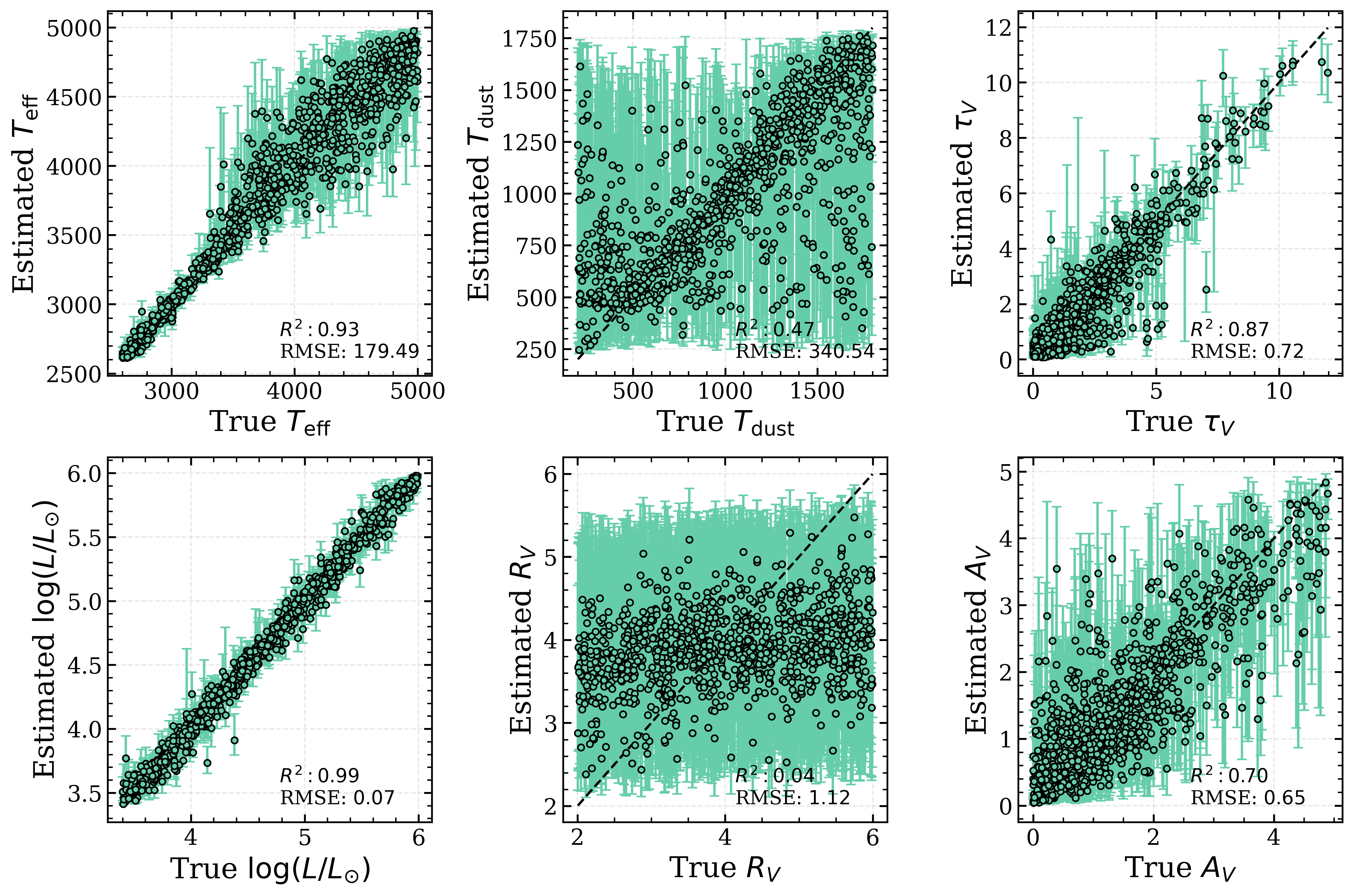}
    \caption{Representative example of test accuracy of our SBI models, illustrated for NGC\,4258 ($d = 6.8\,{\rm Mpc}$), using a withheld simulated test set of 1000 SEDs. We recover our primary parameters of interest, $T_{\rm eff}$, $\log(L/L_{\odot})$ and $\tau_V$ with high accuracy as indicated by the $R^2$ values. $T_{\rm dust}$ is poorly constrained using our NIR data, while $R_V$ has a minimal impact on the SED, largely returning the prior.
    }\label{fig:sbi_acc}
\end{figure*}

\begin{figure*}[tp]
    \centering
    \includegraphics[width=\textwidth]{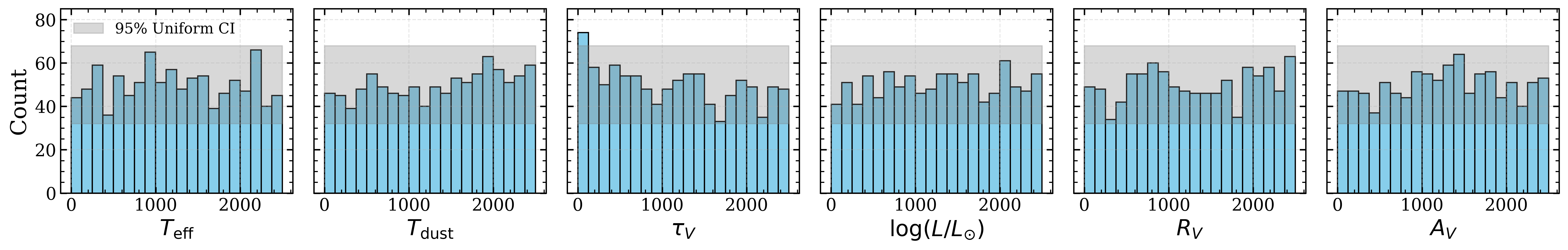}
    \caption{Simulation-Based Calibration (SBC) ranks for the SBI model of NGC\,4258 ($d=7.6\,{\rm Mpc}$). Ranks are computed as the number of sampled posterior observations that are lower than the true simulated value for each parameter and quantify the bias and variance calibration of the model. For an ideal model, we expect uniform ranks, as shaded in the $95\%$ confidence interval for a uniform distribution. We find that the ranks of our posteriors follow the expected ideal distribution.
    }\label{fig:sbc_ranks}
\end{figure*}

We compute ranks for $1000$ simulated observations, with parameters drawn from the prior (Table \ref{tab:sbi_prior}), and $2500$ posterior samples drawn for each observation. We find our SBI models to be well-calibrated using SBC, with the empirical CDF of the ranks shown in Figure \ref{fig:sbc_tarp} and histogram of the ranks shown in Figure \ref{fig:sbc_ranks}. Our posterior rank distribution falls within the expected $95\%$ credible interval for a uniform distribution.

While SBC is a good check of variance and bias calibration in the posterior marginal distributions, TARP checks calibration of the joint posterior. TARP computes the number of inferred posterior samples $\theta_*$ within a distance $r$ of the true simulated parameters $\theta_r$, at various ranks \citep{Lemos23}. The expected coverage probability should follow the distribution of the coverage level $\alpha$. TARP is a necessary and sufficient condition for posterior accuracy \citep{Lemos23} and can also characterize biased or dispersed posteriors. We use the same simulated test set as for SBC to run TARP, show a representative example of our model performance on TARP in Figure \ref{fig:sbc_tarp}, and find it to be well-calibrated. Using SBC and TARP, we show that our model can produce reliable posterior estimates and capture the uncertainty distributions of our data well. 

We also inspect the accuracy of the model inference by computing the coefficient of determination $R^2$ and the root-mean-squared error (RMSE) between the median inferred parameters and the true simulated values for each parameter, defined as

\begin{equation}
\label{equation: r2}
    R^2 = 1 - \frac{\sum_{n=1}^{N}(x_n - \hat{x}_n)^2}{\sum_{n=1}^{N} (x_n - \bar{x})^2}
\end{equation}

\begin{equation}
    \mathrm{RMSE} = \sqrt{\frac{1}{N}\sum_{n=1}^{N}\left(x_n - \hat{x}_n\right)^2}\,.
\end{equation}

We show an example of parameter recovery in Figure \ref{fig:sbi_acc}.

We note that given our NIR data, $T_{\rm dust}$ is poorly constrained, with large uncertainties in inferred values. The same is true for $R_V$, which largely returns the prior $\mathcal{U}[2, 6]$ as it has a minor impact on the SED morphology. We anticipate adding mid-infrared data from the Mid-Infrared Instrument (MIRI) on {\it JWST} wherever available, which will improve inference of $T_{\rm dust}$, and will also improve constraints on $\tau_V$ and dust composition. We marginalize over both these parameters, as well as $A_V$. Our primary parameters of interest $T_{\rm eff}$, $\log(L/L_{\odot})$, and $\tau_V$ are all recovered well, with uncertainties consistent with expectations from fitting photometric samples \citep{Beasor22}. The parameter space in our case has degeneracies between different parameters (e.g., $T_{\rm eff}$ and $A_V$; Figure \ref{fig:sbi_mcmc_compare}), which limits the precision to which we can measure these quantities. We note that our model performs very well in recovering the bolometric luminosity ($R^2 = 0.99$).

The expected metrics from each test for a well-calibrated model are as below:
\begin{enumerate}
    \item SBC tests uncertainty calibration across marginal estimates for each parameter \citep{talts20}. A well-calibrated model will produce uniform ranks for each parameter as explained in Section \ref{sec:sbi}. We evaluate the $p-$value of KS tests comparing the measured ranks to those from a uniform distribution in Table \ref{tab:sbi_diagnostics}. We expect $p-$value $>0.05$ where we cannot reject the null hypothesis that the ranks follow a uniform distribution. Models, or specific marginals, that do not meet this criterion may be over- or under-confident. We note that the majority of our physical parameter inferences have well-calibrated posterior uncertainties based on Table \ref{tab:sbi_diagnostics}.
    \item TARP checks the posterior inference jointly across all parameters, producing a single calibration curve: the expected coverage probability (ECP) as a function of credibility level $\alpha$ \citep{Lemos23}. The ECP calculates the probability that the true parameter value lies within the inferred credibility interval, and is used widely to assess the performance of machine learning models where uncertainty quantification is critical \citep{Lemos23}. Similar to SBC, we expect the ECP vs. $\alpha$ curve to follow a uniform distribution, and quantify the $p-$-value of a KS test comparing the measured distribution to a uniform distribution. We expect $p > 0.05$, and find this to be true for all models. We also report area to curve (ATC) for all models, which measures the deviation from the ideal $y=x$ curve, and find this value is close to zero for our models.
    \item Finally, we quantify the accuracy of the model by running inference on a withheld test set that the model does not access during training. The test set is simulated using the same process used to generate the training set. We measure the accuracy using $R^2$ (Equation \ref{equation: r2}) for each marginal, with the ideal value being 1. We note good accuracy for $T_{\rm eff}$, $\tau_V$, and $\log(L/L_{\odot})$, which are the main parameters of interest for source classification (Section \ref{sec:cat}), while we marginalize over the other parameters. We anticipate adding MIR data will improve estimates of $T_{\rm dust}$ and constrain $\tau_V$ further.
\end{enumerate}

\startlongtable
\begin{deluxetable*}{lcccccc|cc|cccccc}
\tablecaption{SBC, TARP, and Test Accuracy Diagnostics for the Galaxy Sample \label{tab:sbi_diagnostics}}
\tablewidth{0pt}
\tablehead{
\colhead{} & 
\multicolumn{6}{c}{SBC KS $p$} & 
\multicolumn{2}{c}{TARP} & 
\multicolumn{6}{c}{Test $R^2$} \\
\cline{2-7} \cline{7-8} \cline{9-15}
\colhead{Galaxy} & 
\colhead{$T_{\rm eff}$} & 
\colhead{$T_{\rm dust}$} & 
\colhead{$\tau_V$} &
\colhead{$\log(L/L_{\odot})$} & 
\colhead{$R_V$} & 
\colhead{$A_V$} &
\colhead{ATC} &
\colhead{KS $p$} &
\colhead{$T_{\rm eff}$} & 
\colhead{$T_{\rm dust}$} & 
\colhead{$\tau_V$} &
\colhead{$\log(L/L_{\odot})$} & 
\colhead{$R_V$} & 
\colhead{$A_V$}
}
\startdata
NGC\,628 & 0.43 & 0.43 & 0.34 & 0.33 & 0.17 & 0.20 & -0.20 & 1.00 & 0.91 & 0.37 & 0.80 & 0.99 & 0.00 & 0.57\\
NGC\,1365 & 0.21 & 0.29 & 0.05 & 0.07 & 0.19 & 0.05 & 0.12 & 0.99 & 0.89 & 0.36 & 0.83 & 0.98 & 0.01 & 0.54\\
NGC\,1637 & 0.55 & 0.51 & 0.58 & 0.96 & 0.62 & 0.38 & -0.27 & 0.99 & 0.83 & 0.27 & 0.78 & 0.98 & 0.00 & 0.31\\
NGC\,3034 & 0.13 & 0.38 & 0.27 & 0.001 & 0.87 & 0.06 & -0.09 & 1.00 & 0.80 & 0.25 & 0.83 & 0.98 & 0.00 & 0.37\\
NGC\,4038 & 0.70 & 0.54 & 0.64 & 0.002 & 0.66 & 0.04 & -0.33 & 0.99 & 0.77 & 0.21 & 0.79 & 0.98 & 0.0 & 0.45\\
NGC\,4258 & 0.39 & 0.02 & 0.03 & 0.43 & 0.61 & 0.42 & 0.24 & 1.00 & 0.93 & 0.47 & 0.87 & 0.99 & 0.04 & 0.70\\
NGC\,4449 & 0.02 & 0.01 & 0.82 & 0.01 & 0.34 & 0.58 & 0.06 & 0.99 & 0.94 & 0.44 & 0.88 & 0.99 & 0.00 & 0.64\\
NGC\,4485 & 0.80 & 0.96 & 0.80 & 0.61 & 0.43 & 0.38 & 0.17 & 1.00 & 0.92 & 0.38 & 0.82 & 0.99 & 0.00 & 0.57\\
NGC\,4536 & 0.08 & 0.92 & 0.91 & 0.64 & 0.42 & 0.79 & -0.06 & 1.00 & 0.86 & 0.07 & 0.77 & 0.97 & 0.00 & 0.36\\
NGC\,4548 & 0.92 & 0.55 & 0.10 & 0.23 & 0.76 & 0.06 & -0.05 & 1.00 & 0.89 & 0.35 & 0.85 & 0.98 & 0.00 & 0.41\\
NGC\,5194 & 0.03 & 0.29 & 0.96 & 0.08 & 0.38 & 0.31 & 0.26 & 0.99 & 0.92 & 0.37 & 0.85 & 0.99 & 0.00 & 0.53\\
NGC\,5236 & 0.38 & 0.58 & 0.40 & 0.01 & 0.02 & 0.01 & 0.18 & 1.00 & 0.91 & 0.39 & 0.87 & 0.99 & 0.00 & 0.58\\
NGC\,5457 & 0.37 & 0.24 & 0.38 & 0.33 & 0.45 & 0.90 & -0.01 & 1.00 & 0.95 & 0.41 & 0.85 & 0.99 & 0.00 & 0.76\\
NGC\,5643 & 0.99 & 0.29 & 0.45 & 0.30 & 0.24 & 0.24 & 0.12 & 1.00 & 0.86 & 0.30 & 0.82 & 0.98 & 0.00 & 0.53\\
\enddata
\end{deluxetable*}

\section{CMD Selection of Seed Stellar Populations and KDE Construction}
\label{appendix: seed}

\begin{figure*}[thp]
\centering
\includegraphics[width=0.49\textwidth]{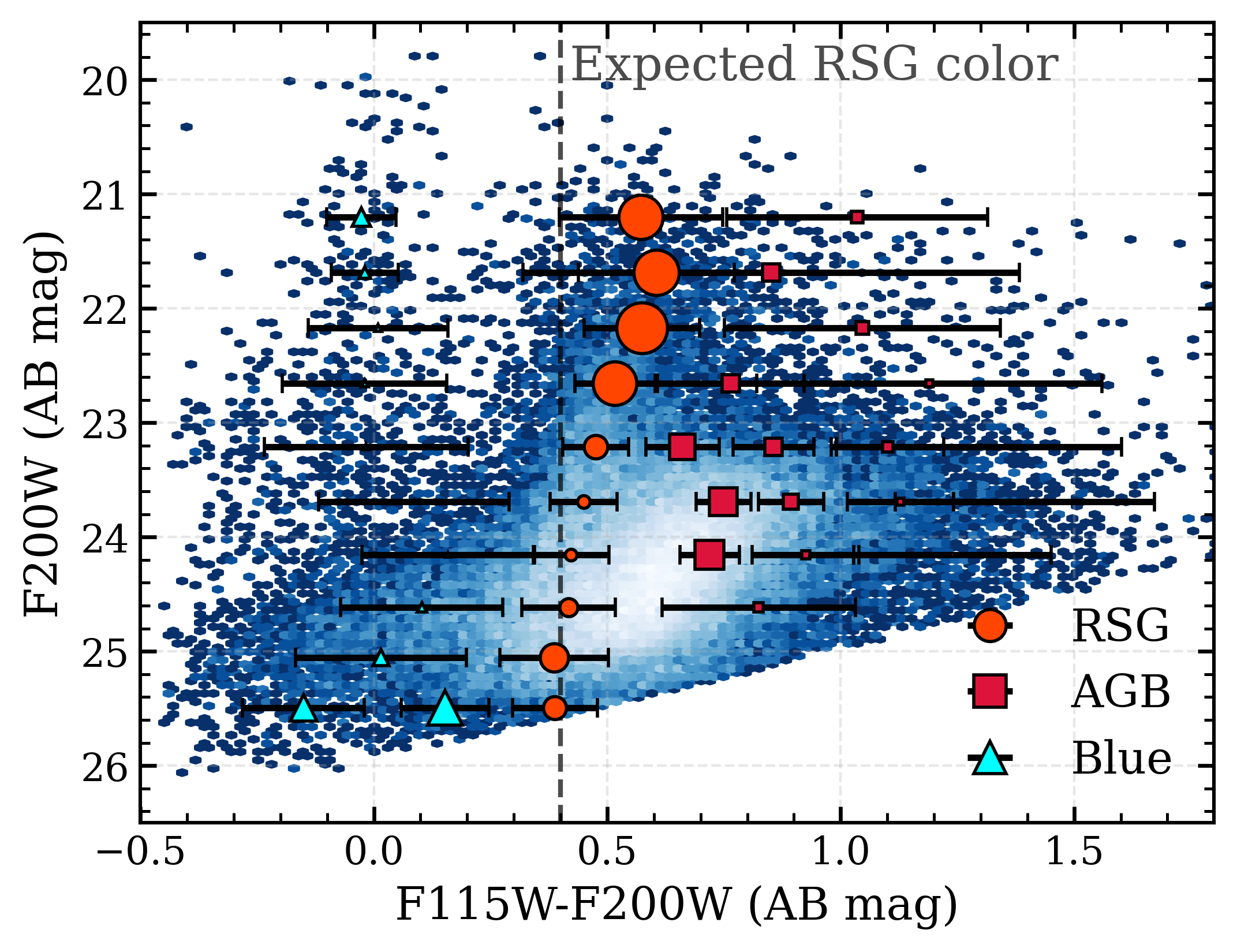}
\includegraphics[width=0.49\textwidth]{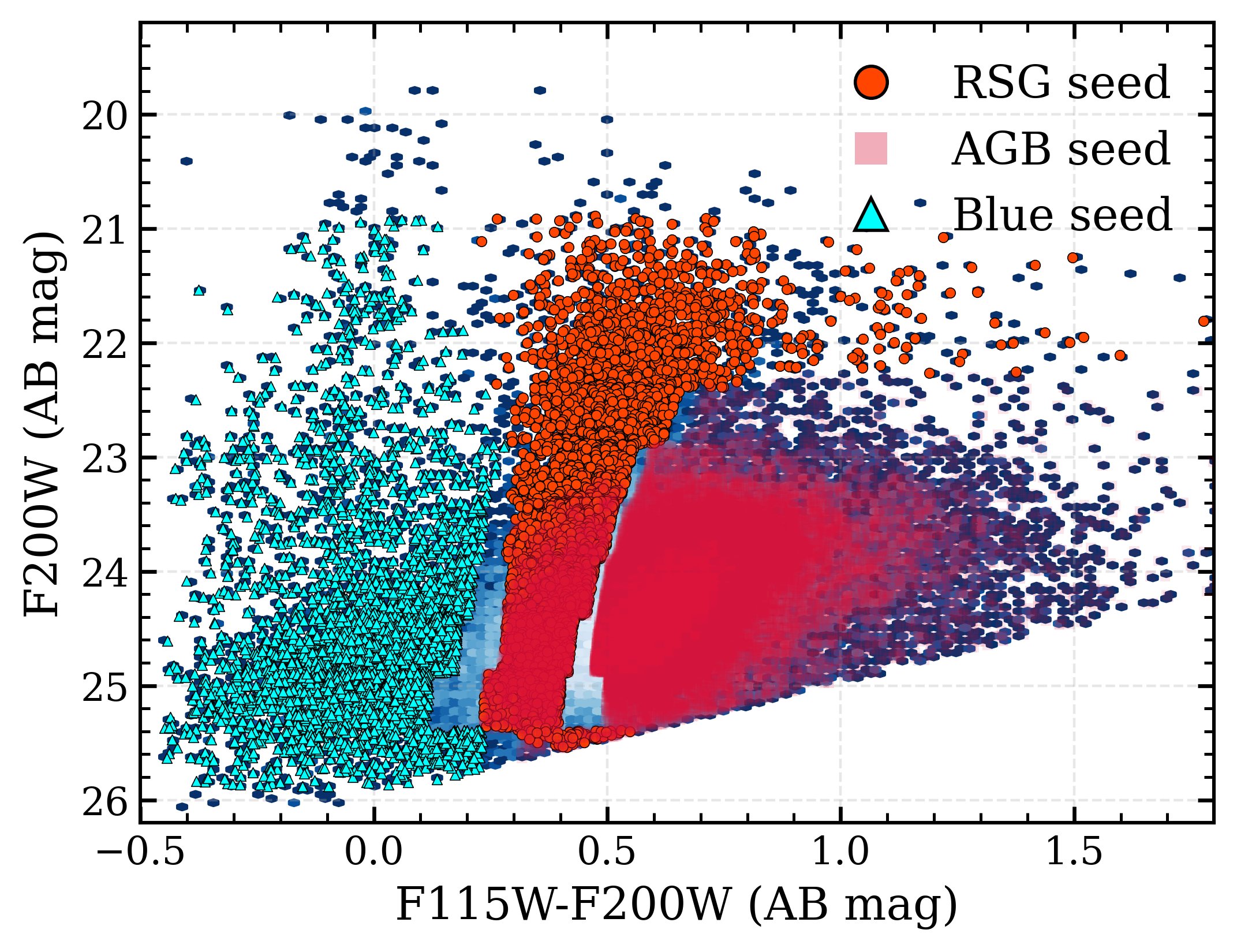}
\caption{Left: We show the split across the RSG, AGB, and blue-sequence stars by the GMM clustering applied to NGC\,5643. The CMD shows all the luminous stars, and we plot the RSG clusters in red, AGB clusters in crimson, and blue-sequence clusters in blue, with the marker size representing the relative weight of each class. The GMM reproduces the expected RSG locus in each magnitude bin, along the expected RSG color (grey dashed line). Right: We correct for AGB contamination at low luminosities using $\tau_V$ to remove very dusty sources inconsistent with being RSGs. We plot the final seed sample for NGC\,5643, colored blue, orange, and red for blue-sequence stars, RSGs, and AGBs, respectively.
}\label{fig:gmm_seed}
\end{figure*}

A priori, the region of the physical parameter phase space ($T_{\rm eff}, \log(L/L_{\odot}), \tau_V$) occupied by RSGs is not well-defined. The effective temperatures of RSGs vary with metallicity, ranging from a median temperature of $\til3600\,{\rm K}$ at solar metallicity, to $\til4000\,{\rm K}$ at LMC/SMC metallicities and as high as $\til4500\,{\rm K}$ at $[Z] = -1.0$ \citep{Gonzalez-Tora21, Davies13}. However, the distribution function of $T_{\rm eff}$ is poorly understood owing to the small sample sizes of measured $T_{\rm eff}$ so far and the small number of host environments/metallicities. The RSG luminosity function is of great importance for the progenitor properties of Type-IIP SNe and has been examined in some detail in small samples \citep{Neugent20, Strotjohann24}. For RSGs identified using NIR photometry or the CMD, this selection method adds a bias to the RSG luminosity function at low luminosities ($\log(L/L_{\odot})$ < 4.0), where it is largely unexplored due to high contamination from AGB stars. Broadly, the RSG luminosity function is expected to peak around $\log(L/L_{\odot})$ \til 4.0 and extend to the HD luminosity limit $\log(L/L_{\odot})$ \til 5.5, which is also not well constrained observationally \citep{Humphreys79, McDonald22}. The dust optical depth $\tau_V$, which can be directly translated into the RSG mass-loss rate, forms the mass-loss relation when combined with $\log(L/L_{\odot})$. RSG mass loss is a highly debated quantity in stellar astrophysics, and several prescriptions for the mass-loss--luminosity relation exist (Section \ref{sec:discussion}). Nevertheless, similar to $T_{\rm eff}$ and $\log(L/L_{\odot})$, the high uncertainty in $\tau_V$ (or its proxy $\dot{M}$) precludes defining a region of parameter space characterizing RSGs across varied environments.

\begin{deluxetable*}{lcccc}
\tablecaption{Summary of Seed Selection Inputs}
\tablewidth{0pt}
\tablehead{
\colhead{Galaxy} & 
\colhead{Red Filter} & 
\colhead{Blue Filter} & 
\colhead{$\alpha$} &
\colhead{Expected RSG Color} 
}
\startdata
NGC\,5236 & F115W & F200W & 0.5 & 0.4 \\
NGC\,5194 & F115W & F200W & 0.5 & 0.4 \\
NGC\,4258 & F115W & F210M & 0.5 & 0.2 \\
NGC\,5643 & F115W & F200W & 0.5 & 0.4 \\
NGC\,628 & F115W & F200W & 0.5 & 0.25 \\
NGC\,1637$^1$ & F150W & F200W & 0.5 & -0.06 \\
NGC\,1365$^2$ & F115W & F200W & 0.3 & 0.4\\
NGC\,4536 & F150W & F200W & 0.5 & 0.0 \\
NGC\,5457 & F115W & F200W & 0.5 & 0.4 \\
NGC\,4449 & F115W & F200W & 0.3 & 0.3 \\
NGC\,4485 & F115W & F200W & 0.3 & 0.3  \\
NGC\,7320$^3$ & F090W & F200W & 0.5 & 0.9  \\
NGC\,4548$^4$ & F150W & F200W & 0.3 & 0.1 \\
NGC\,4038$^5$ & F115W & F150W & 0.5 & 0.4 \\
\enddata
\tablecomments{The expected RSG color corresponds to the red filter - blue filter color (e.g., F115W - F200W = 0.4 for NGC\,5236) \\
$^1$ GMM probability threshold 0.5 for all classes; minimum class weight of 0.1 \\
$^2$ GMM probability threshold 0.5 for all classes \\
$^3$ Only stars with $21 < {\rm F200W} < 26$ are used in seed selection due to a brighter blue-sequence star population; minimum of 6 GMM components \\
$^4$ Only stars with ${\rm F200W} < 24.5$ are used in seed selection due to heavy AGB contamination in lower luminosities; minimum class weight of 0.1 \\
$^5$ Stars with ${\rm F150W} > 25$ are classified as AGBs in the seed due to heavy AGB contamination in lower luminosities
}\label{tab:seed_filts}
\end{deluxetable*}

We require three clusters of luminous stars per galaxy, one for each of RSGs, AGBs, and blue-sequence stars, which includes YSGs/BGSs/foreground stars.AGBs and RSGs form a nearly continuous density in the 3D $T_{\rm eff} - \log(L/L_{\odot}) - \tau_V$ phase space, while the blue-sequence stars are relatively well separated based on temperature, with a small contamination from RSGs. To distinguish these classes with good accuracy, we utilize clustering in the CMDs to select a seed sample acting as a ``prior'' for the physical parameters characterizing each stellar class. We expect the CMD-selected samples to be contaminated \citep{Boyer11}. Hence, we combine the samples selected from the CMD and apply simple cuts based on physical parameters to maximize purity in the seed sample. If the seed sample is reasonably representative of the underlying stellar population, our downstream KDE analysis using this seed sample will reassign classes for contaminants, although a higher-quality seed ensures more realistic classification probabilities.

\begin{figure}[htp]
    \centering
    \includegraphics[width=0.45\textwidth]{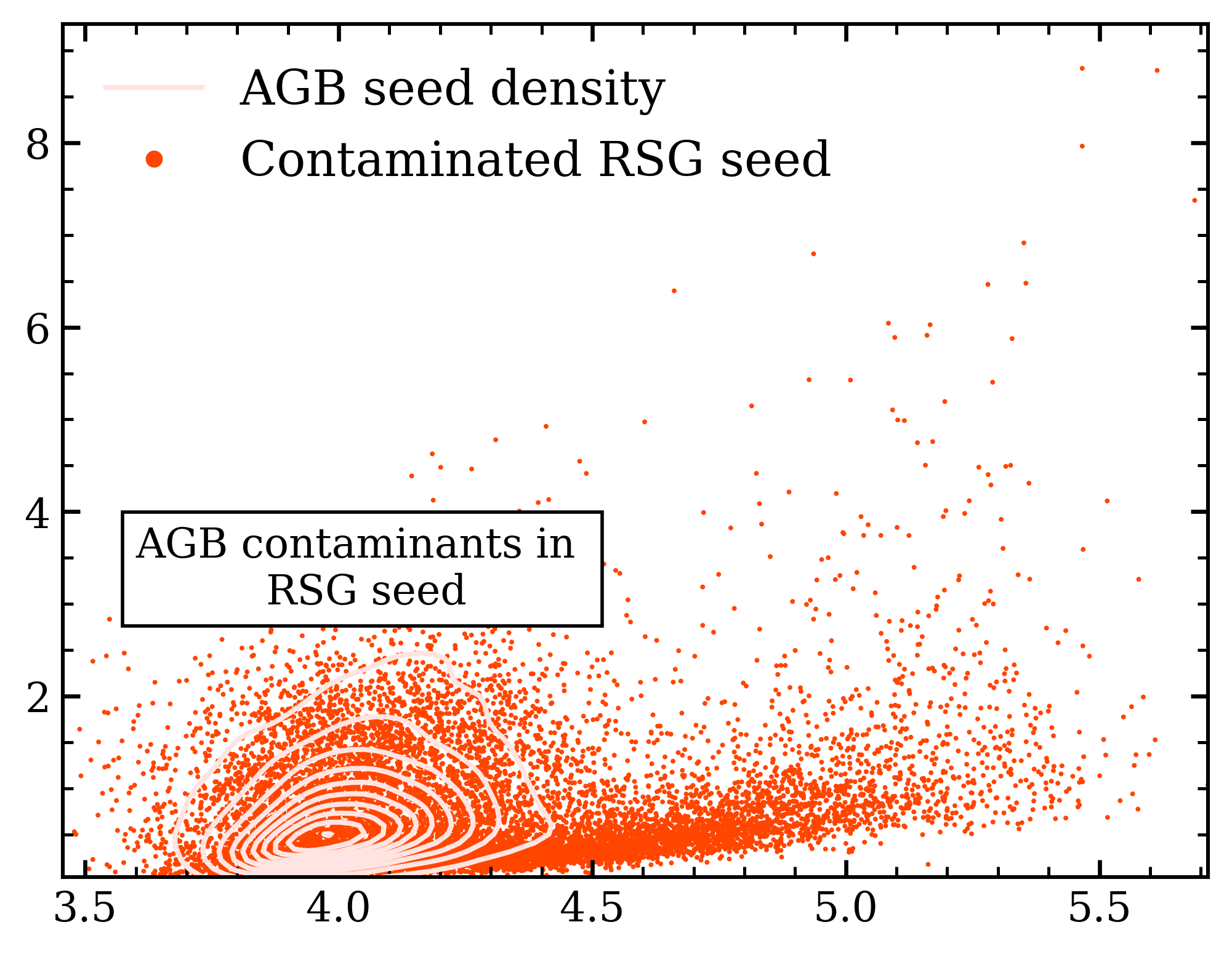}
    \caption{Contamination in the CMD from AGB stars. We find an excess of very dusty sources at low luminosities in the original RSG seed (orange circles), which is inconsistent with RSG evolution models. However, we find these stars fall right in the highest-density region of AGB physical parameter phase space (off-white contour). Thus, we classify such sources as AGBs in the seed sample.
    }\label{fig:lowL_agb_seed}
\end{figure}

To minimize contamination in the seed sample, we fit a Gaussian Mixture Model (GMM) to color (e.g., F115W-F200W or F150W-F200W) across discrete magnitude bins to identify the RSG branch in the CMD. We use this branch to assign redder stars to the AGB class and the bluer stars to the blue-sequence class. A similar analysis has been done in \citep{Hirschauer20}, which used KDEs in magnitude bins to define the boundaries between RSGs and AGBs. GMMs are a versatile extension to this technique, offering improved capabilities in disentangling mixed populations as we have here. GMMs require the number of Gaussian components used in the fit as an input. In our case, this is often more than three, as the AGB population is not a clean Gaussian and often requires multiple sub-components. In each magnitude bin, we fit a set of models with the number of components ranging from three to seven and pick the best fit based on the Bayesian Information Criterion. We find that the RSG branch is typically identified in each bin. Since the number of Gaussian components can vary per bin, we set an expected RSG color based on the CMD and pick the GMM mean closest to this value as the RSG sample (Figure \ref{fig:gmm_seed}). We note, however, that to reliably identify the various stellar classes, each bin needs a sufficient sample size of each class, especially of the sparse blue-sequence class. This necessitates picking a relatively large value for the bin spacing, at $\Delta{m} = 0.5\,{\rm mag}$.

The RSG branch in CMDs is nearly vertical but sloped. Since we require a large $\Delta{m}$, we account for this slope by fitting a quadratic polynomial to the GMM mean color of the RSG seed and detrending the entire CMD using this polynomial, such that the RSG branch is now vertical with an expected color of zero. We run the GMM selection algorithm two more times to converge, and translate the classifications back to the original colors, with the seed sample now reflecting the slope in the stellar population densities. We assign all red stars with $\log(L/L_{\odot}) > 5$ as RSGs, as this is above the maximum expected AGB luminosity, and ensure that stars bluer than $\mu_{RSG} - 3\sigma_{RSG}$ are assigned to the blue-sequence class in each bin, in case the GMM fails to account for the sparse blue-sequence population. Since the stellar populations overlap in regions of the CMD, we apply a probability cut using the GMM classifications, requiring $p \geq 0.75$ for RSGs and blue-sequence seeds and $p \geq 0.5$ for the AGB seed. 

Finally, we minimize contamination in the RSG seed by restricting $T_{\rm eff}$ to be within $3200\,{\rm K}$ to $4500\,{\rm K}$. On inspecting the seed, we find a population of stars with $\log(L/L_{\odot}) < 4.5$ with high optical depths $\tau_V > 0.5$ classified as RSGs based on their location in the CMD. However, we note that they fall cleanly within the AGB phase space in physical parameters and form a distinct group compared to the broader RSG population (Figure \ref{fig:lowL_agb_seed}). We change the seed classification of these stars from RSG to AGB, noting that these are very likely sources of contamination in the CMD-based RSG selection criteria. We show an example of our final selected seed population in Figure \ref{fig:gmm_seed}. 

We use a Gaussian KDE with the seed sample to build separate PDFs for the RSG, AGB, and hot blue-sequence classes, directly in the 3D parameter phase space. Beyond 1D, bandwidth estimation becomes an important factor for KDE performance. The bandwidth dictates the scale of variance of the PDF and controls the smoothness of the underlying density. Due to the sparsity of our data in certain regions of parameter space (e.g., at luminosities $\log(L/L_{\odot}) > 5.5$; Figure \ref{fig:kde_phase}), we use Abramson scaling \citep{Abramson1982} to reduce bias in the tail of the PDF and represent densities as accurately as possible in sparse regions. We implement this by weighting the seed sample points using

\begin{equation}
w_i = \left(\frac{\rho_i}{\bar{\rho}}\right)^{-\alpha}
\end{equation}

where $\rho_i$ is the initial density of points estimated using a KDE with its bandwidth set by Scott's rule \citep[default in \texttt{SciPy};][]{Scott1979}, $\bar{\rho}$ is the average density of the population, and $\alpha$ is a smoothing factor, typically set to $0.5$. Larger values of $\alpha$ lower the probability at the intersection of stellar populations by increasing the density at the edges, while lower values of $\alpha$ tend to overfit the seed sample parameters, making the classifier brittle to sources not in the seed sample. We run this process independently for each stellar class and calculate the average cluster density per class, resulting in one three-dimensional PDF across $T_{\rm eff}$, $\log(L/L_{\odot})$, and $\tau_{V}$ per galaxy, computed using the weighted ``adaptive'' KDEs. We scale $T_{\rm eff},\,\log(L/L_{\odot}),$ and $\tau_V$ to have zero mean and unit variance as $z = (x - \mu)/\sigma$ before passing them as an input to the KDE. The mean and standard deviation are calculated for the sources in the seed sample, and the same values are used to scale the remaining sources.

\bibliographystyle{apj}
\bibliography{references}

\label{lastpage}
\end{document}